\documentclass[a4paper,10pt]{article}
\usepackage[utf8]{inputenc}
\usepackage{amsthm}
\usepackage{amsfonts}
\usepackage{amssymb}	
\usepackage{amsmath}
\usepackage{bbold}
\allowdisplaybreaks
\usepackage{mathtools}
\usepackage[english]{babel}
\usepackage{color}
\usepackage{slashed}
\usepackage{enumerate}
\usepackage{graphicx}
\usepackage{systeme}
\usepackage{dsfont}
\usepackage{relsize}
\usepackage[hidelinks]{hyperref}
\usepackage[dvipsnames]{xcolor}

\usepackage{float}
\usepackage{tikz}
\newcommand\Alpha{\mathrm{A}}
\newcommand\Beta{\mathrm{B}}
\usepackage[labelformat=simple]{subcaption}

\usepackage{graphicx}
\usepackage{bmpsize}
\usepackage{epstopdf}
\usepackage{bbm}
\usepackage{xcolor,etoolbox}
\usepackage{titling}
\thanksmarkseries{arabic} 
\usepackage{authblk}
\usepackage{blindtext,graphicx}
\usepackage[absolute]{textpos}
\usepackage[hang,flushmargin]{footmisc}

\usepackage{xfrac}
\usepackage{nicefrac}
\usepackage{esvect}
\usepackage{cases}
\usepackage{empheq}
\usepackage{stmaryrd}
\usepackage{cancel}
\usepackage{url}
\usepackage[font=small]{caption}
\usepackage{multirow}
\usepackage[section]{placeins}
\usepackage[percent]{overpic}
\usepackage{tikz}
\usetikzlibrary{calc}

\usepackage[margin=0.90in]{geometry}
\usepackage{setspace}
\newcommand{\comsol}{\textit{Comsol Multiphysics}\textsuperscript{\textregistered}}

\renewcommand{\.}{\hspace*{0.07em}}
\DeclareMathOperator{\sym}{sym}
\DeclareMathOperator{\Curl}{Curl}
\DeclareMathOperator{\dev}{dev}
\DeclareMathOperator{\Div}{Div}

\renewcommand{\skew}{\mathop{\mathrm{skew}}\nolimits}
\newcommand{\tr}{\mathop{\mathrm{tr}}\nolimits}
\newcommand{\abs}[1]{\lvert #1 \rvert}

\DeclareMathAlphabet{\mathbbold}{U}{bbold}{m}{n}
\DeclareMathOperator{\avg}{avg}
\newcommand{\id}{\mathbbold{1}}
\newcommand{\rev}{\textcolor{black}}

\graphicspath{{images/}}

\usepackage[giveninits=true,sortcites=false,date=year,maxbibnames=99,doi=false,isbn=false,url=false,eprint=false]{biblatex}
\renewbibmacro{in:}{}
\DeclareFieldFormat{pages}{#1}
\AtEveryBibitem{\clearlist{language}}

\usepackage{csquotes}

\usepackage{setspace}
\title{Effective interface forces to model boundary effects in a finite-size metamaterial through the reduced relaxed micromorphic model}
\newcommand*\samethanks[1][\value{footnote}]{\footnotemark[#1]}
\author{
Plastiras~Demetriou\thanks{Institute for Structural Mechanics and Dynamics, Technical University Dortmund, August-Schmidt-Str.~8, 44227 Dortmund, Germany. Corresponding author's email: plastiras.demetriou@tu-dortmund.de},
\quad Jendrik~Voss\samethanks[1]
\quad and
\quad Angela Madeo\samethanks[1]
}

\thanksmarkseries{arabic}
\date{\today}
\begin{document}
\maketitle
\begin{abstract}
\noindent
We use the reduced relaxed micromorphic model (RRMM) to capture the effective ``bulk" dynamical response of finite size metamaterial specimens made out of a Labyrinthine unit cell.
We show that for small finite-size specimens, boundary effects can play a major role, so that the RRMM needs an enrichment to capture the metamaterial's bulk response, as well as the boundary effects.
A benchmark test is introduced to show that different metamaterial/ homogeneous material interfaces can drive completely different responses even if the bulk metamaterial remains the same.
We show with no remaining doubts that the concept of ``interface forces" must necessarily be introduced if one wants to model finite-size metamaterials in a homogenized framework.
\end{abstract}\textbf{Keywords}: metamaterials, metastructure, generalized continua, wave propagation, enriched continua, relaxed micromorphic model, dispersion curves, boundary effects, interface forces, unit cell cuts,  finite-size specimen \\[.65em]

{\parskip=1mm\tableofcontents}

\section*{Introduction}
In the last decades, mechanical metamaterials have gathered significant attention, owing to their exotic capabilities regarding mechanical wave propagation.
Metamaterials are materials with a microstructure specifically tailored to grant them unique properties, such as a negative Poisson’s ratio \cite{lakes1987foam,mousanezhad2015hierarchical,d20183d,}, negative refraction \cite{willis2016negative,bordiga2019prestress,zhu2015study,srivastava2016metamaterial,lustig2019anomalous,morini2019negative}, chiral effects \cite{movchan2022waves,garau2018interfacial,frenzel2017three,rizzi2019identification1,rizzi2019identification2}, band-gaps \cite{liu2000locally,wang2014harnessing,bilal2018architected,celli2019bandgap}, cloaking \cite{buckmann2015mechanical,misseroni2016cymatics,norris2014active,misseroni2019omnidirectional}, and others \cite{carta2019flexural,frecentese2019dispersion,movchan2022wave,movchan2022wave2,lakes2023experimental,giorgio2017continuum,zheng2022modeling}.

The potential applications of mechanical metamaterials are promising, specifically in regards to acoustic isolation and passive vibration control.
While significant progress has been made in understanding, designing, manufacturing, and modeling such materials, open problems still exist.
One of the key problems, is modeling metamaterials at large scales, which is crucial for real-world engineering applications.
This problem implies the need for a homogenized description of metamaterials.
The importance of micromorphic models for modeling metamaterials and heterogeneous media though a homogenized description has been acknowledged through the development of homogenization techniques in the quasi-static regime \cite{alavi2021construction,alavi2022chiral,alavi2022continualization,skatulla2021local,van2020newton,rokovs2020extended,sridhar2020frequency,rokovs2019micromorphic,liu2021computational,bensoussan2011asymptotic,sanchez1980non,allaire1992homogenization,milton2002theory,hashin1963variational,willis1977bounds,pideri1997second,bouchitte2002homogenization,camar2003determination,suquet1985elements,miehe1999computational,geers2010multi}
as well as, more recently, in the dynamic regime~\cite{hill1963elastic,bacigalupo2014second,chen2001dispersive,boutin2014large,craster2010high,andrianov2008higher,hu2017nonlocal,willis2009exact,willis2011effective,willis2012construction,srivastava2014limit,sridhar2018general,srivastava2017evanescent,schwan2021extended}.
In recent papers~\cite{aivaliotis2020frequency,demore2022unfolding,ramirez2023multi,voss2023modeling,rizzi2022boundary,rizzi2021exploring,rizzi2022metamaterial,rizzi2022towards} we have shown that the reduced relaxed micromorphic model (RRMM) performs well in describing the response of infinite size metamaterials and also for some simple finite size problems.

We have shown in~\cite{demetriou2024reduced}, that the RRMM captured adequately the response of two finite-size metamaterials constructed from two different tetragonal unit cell ``cuts" except close to the boundary, where ``cut" related boundary effects took place.
In the same paper, we argued that enriching the boundary conditions of the model would be a possible way to capture these boundary effects.
This has been proven in~\cite{ramirez2024effective}, where the enrichment of boundary conditions was introduced, through the new concept of interface forces, where the authors showed that the different boundary of a metamaterial (direct consequence of the choice of unit cell ``cut") must lead to the activation of a different surface force in the homogenized framework, if one wants to capture boundary effects using an effective ``homogenized" model.
This method, inspired by the elastic interface model, is the first application of an interface model on macroscopic boundaries, that addresses and solves the problem of boundary effects arising in metamaterials of finite size.

In this work, we study the response of four finite-size specimens constructed from four different ``cuts".
We capture their behavior using the RRMM coupled with the new concept of interface forces at metamaterial's boundaries.
We show what ansatz these forces should follow, for the particular test under consideration, and how this ansatz scales when the magnitude of the excitation force is scaled.

\underline{In particular:}

In \textbf{Section}~\ref{sec:cut_imp} we stress the importance of the choice of the metamaterial's unit cell ``cut" for finite size applications and we introduce the four ``cuts" that will be used in the benchmark tests in this work.

In \textbf{Section}~\ref{sec:model} the reduced relaxed micromorphic model (RRMM) is introduced, which is the enriched model we use in this work to model metamaterial specimens in a homogenized framework. We present the equilibrium equations and boundary conditions derived using variational principles and we recall the parameter identification procedure used to fit the parameters of the RRMM for the specific metamaterial.

In \textbf{Section}~\ref{Sec:FE_S} we explain in detail both the full microstructured simulations and those implemented by modeling the metamaterial with the RRMM coupled with interface forces and show explicitly the boundaries where the activation of an interface/surface force can occur.

In \textbf{Section}~\ref{sec:material_interfaces} we introduce the method of interface forces in a reduced relaxed micromorphic (RRM) framework, stressing the fact that it is essentially the first application of an interface model on macroscopic metamaterial interfaces and also (to our knowledge) the first attempt to tackle the problem of modeling boundary effects in metamaterials using a homogenized model, where boundary effects arise from the choice of unit cell ``cut" in the finite-size metamaterial.

In \textbf{Section}~\ref{sec:sim_results} we introduce a specific ansatz for our interface forces, and show that this ansatz holds for the frequencies under consideration except for a small frequency range in which the corresponding wavelengths are comparable to the specimen's size.
We show the results of our simulations using vanishing and non-vanishing interface forces, for the RRMM and for a Cauchy material which is the long-wavelength limit of the RRMM.


In \textbf{Section}~\ref{sec:transm} we present the transmissibility plot for the four finite-sized metamaterials under consideration to stress once again the importance of the choice of unit cell ``cut" and we show a particular case where a region of low transmissibilty appears outside of a band-gap, owing to the finite size and the choice of unit cell ``cut". 
Lastly, we present some independent tests with increasingly bigger metamaterial specimens using the same four unit cell ``cuts", in order to stress that the infinite metamaterial assumption of Bloch-Floquet analysis has limits. We explain that this assumption can only hold when boundary effects can be neglected for any possible unit cell ``cut" for the specfic test under consideration.
%
%
%
\section{\label{sec:cut_imp}Importance of the choice of the metamaterial's unit cell ``cut" for finite-sized problems}
\begin{figure}[h!]
\centering
\includegraphics[width=1\textwidth]{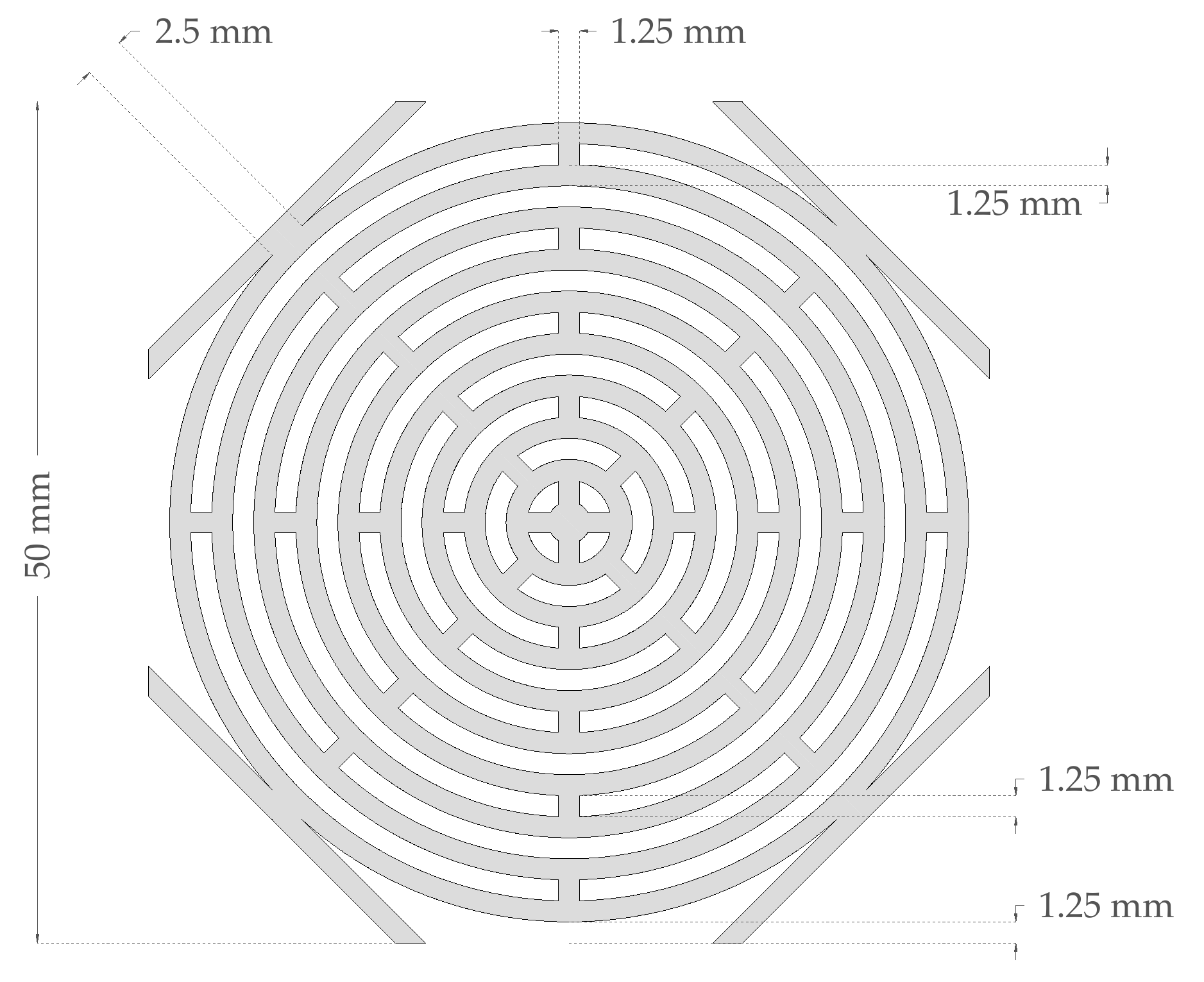}
\\*
\renewcommand{\arraystretch}{1.35}
\centering
\begin{tabular}{cccccc}
\hline
$L_{\rm c}$ [m]& $\rho_{\rm poly}$  [kg/m$^3$] & $E_{\rm poly}$ [Pa] & $\nu_{\rm poly}$ [-]
\\
\hline
0.05 & 1220 & 1.45$\times$10$^9$ & 0.4
\\
\hline
\end{tabular}
\caption{Unit cell and material properties of the metamaterial studied in this paper. The base material is Polyethylene currently used in 3D printing.}
\label{fig:dimensions}
\end{figure}

We start with presenting the geometrical and material characteristics of the metamaterial under study and whose infinite-size dynamical response will be described through the reduced relaxed micromorphic model.
The considered metamaterial is built via the periodic repetition in space of the Labyrinthine unit cell presented in Fig.~\ref{fig:dimensions}.

This specific unit cell presents wide band-gap behavior in the acoustic range and has already been successfully tested in experimental campaigns \cite{hermann2024design}.
An effective reduced relaxed micromorphic modeling of finite-size metamaterials stemming from this unit cell is particularly important in view of upscaling towards larger scales. 
Indeed, mastering the homogenized response of this metamaterial through the RRMM at small scales will allow the design of larger scale structures as it would not be otherwise possible via fully detailed microstructured simulations.

Metamaterial specimens based on the unit cell given in Fig.~\ref{fig:dimensions} can be manufactured using 3D printing techniques \cite{hermann2024design}.
All the geometrical and material properties needed to characterize the considered unit cell are given in Fig.~\ref{fig:dimensions}.

\begin{figure}[h!]
    \centering
    \includegraphics[width=1\textwidth]{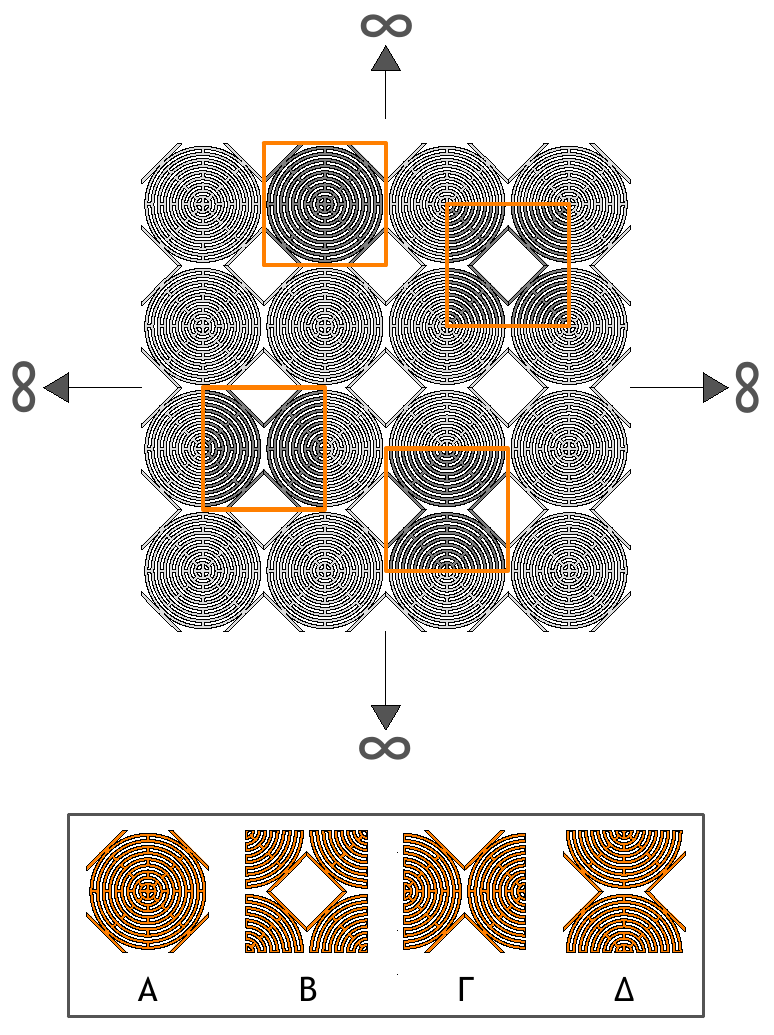}
    \caption{Infinite size periodic metamaterial (top) and the four unit cell cuts of the labyrinthine metamaterial used in our simulations (bottom). 
    Cells $\Alpha$ and $\Beta$ are of tetragonal symmetry and cells $\Gamma$ and $\Delta$ are of orthotropic symmetry.}
    \label{fig:4_cuts}
\end{figure}

The periodic repetition in space of the unit cell of Fig.~\ref{fig:dimensions}, gives rise to the infinite-size metamaterial depicted in Fig.~\ref{fig:4_cuts}.

\underline{There is one important point to make:} Although we build our infinite metamaterial from the unit cell described in Fig.\ \ref{fig:dimensions}, the base unit cell is not inclusive.
Instead, the same infinite metamaterial can be equivalently obtained by the periodic repetition in space of any of the unit-cell ``cuts" $\Alpha$, $\Beta$, $\Gamma$ and $\Delta$, as shown in Fig.~\ref{fig:4_cuts}, as well as of infinitely many other ``cuts" (some of which are less symmetric) which are not shown in Fig.~\ref{fig:4_cuts}.
While the choice of the unit cell's ``cut" doesn't play a role for retrieving the resulting infinite-sized metamaterial and corresponding bulk behavior (Bloch-Floquet analysis), it has a huge effect as soon as metamaterials of finite size, i.e.\ with boundaries, are considered.

It has been shown in \cite{hermann2024design} that the choice of the unit cell's cut can play a major role as soon as small finite-sized metamaterial specimens are considered.
The impact of such boundary effects on finite-sized metamaterials' structures has also been validated experimentally \cite{hermann2024design}.

It is the ambition of this paper to enrich the reduced relaxed micromorphic model with suitable boundary and interface conditions to show that this model can be safely used also when boundary effects cannot be neglected due to the finite size of the considered structure.
The use and calibration of the reduced relaxed micromoprhic model on small finite-sized metamaterials' specimens is a necessary step if one wants to reach the major challenge of designing large-scale albeit finite-sized metamaterial's structures. This necessity arises from the fact that an enriched homogeneous model cannot account for different shaped boundaries on its own, i.e.\ it cannot distinguish between the different unit cell's ``cut" of the finite sized metamaterial.
As we will show in detail in the remainder of this paper, to reach this goal, the concept of ``macroscopic elastic interface" and of ``interface forces" must be introduced in the reduced relaxed micromorphic framework.
Such concepts have been introduced for the first time in \cite{ramirez2024effective} to describe, through the reduced relaxed micromorphic model, the boundary effects arising in finite-sized metamaterial's specimens based on a different unit cell with respect to the one studied in the present paper.
However, due to the novelty of the introduced concepts, a thorough study is needed to address all the challenges associated to this new idea. 

In this paper, we will unveil which type of interface forces are needed in the reduced relaxed micromoprhic framework to reproduce the dynamic behavior of four types of small metamaterials' specimens which are obtained by differently ``cutting" the base unit cell of Fig.~\ref{fig:4_cuts} which, in turn, implies that the macroscopic boundaries of the considered specimens are of different type (see Section \ref{Sec:FE_S} for more details).
We will show that if one wants to catch the different response of these four specimens (predominantly driven by boundary effects) in a reduced relaxed micromoprhic framework, then the concept of macroscopic interface forces must necessarily be introduced.

Moreover, we will start unveiling, for the first time, some general properties that these macroscopic interface forces should possess, such as scaling with respect to the intensity of the externally applied load.
\section{The relaxed micromorphic model: a reduced version for dynamics}\label{sec:model}
We introduce here the equilibrium equations, the associated boundary conditions, and the constitutive relations for a reduced version  of the relaxed micromorphic model for dynamic applications\footnote{The adjective ``relaxed" was introduced by some of the authors for the specific micromorphic-type continuum model they pioneered some years ago with the following Strain Energy density
\begin{align*}
    W (\nabla u, P, \Curl P)&=\dfrac{1}{2} \langle \mathbb{C}_{\rm e} \, \sym \left(\nabla u -  \, P \right), \sym \left(\nabla u -  \, P \right) \rangle
    + \dfrac{1}{2} \langle \mathbb{C}_{\rm c} \, \skew \left(\nabla u -  \, P \right), \skew \left(\nabla u -  \, P \right) \rangle
    + \dfrac{1}{2} \langle \mathbb{C}_{\rm micro} \, \sym   \, P,\sym   \, P \rangle\\
    &\phantom=\; + \dfrac{1}{2} \langle \mathbb{L}_{\rm s} \, \sym   \, \Curl \,P,\sym   \, \Curl \,P \rangle
    + \dfrac{1}{2} \langle \mathbb{L}_{\rm a} \, \skew   \, \Curl \,P,\skew   \, \Curl \,P \rangle.
\end{align*}
The term ``relaxed" is related to: $(i)$ the fact that the curvature term in the strain energy term of the full model is related to the Curl of the micro-distortion $P$ instead than of its entire gradient and $(ii)$ contrarily to classical Mindlin-type models, there are no mixed terms of the type $\langle \nabla u-P,\sym \,P\rangle$.}
\footnote{In the present paper the relaxed micromorphic model is used in a reduced form which considers $(i)$ the non-local term related to $\Curl P$ to be vanishing and $(ii)$ the term $\dfrac{1}{2} \langle \mathbb{T}_{\rm c} \, \skew \nabla \dot{u}, \skew \nabla \dot{u} \rangle$ to be vanishing as well.
The first simplification is related to the fact that these non-local terms play negligible role in dynamic problems\cite{ramirez2024effective,demetriou2024reduced}, while the second simplification comes form the results presented in \cite{erel2025null} that show that the dispersion curves are not affected by this term.}.
The equilibrium equations and the boundary conditions are derived with a variational approach thanks to the associated Lagrangian
\begin{equation}
\mathcal{L} (\dot{u},\nabla \dot{u}, \dot{P}, \nabla u, P)
\coloneqq
K (\dot{u},\nabla \dot{u}, \dot{P})
-
W (\nabla u, P)
\label{eq:lagrangian}
\end{equation}
where $K$ and $W$ are the kinetic and strain energy, respectively, defined as
\begin{align}
K (\dot{u},\nabla \dot{u}, \dot{P}) &\coloneqq
\dfrac{1}{2}\rho \, \langle \dot{u},\dot{u} \rangle
+ \dfrac{1}{2} \langle \mathbb{J}_{\rm m}  \, \sym  \, \dot{P}, \sym  \, \dot{P} \rangle
+ \dfrac{1}{2} \langle \mathbb{J}_{\rm c} \, \skew  \, \dot{P}, \skew  \, \dot{P} \rangle
\label{eq:Ene_kine}
\\
&\phantom=\;
+ \dfrac{1}{2} \langle \mathbb{T}_{\rm e} \, \sym \nabla \dot{u}, \sym \nabla \dot{u} \rangle
\,,
\notag
\\
W (\nabla u, P) &\coloneqq
\dfrac{1}{2} \langle \mathbb{C}_{\rm e} \, \sym \left(\nabla u -  \, P \right), \sym \left(\nabla u -  \, P \right) \rangle
+ \dfrac{1}{2} \langle \mathbb{C}_{\rm c} \, \skew \left(\nabla u -  \, P \right), \skew \left(\nabla u -  \, P \right) \rangle
\label{eq:Ene_strain}
\\
&\phantom=\;
+ \dfrac{1}{2} \langle \mathbb{C}_{\rm micro} \, \sym   \, P,\sym   \, P \rangle
\notag
\end{align}
where $\langle\cdot , \cdot\rangle$ denote the scalar product, the dot represents a derivative with respect to time, $u \in \mathbb{R}^{3}$ is the macroscopic displacement field, $P \in \mathbb{R}^{3\times 3}$ is the non-symmetric micro-distortion tensor, $\rho$ is the macroscopic apparent density, $\mathbb{J}_{\rm m}$, $\mathbb{J}_{\rm c}$, $\mathbb{T}_{\rm e}$ are 4th order micro-inertia tensors, and $\mathbb{C}_{\rm e}$, $\mathbb{C}_{\rm m}$, $\mathbb{C}_{\rm c}$ are 4th order elasticity tensors (see \cite{rizzi2022boundary,voss2023modeling} for more details).

In particular, we report here the structure of the micro-inertia and the elasticity tensors for the tetragonal class of symmetry and in Voigt notation\footnote{\rev{The dimensions of the matrix representation of elastic and micro-inertia tensors are ${(\mathbb{C}_{\rm e},\mathbb{C}_{\rm micro},\mathbb{J}_{\rm m},\mathbb{T}_{\rm e},\mathbb{L}_{\rm s},\mathbb{M}_{\rm s})\in \mathbb{R}^{6\times 6}}$ and ${(\mathbb{C}_{\rm c},\mathbb{J}_{\rm c},\mathbb{T}_{\rm c},\mathbb{L}_{\rm a},\mathbb{M}_{\rm a})\in \mathbb{R}^{3\times 3}}$.}}
, \rev{where only the parameters involved under the plane-strain hypothesis are explicitly presented. Thereby, the symbol $\star$ indicates that the specific entry do not intervene under the plane-strain hypothesis.
The class of symmetry has been chosen accordingly with the symmetry of the unit cell presented in Figure~\ref{fig:dimensions}, under the assumption that the same class of symmetry applies both at the micro- and the macro-scale.}
\begin{align}
\mathbb{C}_{\rm e}
&=
\begin{pmatrix}
\kappa_{\rm e} + \mu_{\rm e}	& \kappa_{\rm e} - \mu_{\rm e}				& \star & \dots	& 0\\ 
\kappa_{\rm e} - \mu_{\rm e}	& \kappa_{\rm e} + \mu_{\rm e} & \star & \dots & 0\\
\star & \star & \star & \dots & 0\\
\vdots & \vdots	& \vdots & \ddots &\\ 
0 & 0 & 0 & & \mu_{\rm e}^{*}
\end{pmatrix},
\qquad
\mathbb{C}_{\rm micro}
=
\begin{pmatrix}
\kappa_{\rm m} + \mu_{\rm m}	& \kappa_{\rm m} - \mu_{\rm m}				& \star & \dots	& 0\\ 
\kappa_{\rm m} - \mu_{\rm m}	& \kappa_{\rm m} + \mu_{\rm m} & \star & \dots & 0\\
\star & \star & \star & \dots & 0\\
\vdots & \vdots	& \vdots & \ddots &\\ 
0 & 0 & 0 & & \mu_{\rm m}^{*}
\end{pmatrix},
\notag
\\*[5pt]
\mathbb{J}_{\rm m}
&=
\rho L_{\rm c}^2
\begin{pmatrix}
\kappa_\gamma + \gamma_{1} & \kappa_\gamma - \gamma_{1} & \star & \dots & 0\\ 
\kappa_\gamma - \gamma_{1} & \kappa_\gamma + \gamma_{1} & \star & \dots & 0\\ 
\star & \star & \star & \dots & 0\\
\vdots & \vdots & \vdots & \ddots &\\ 
0 & 0 & 0 & & \gamma^{*}_{1}\\ 
\end{pmatrix},
\qquad
\mathbb{T}_{\rm e}
=
\rho L_{\rm c}^2
\begin{pmatrix}
\overline{\kappa}_{\gamma} + \overline{\gamma}_{1} & \overline{\kappa}_{\gamma} - \overline{\gamma}_{1} & \star & \dots	& 0\\ 
\overline{\kappa}_{\gamma} - \overline{\gamma}_{1} &   \overline{\kappa}_{\gamma} + \overline{\gamma}_{1} & \star & \dots & 0\\ 
\star & \star & \star & \dots & 0\\
\vdots & \vdots & \vdots & \ddots &\\ 
0 & 0 & 0 & & \overline{\gamma}^{*}_{1}
\end{pmatrix},
\label{eq:micro_ine_1}
\\*[5pt]
\noalign{\centering
$\mathbb{J}_{\rm c}
=
\rho L_{\rm c}^2
\begin{pmatrix}
\star & 0 & 0\\ 
0 & \star & 0\\ 
0 & 0 & 4\,\gamma_{2}
\end{pmatrix},
\qquad
\mathbb{C}_{\rm c}
= 
\begin{pmatrix}
\star & 0 & 0\\ 
0 & \star & 0\\ 
0 & 0 & 4\,\mu_{\rm c}
\end{pmatrix}.
$}
\notag
\end{align}
Only the in-plane components are reported since these are the only ones that play a role in the plane-strain simulations presented in the following sections.
The choice of this particular class of symmetry is related to the fact that a macroscopic block of the metamaterial considered in this paper clearly shows tetragonal symmetry (see macroscopic metamaterial block in Fig.~\ref{fig:4_cuts}).
The action functional $\mathcal{A}$ is defined as
\begin{align}
\mathcal{A}=\iint\limits_{\Omega \times \left[0,T\right]} 
\mathcal{L} \left(\dot{u},\nabla \dot{u}, \dot{P}, \nabla u, P\right)
\, \text{d}x \, \text{d}t
\label{eq:action_func}
\end{align}
and its first variation $\delta \mathcal{A}$ is taken with respect to the kinematical fields $(u,P)$.
We consider that $\delta \mathcal{A}$ is equal to the internal work $W^{\rm int}$ of the reduced relaxed micromorphic continuum ($W^{\rm int}=\delta \mathcal{A}$) and that the external work is given only by externally applied boundary forces, i.e.\ $W^{\rm ext}=\displaystyle \int\limits_{\partial\Omega} \langle f^{\rm ext},\delta u\rangle \,ds$.
The principle of virtual work $W^{\rm int}+W^{\rm ext}=0$ thus implies the strong form governing equations for the reduced relaxed micromorphic model
\begin{align}
\rho\.\ddot{u} - \Div \widehat{\sigma} = \Div \widetilde{\sigma}
\,,
&&
\overline{\sigma} = \widetilde{\sigma} - s
\label{eq:equi_equa_RRMM}
\end{align}
with
\begin{align}
\widetilde{\sigma}
&
\coloneqq
\mathbb{C}_{\rm e}\.\sym (\nabla u-P) + \mathbb{C}_{\rm c}\.\skew (\nabla u-P)
\,,
&
\widehat{\sigma}
&
\coloneqq
\mathbb{T}_{\rm e}\,\sym \nabla\ddot{u} 
\,,
\label{eq:equiSigAll}
\\*[5pt]
s
&
\coloneqq
\mathbb{C}_{\rm micro}\.\sym P
\,,
&
\overline{\sigma}
&
\coloneqq
\mathbb{J}_{\rm m}\.\sym \ddot{P} + \mathbb{J}_{\rm c}\.\skew \ddot{P}
\end{align}
as well as the Neumann boundary conditions
\begin{align}
\widetilde{t} \coloneqq \left(\widetilde{\sigma} + \widehat{\sigma} \right)n = f^{\rm ext}
\label{eq:tractions}
\end{align}
where $\widetilde{t}$ is the generalized traction, $n$ is the normal to the boundary and ${f}^{ext}$ is the externally applied boundary force per unit area.
Since we will use it in the following, we also recall here the expression of the traction for a classical \emph{isotropic Cauchy model}
\begin{align}
    t \coloneqq \sigma \. n =f^{\rm ext}
    \qquad\text{with}\qquad
    \sigma \coloneqq \kappa \. \tr \left(\sym (\nabla u) \right) \id + 2\mu \. \dev\sym\nabla u
    \label{eq:tractions_cau}
\end{align}
where $\kappa$ and $\mu$ are the classical bulk and shear moduli respectively, $\dev X = X - \frac13 \tr X$ and again $f^{\rm ext}$ is any externally applied force per unit area.
\subsection{Identification of the RRMM parameters via dispersion curves fitting}
In this section we briefly recall the reduced relaxed micromorphic parameters identification procedure that is done by the means of fitting the dispersion curves as obtained via classical Bloch-Floquet analysis.
The dispersion curves (Fig.~\ref{fig:DC_freq}) of the microstructured material can be obtained by employing a standard Bloch-Floquet analysis on any unit cell of the four in Fig.~\ref{fig:4_cuts}.
In our case the employment of standard Bloch-Floquet analysis is done using \comsol.
All four unit cells give rise to the same dispersion curves since a Bloch-Floquet analysis employs periodic boundary conditions, thus mimicking an infinite domain.
All four unit cells shown in Fig.~\ref{fig:4_cuts} are equivalent in the sense that they give rise to the same infinite metamaterial. 

In the next step, the data from Bloch-Floquet analysis (performed on any of the four cuts in Fig.~\ref{fig:4_cuts} to obtain the dispersion curves) is used to fit the bulk material parameters of the reduced relaxed micromorphic model. 
Thus these parameters correspond to the infinite metamaterial in Fig.~\ref{fig:4_cuts}. 
To perform the reduced relaxed micromorphic fitting, dispersion curves for the RRMM are obtained analytically by finding the roots of the determinant of the acoustic tensor associated to the homogeneous equilibrium equations (\ref{eq:equi_equa_RRMM}) under a plane-wave ansatz (for more details see \cite{voss2023modeling}).
The number of independent parameters in the reduced relaxed micromorphic model is 15: 8 of them can be analytically evaluated or (analytically reduced by the other remaining parameters \eqref{eq:analytical_parameters} using the expression of the cutoffs and long-wavelength limit\footnote{As proved in previous papers, when considering the full relaxed micromorphic model in the static case, rigorous homogenization formula relating the micromorphic coefficient to the macro Cauchy limit can be derived when letting the static characteristic length tend to zero (see \cite{d2020effective}).
We recall here these formulas relating the static micromorphic coefficients to the ones associated with the macroscopic Cauchy limit. The apparent density is computed through the standard procedure used to compute the density of a heterogeneous material.}, while the remaining free parameters are obtained with a quadratic error minimization procedure so that the dispersion curves issued via the reduced relaxed micromorphic model are the closest possible to those issued via Bloch-Floquet analysis (see Fig.~\ref{fig:DC_freq}).
The parameters with an analytical expression are \cite{d2020effective}
\begin{align}
\rho
&=
\rho_{\rm Ti}\,\frac{A_{\rm Ti}}{A_{\rm tot}}
\,,
&
\kappa_{\rm e}
&=
\frac{\kappa_{\rm m} \, \kappa_{\rm M}}{\kappa_{\rm m}-\kappa_{\rm M}}
\,,
&
\mu_{\rm e}
&=
\frac{\mu_{\rm m} \, \mu_{\rm M}}{\mu_{\rm m}-\mu_{\rm M}}
\,,
\notag
\\*
\mu_{\rm e}^*
&=
\frac{\mu_{\rm m}^* \, \mu_{\rm M}^*}{\mu_{\rm m}^*-\mu_{\rm M}^*}
\,,
&
\kappa_\gamma
&=
\frac{\kappa_{\rm e}+\kappa_{\rm m}}{\rho\,L_{\rm c}^2\,\omega_{\rm p}^2}
&
\gamma_1
&=
\frac{\mu_{\rm e}+\mu_{\rm m}}{\rho\,L_{\rm c}^2\,\omega_{\rm s}^2}
\,,
\label{eq:analytical_parameters}
\\*
\gamma_1^*
&=
\frac{\mu_{\rm e}^*+\mu_{\rm m}^*}{\rho\,L_{\rm c}^2\,\omega_{\rm ss}^2}
\,,
&
\gamma_2&=\frac{\mu_{\rm c}}{\rho\,L_{\rm c}^2\,\omega_{\rm r}^2}
\notag
\end{align}
where $A_{\rm Ti}$ and $A_{\rm tot}$ are respectively the area of Polyethylene and the total area (including the voids) of the unit cell, $\omega_{\rm p}$, $\omega_{\rm r}$, $\omega_{s}$, and $\omega_{ss}$ are the cut-off frequencies, namely the frequencies for a vanishing wavenumber $k=0$, and $\kappa_{\rm M}$, $\mu_{\rm M}$, and $\mu_{\rm M}^*$ are the \textit{macro-parameters}, which represent the stiffness of the microstructured material for the long-wavelength limit: i.e.\ the stiffness of the Cauchy continuum $\mathbb{C}_{\rm Macro}$ which comes from the periodic homogenization of the considered metamaterial, while $L_{\rm c}$ is the length of the side of the unit cell, here $L_{\rm c}=0.05$ m.

As already remarked, the remaining 7 free parameters $\kappa_{\rm m}$, $\mu_{\rm m}$, $\mu_{\rm m}^*$, $\mu_{\rm c}$, $\overline{\kappa}_\gamma$, $\overline{\gamma}_1$ and $\overline{\gamma}_1^*$ are obtained by minimizing the distance between the dispersion curves obtained via Bloch-Floquet analysis and the ones of the equivalent reduced relaxed micromorphic model through a fitting procedure (for more details see \cite{voss2023modeling}).
All the material parameters of the reduced relaxed micromorphic model characterising the infinitely big microstructured material in Fig.~\ref{fig:4_cuts} are summarised in Table~\ref{table:microp}, and the plots of the two sets of curves are shown in Fig.~\ref{fig:DC_freq}. The dispersion curves depicted in Fig.~\ref{fig:DC_freq} represent the first 6 modes of an infinite metamaterial constituted by a periodic repetition of any of the four unit cells given in Fig.~\ref{fig:4_cuts}: \rev{green} curves are associated to pressure modes, while red curves represent shear modes.\footnote{The band-gap region is highlighted with a light brown color. 
Clearly, the full microstructured system may exhibit an infinite number of modes when increasing frequency. On the other hand, the considered micromorphic model can only reproduce the first six of these modes. 
To introduce higher frequency modes  also in the micromorphic modeling framework, extra microstructure-related degrees of freedom should be introduced in addition to the micro-distortion tensor $P$ alone. 
However, we are interested in this paper to the lower frequency response of the considered metamaterial and, in particular, to the possibility of adequately describing dispersion and band-gaps in this frequency interval.}

\begin{table}[h!]
\renewcommand{\arraystretch}{1.35}
\centering
\begin{tabular}{cccccc}
\hline
$L_{\rm c}$ [m]& $\kappa_{\rm e}$ [Pa]& $\mu_{\rm e}$ [Pa]& $\mu_{\rm e}^{*}$ [Pa]
\\
\hline
0.05 & 2.56$\times$10$^6$ & 1.18$\times$10$^6$ & 5.38$\times$10$^5$
\\
\hline
\hline
$\mu_{\rm c}$ [Pa]& $\kappa_{\rm m}$ [Pa]& $\mu_{\rm m}$ [Pa]& $\mu_{\rm m}^{*}$ [Pa]
\\
\hline
243.1 & 5.81$\times$10$^6$ & 5.13$\times$10$^8$ & 5.13$\times$10$^8$
\\
\hline
\hline
$\kappa_{\gamma}$ [-]& $\gamma_1$ [-]& $\gamma^{*}_1$ [-]& $\gamma_2$ [-]
\\
\hline
2.68 & 165 & 164.84 & 0.001
\\
\hline
\hline
$\overline{\kappa}_{\gamma}$ [-]& $\overline{\gamma}_1$ [-]& $\overline{\gamma}^{*}_1$ [-]& $\left[-\right]$ 
\\
\hline
3.53 & 2.63 & 1.78 & -
\\
\hline
\hline
$\kappa_{\rm Macro}$ [Pa]& $\mu_{\rm Macro}$ [Pa]& $\mu^{*}_{\rm Macro}$ [Pa]& $\rho$ [kg/m$^3$]
\\
\hline
1.78.  $\times$10$^6$ & 1.17$\times$10$^6$ & 5.37$\times$10$^5$ & 540.7
\\
\hline
\end{tabular}
\caption{
Values of the elastic parameters, the micro-inertia parameters, the size of the unit cell $L_{\rm c}$, and the apparent density $\rho$ for the reduced relaxed micromorphic model calibrated on the metamaterial whose building block is any of the four unit cells in Fig.~\ref{fig:4_cuts}. In the last row, we give the associated \textit{macro-parameters}, i.e.\ the corresponding long-wavelength limit Cauchy material coefficients \cite{neff2020identification,rizzi2021exploring}.
}
\label{table:microp}
\end{table}
\section{Full-microstructured and reduced-relaxed-micromorphic finite element simulations for selected benchmark tests}\label{Sec:FE_S}
In this section, we present the 4 benchmark tests that have been chosen to unveil the importance of the concept of interface forces for the homogenized modeling of finite-size metamaterials.
We present the setting-up of the numerical simulations on a finite-size metamaterial both with a microstructured Cauchy model (full detail of the unit-cells microstructure is coded in the numerical simulations) and the reduced relaxed micromorphic model (a homogeneous domain is used in the numerical simulations in which the constitutive laws are those of the reduced relaxed micromoprhic model and are eventually enriched with the concept of interface forces). 
For both the microstructured and the micromorphic simulations, we will take advantage of the symmetry of the problem, allowing us to reduce the computational time by simulating half of the structure while applying the appropriate symmetry conditions on the cut boundaries.

\begin{figure}[h!]
\centering
\includegraphics[width=\textwidth]{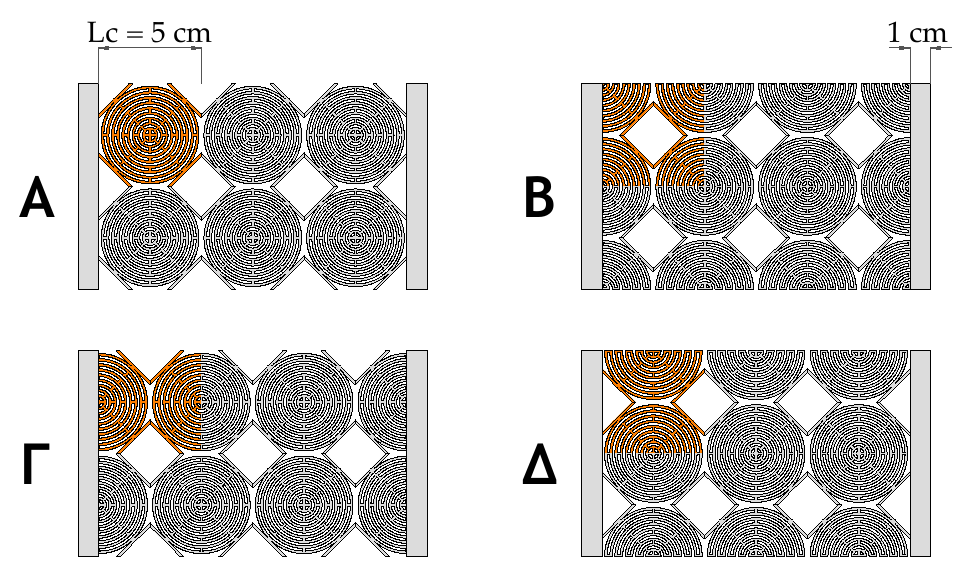}
\caption{
The four 3x2 specimens, each constructed from one of the four unit cell cuts of Fig.~\ref{fig:4_cuts}.
The specimens are embedded between two Cauchy plates made out of the same material (Polyethylene). 
For simplicity we keep the nomenclature $\Alpha, \Beta, \Gamma, \Delta$, also to indicate these different macroscopic finite-sized specimens.
}
\label{fig:4_3x2_whole}
\end{figure}
 
%
%
%
\subsection{Full-Microstructured simulations set-up}
\begin{figure}[h!]
\centering
\includegraphics[width=0.99\textwidth]{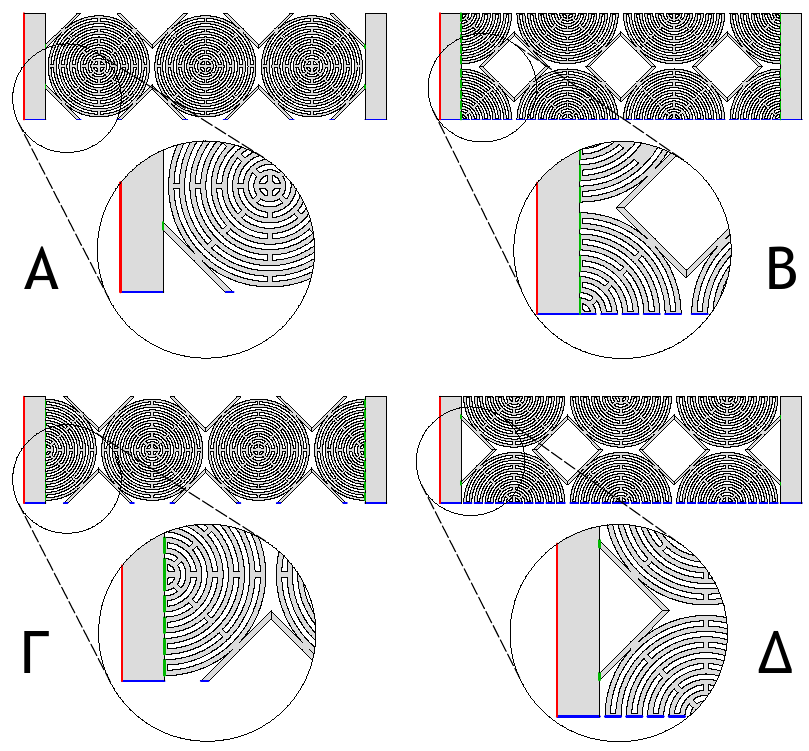}
\caption{Schematic view of the geometry and the labeling of the boundaries and interfaces for the four microstructured specimens: 
 $\Alpha$ (top left), $\Beta$ (top right), $\Gamma$ (bottom left) and $\Delta$ (bottom right).}
\label{fig:BC_MS}
\end{figure}

The full-microstructured simulations take into account the detailed specimen's geometry and the metamaterial's behavior is simply given by classical Cauchy elasticity. This process is typically computationally expensive due to the complex interior geometry of larger specimens, but it is manageable here because of the low number of unit cells used.
All the 2D simulations presented here have been performed under a plane-strain assumption and with a \textit{time-harmonic} ansatz: this means that instead of solving a time-dependent problem, the corresponding eigenvalue problem is numerically solved for each frequency value considered.
The four microstructured materials presented in this work have been built as a regular grid of finite-size ($3\times 2$ unit cells of size $L_{\rm c}=0.05$ m each) as can be seen in Fig.\ \ref{fig:4_3x2_whole}. Their building blocks are the unit cells made up of Polyethylene shown in Fig.~\ref{fig:4_cuts}.
The resulting metamaterials are in perfect contact on the left and right side with homogeneous Cauchy plates also made up of Polyethylene. 
The thickness of the plates is set to be $1$ cm.
The following boundary and interface conditions have been enforced on the relevant boundaries (see Fig.~\ref{fig:BC_MS} to identify the different interfaces):
\begin{align}
    \sigma \. n  &= \overline{F} &&\text{prescribed force - \textcolor{red}{red}}\notag\\
    u^{-}  &= u^{+}  \quad\text{and}\quad (\sigma \. n)^{-}  = (\sigma \. n)^{+} &&\text{perfect contact (continuity of displacement and traction) - \textcolor{Green}{green}\footnotemark}\notag\\
    \sigma \. n &= 0 &&\text{stress free - black}\label{eq:BC_MS}\\
    u_{y} &= 0 && \text{symmetry - \textcolor{blue}{blue}}\notag
\end{align}
\footnotetext{these jump conditions can be written in a more compact form as $\left[\left[u\right]\right]=0$ and $\left[\left[t\right]\right]=0$, $t=\sigma n$ being the Cauchy traction on each side.}where $\sigma$ is the classical Cauchy stress tensor given in eq.~(\ref{eq:tractions_cau}), $n$ is the outward unit normal to the interface and the expression of the externally applied force per unit area is: $\overline{F}=10\. \widehat{e_1}+ 0\. \widehat{e_2}$ [N/m$^{2}$].\footnote{The value of the externally applied load is compatible with the experimental one used in \cite{hermann2024design}.}
This problem thus corresponds to a compression test along the horizontal direction, mimicking, in a 2D setting, the experimental testing performed in \cite{hermann2024design}.
The simulations have been performed by using the \textit{Solid Mechanics} physics package of \comsol.
In order to ease the numerical convergence of the analysis, and also to represent appropriately the material damping of Polyethylene, we introduced some damping using an isotropic loss factor with a value of $\eta=0.1$ in all our calculations.\footnote{This value of the damping has been found to be realistic when comparing numerical simulations to experimental tests \cite{hermann2024design} \rev{and using a lower loss factor in our simulations does not show significant changes especially in the qualitative behavior}.}

%
%
%
\subsection{Reduced-relaxed-micromorphic simulations set-up}
\begin{figure}[h!]
\centering
\includegraphics[width=0.5\textwidth]{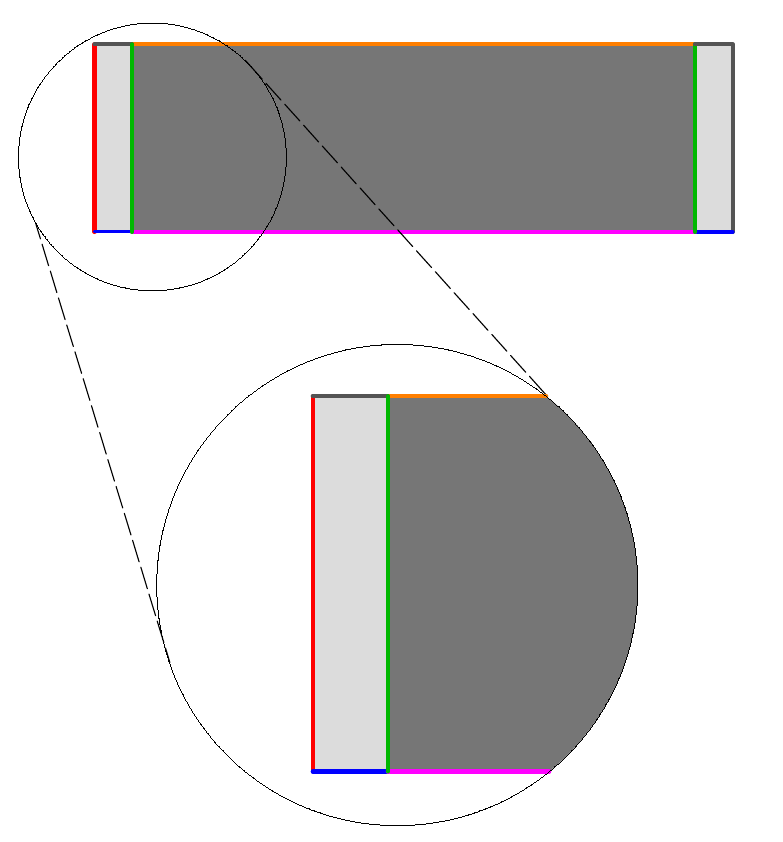}
\caption{
Schematic view of the geometry and the labeling of the boundaries and interfaces for the equivalent reduced relaxed micromorphic modeling of the considered metamaterial's specimens: Reduced relaxed micromorphic governing equations~(\ref{eq:equi_equa_RRMM}) are enforced in the darker gray bulk region, while classical isotropic Cauchy governing equations $\Div \sigma=\rho\.\ddot{u}$ with $\sigma$ given in eq.~(\ref{eq:tractions_cau}) are enforced in the two thin plates (lighter gray region). The enforced boundary and interface conditions are detailed in Eqs.~(\ref{eq:BC_MM}).
}
\label{fig:BC_rrmm}
\end{figure}

The microstructured metamaterial is here modeled with the reduced relaxed micromorphic model, which is characterised by the material parameters in Table~\ref{table:microp}.
This means that the microstructured (heterogeneous) domains in Fig.\ \ref{fig:4_3x2_whole} are replaced with a homogeneous domain of the same size (see Fig.\ \ref{fig:BC_rrmm}) whose constitutive behavior is set to be that of a reduced relaxed micromorphic continuum (i.e.\ governing equations (\ref{eq:equi_equa_RRMM}) hold in this bulk domain with the kinematic variables $u$ and $P$). 
In addition, the following boundary and interface conditions have been enforced (see Fig.~\ref{fig:BC_rrmm} for the definition of the different interfaces).
\begin{align}
    \sigma \. n &= \overline{F} & && & \text{prescribed force - \textcolor{red}{red} (Cauchy)}\notag\\
    \sigma \. n &= 0 &&&& \text{stress free - black (Cauchy)}\notag\\
    u_{y} &= 0 &&&& \text{symmetry - \textcolor{blue}{blue} (Cauchy)}\label{eq:BC_MM}\\
    \left(\widetilde{\sigma} + \widehat{\sigma} \right) n &= 0  &&&& \text{stress free - \textcolor{orange}{orange} (RRMM)}\notag\\*
    u_{y} &= 0 && \text{and}\quad P_{12}=P_{21}=0 && \text{symmetry - \textcolor{magenta}{magenta} (RRMM)}\notag\\
    u_{Cauchy} & = u_{\rm RRMM} &&\text{and}\quad \sigma \. n  = \left(\widetilde{\sigma} + \widehat{\sigma} \right) n && \text{perfect contact - \textcolor{Green}{green} (Cauchy/RRMM interface)}\footnotemark\notag
\end{align}
\footnotetext{These jump conditions can be also written in compact from as $\left[\left[u\right]\right]=0$ and $\left[\left[t\right]\right]$=0 where $\left[\left[u\right]\right]$ represents the jump of displacement and $\left[\left[t\right]\right]=t^{+}-t^{-}=\sigma n-\left(\widetilde{\sigma} + \widehat{\sigma} \right) \, n$ is the jump of the generalized traction across the green interface.}where the Cauchy stress $\sigma$ is given in eq.~(\ref{eq:tractions_cau}), the enriched stress tensors $\widetilde{\sigma}$ and $\widehat{\sigma}$ are given in eq.~(\ref{eq:equiSigAll}), the expression of the applied force per unit area is: $\overline{F}=10\. \widehat{e_1}+ 0\. \widehat{e_2}$ [N/m$^2$] and the symmetry conditions are calculated according to Appendix \ref{symmetry}.
The effective homogeneous material modeled with the reduced relaxed micromorphic model is also embedded between two slender homogeneous Cauchy plates made out of Polyethylene.
The simulations have been performed by using the \textit{Weak Form PDE} physics package of \comsol.
This package requires the implementation of the expression of the Lagrangian~(\ref{eq:lagrangian}) and the appropriate boundary and interface conditions.
To have a consistent comparison with the results from the microstructured simulations, we have introduced the same damping value of the isotropic loss factor ($\eta=0.1$) also in this case.

%
%
%
%
%
%
\section{Material interfaces: from microscopic towards macroscopic interfaces}\label{sec:material_interfaces}
In the context of mechanical problems, various interface models can be categorized depending on the quantities that may suffer a jump across the interface itself:
\begin{itemize}
  \item \textbf{Perfect interface model}: continuity of both displacement and traction is prescribed across the interface \cite{ramirez2023multi,demetriou2024reduced,demore2022unfolding,javili2017micro}.
  \item \textbf{Cohesive interface model}: continuity of traction is respected on the interface but the displacement can suffer a jump \cite{giusti2023cohesive, campilho2013modelling,volokh2004comparison,javili2017micro}.
  \item \textbf{Elastic interface model}: continuity of displacement is respected on the interface but the traction can suffer a jump \cite{murdoch1976thermodynamical,gurtin1975continuum,moeckel1975thermodynamics,javili2017micro}.
  \item \textbf{General interface model}: a jump of both displacement and traction is prescribed on the interface \cite{javili2017micro}.
\end{itemize}
These various interface models are well known in the literature and widely used in the context of periodic homogenization of heterogeneous media \cite{javili2017micro}.
This means that, usually, the discontinuity of the material properties is considered at the scale of the unit cell (e.g.\ the microscopic interface that separates two different materials inside the unit cell).
These macroscopic material interfaces are then taken into account in the homogenization procedure to show that the obtained homogenized (infinite-size) continuum exhibits different macroscopic bulk properties if one considers different interface properties at the microscopic level.
For the latter, a finite thickness interface can be approximated by a zero-thickness interface for practical purposes,  when its thickness is relatively small compared to other length scales \cite{javili2018aspects}.

The viewpoint adopted in the present paper is quite different, since we are stating that, when considering finite-size (macroscopic) metamaterials' specimens, also the macroscopic specimen's boundaries should be treated as material interfaces carrying their own material properties.
In particular, the macroscopic interfaces considered in this paper will be treated as (macroscopic) elastic interfaces across which the displacement field remains continuous, while the traction may suffer a jump.
Indeed this choice is justified by the type of metamaterials that we want to describe through our homogenized model.
The macroscopic interfaces occurring in our problem are highlighted in Fig.~\ref{fig:BC_rrmm} with a green (RRM/Cauchy interface) or an orange (free RRM interface) color.
When considering the green interface connecting the RRM continuum to the Cauchy plate, we want that this interface models the transition from the metamaterial in Fig.~\ref{fig:BC_MS} to the Cauchy continuum.
On the one hand, given the type of connections shown in Fig.~\ref{fig:BC_MS}, under the hypothesis that there are no defects in the solid-solid connections between the metamaterial and the Cauchy plate, it is natural to consider that at the macroscopic (homogenized) level one should impose continuity of the macroscopic displacemenmt across the green line in Fig.~\ref{fig:BC_rrmm}.
On the other hand, it is also evident that if one would also impose the continuity of traction across the green line in Fig.~\ref{fig:BC_rrmm}, then the RRMM would not be able to discriminate between the 4 different cases presented in Fig.~\ref{fig:BC_MS}, since the homogenized solution imposing $\left[\left[u\right]\right]=0$ and $\left[\left[t\right]\right]=0$ is unique.

It becomes thus clear that, in order to discriminate between the 4 different cases in Fig.~\ref{fig:BC_MS}, the homogenized counterpart must account for the possibility of a jump of the traction along the green lines.
Across these lines, the last equations in (\ref{eq:BC_MM}) should then be modified to
\begin{equation}
    \left[\left[u\right]\right]=0\qquad\text{and}\qquad\left[\left[t\right]\right]=f^{\rm interface}
    \label{eq:interface_force}
\end{equation}
where $f^{\rm interface}$ is an interface force that takes different expressions depending on the type of metamaterial/Cauchy connection depending on the underlying ``structure" given in the ``real" microstructured metamaterial, i.e.\ one of the different metamaterial/Cauchy connections given in Fig.~\ref{fig:BC_MS}).
The orange interface in Fig.~\ref{fig:BC_rrmm} is a particular limiting case of the green interface just described.
The orange line represents a ``free" RRM interface, in the sense that a RRM continuum is present on one side of the interface, while no other material is present on the other side.
In this particular case, no condition must be imposed on the macroscopic displacement of the RRM continuum which remains arbitrary.
On the other hand, the generalized traction that can arise at this orange interface might not be vanishing, so that the 4th condition in (\ref{eq:BC_MM}) should be modified to
\begin{equation}
   \text{t}:=\left(\widetilde{\sigma} + \widehat{\sigma} \right) n=f^{\rm interface}
    \label{interface_force_free_boundary}
\end{equation}
where $f^{\rm interface}$ is a macroscopic interface force that might take different (in general non-vanishing) values depending on the underlying ``structure" given in the ``real" microstructured metamaterial close to the upper interface, i.e.\ one of the 4 upper unit cells' cuts shown in Fig.~\ref{fig:BC_MS}).
It becomes apparent that when considering a homogenized (macrosopic) model as the RRMM for the description of the heterogeneous materials which have different properties close to the ``free" microscopic interface, then the associated homogenized (macroscopic) interface does not have, in general, a vanishing generalized traction.
This is due to the fact that small layers of the metamaterial across the free interface can create some stress concentrations which are entirely related to the specific properties of the underlying microstructure close to the free interface (boundary effects).
Such stress concentrations, (typically of the order of 1/2 to 1 unit cell size when considering a ``free" interface) are concentrated close to the free interface and might be very different depending on the unit cell ``cut".
This diversity of response must be accounted for at the homogenized level through the interface force $f^{\rm interface}$.

Up to this day, interface models have been used in the modeling of a number of phenomena such as adhesives and their fracture \cite{spannraft2022generalized,katsivalis2020development,yang2021modified}, crack growth \cite{roth2014simulation,lorentz2012modelling,li2002analysis}, damage \cite{alfano2006combining,abrate2015cohesive,khoramishad2010predicting}, surface effects between matrix and inclusion in nano-materials and composites \cite{javili2017micro,sharma2003effect}, grain boundary microcracking \cite{simonovski2015cohesive,glaessgen2006multiscale,pezzotta2008cohesive}.
To our knowledge, as has been extensively explained in \cite{ramirez2024effective}, no application exists for the modeling of macroscopic (homogenized) boundaries in the area of metamaterials and metastructures, which is where we want to address some of the open challenges in the present paper.
We stress that interface methods have been used to model size effects in nanomaterials \cite{javili2017micro} or composites with fibers \cite{hashin2002thin}, but only to model the interphase between the matrix and the inclusion in the RVE of the composite an thus, as stated before, only at the microscopic level. 
\section{Macroscopic elastic interfaces in the homogenized modeling of mechanical metamaterials through the RRMM: a case study}\label{sec:sim_results}
In the context of modeling metamaterials as homogenized media, the concept of considering macroscopic interface forces is practically disregarded in the literature.
This concept of macroscopic interface force has been introduced in \cite{ramirez2024effective} for the first time. 
However, given the novelty of this concept many challenges remain open and must be addressed with targeted case studies so as to unveil the specific properties that such interface forces should have and to draw some general conclusions.
It is the ambition of this paper to thoroughly study the properties of interface forces in the context of RRM elasticity for the case study presented in Figs.~\ref{fig:BC_MS} and~\ref{fig:BC_rrmm}.
As we have stated in Section \ref{sec:material_interfaces}, introducing the concept of interface forces at macroscopic interfaces separating a homogenized RRM continuum from another continuum is an absolute priority if one wants to extend the use of homogenized models from a purely infinite-size framework to the study of more realistic ``finite-size" structures.
In this section we will describe in more detail the specific form that such interface forces should take so that the RRMM can be safely used to describe the 4 different metamaterial's specimens of Fig.~\ref{fig:BC_MS} which are all issued from the same infinite metamaterial which is differently ``cut" at the corresponding macroscopic interfaces.

For the case study of \rev{this} paper, introduced in Fig.~\ref{fig:BC_MS} (Fig.~\ref{fig:BC_rrmm} for its RRM counterpart), we consider the following traction jump conditions across the green lines in Fig.~\ref{fig:BC_rrmm} separating the RRM continuum from the Cauchy plates:
\begin{align}
    t_{({\rm Cauchy})_{i_n}}=\alpha_{i_n}t_{({\rm RRMM})_{i_n}}+\beta_{i_n}
    \label{eq:jumpin}
\end{align}
where $i=L,R$ for left or right interface respectively and $n=x,y$ for $x$ or $y$ component of the tractions respectively.\footnote{We remark that the jump condition (\ref{eq:jumpin}) $t_{({\rm Cauchy})_{i_n}}=\alpha_{i_n}t_{({\rm RRMM})_{i_n}}+\beta_{i_n}$ can be rewritten setting $\gamma=\alpha-1$, so that $t_{({\rm Cauchy})_{i_n}}=t_{({\rm RRMM})_{i_n}}+\gamma_{i_n}t_{({\rm RRMM})_{i_n}}+\beta_{i_n}$ or
$\left[\left[t\right]\right]=\gamma_{i_n}t_{({\rm RRMM})_{i_n}}+\beta_{i_n}$. With reference to eq.~(\ref{eq:interface_force}) this means that we are setting 
\begin{equation}
    f^{\rm interface}_{i_{n}}=(\alpha_{i_{n}}-1)\.t_{({\rm RRMM})_{i_n}} + \beta_{i_n}.
    \label{eq:interface_force_a-1}
\end{equation}
In other words, we assign that the interface force which is stemming from the heterogeneity of the underlying microstructure close to the Cauchy/RRM interface is a linear function of the RRM traction $t_{\rm RRMM}=\left(\widetilde{\sigma} + \widehat{\sigma} \right) n$.}
\rev{Note that $\alpha$ is dimensionless while $\beta$ is in $\frac{\rm N}{\rm{m}^2}$. However, we omit the unit in the captions of the following figures to shorten their text.}

We thus note, that here the jump of the traction at the interface takes the simple form of a linear function of the RRM traction.
However, depending on the complexity of the problem (e.g.\ more complex loading conditions or different geometries of the macroscopic specimens), this assumption could change and the jump of traction could be allowed to take more complex forms, e.g.\ $\alpha$ and $\beta$ being functions of the $x$ or $y-$coordinates along the interface, instead of being constants.
\subsection{Reduced Relaxed Micromorphic Model with and without interface forces vs Microstructured and long-wavelength limit Cauchy simulations}
In this section, we show the importance of the concept of interface forces when a homogenized model like the RRMM has to be used for the modeling of finite-sized metamaterials' structures.

\begin{figure}[h!]
\centering
\includegraphics[width=1\textwidth]{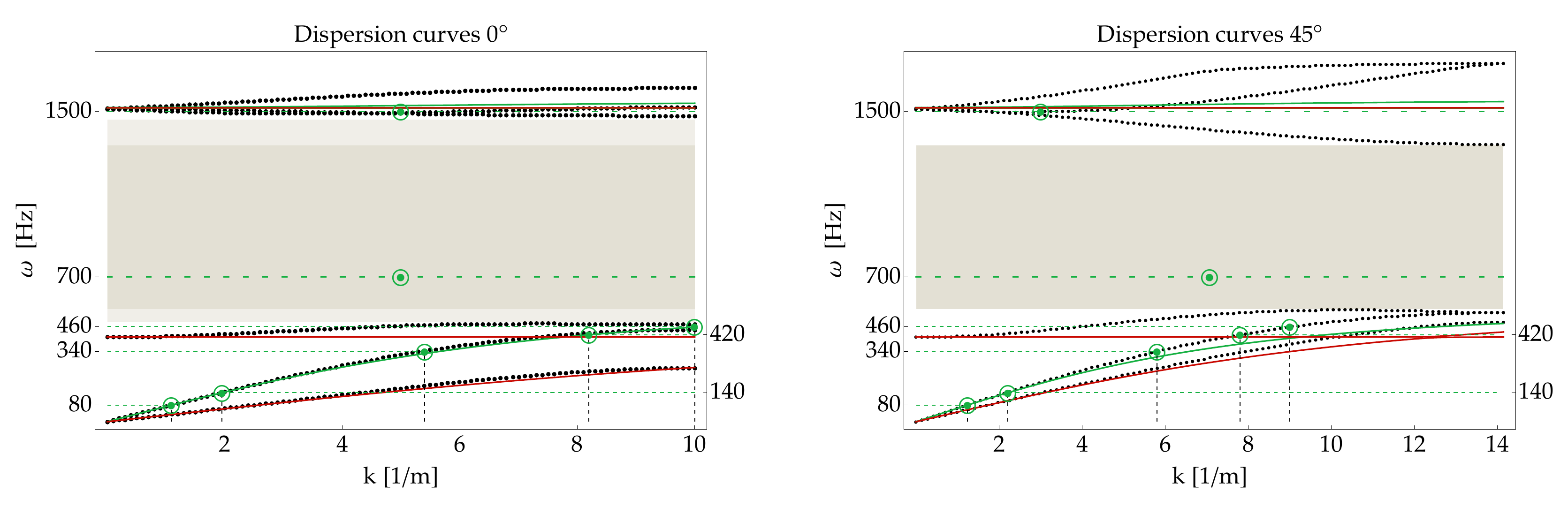}
\caption{Dispersion curves for $0^\circ$ (left),  and for $45^\circ$ (right).The dots correspond to the solution of the standard Bloch-Floquet analysis performed on any of the four unit cells in Fig.~\ref{fig:4_cuts} by using \comsol,  while the solid lines represent the analytical expression of the dispersion curves for the RRRM. 
Green circular points indicate the frequencies and corresponding wavenumbers for which our finite size RRM simulations were implemented (the points in the band-gap show only the frequency used, since there exists no corresponding real wavenumber).}
\label{fig:DC_freq}
\end{figure}

To this aim, we consider the comparison between the RRM simulations and the microstructured ones for the $\Alpha$, $\Beta$, $\Gamma$, $\Delta$ ``cuts" of different frequencies.
When the solution of the RRMM does not match the microstructured solution, we calibrate the corresponding interface forces arising at the interfaces between the RRM domain (corresponding to the metamaterial) and the Cauchy plates (see eq.~(\ref{eq:interface_force_a-1})), until the RRM solution matches the microstructured one.
We also provide a comparison with a simulation in which the metamaterial domain is modeled through a homogeneous Cauchy continuum which is the long-wavelength limit of the RRMM.
In this way, we are able to clearly unveil the advantages brought by the RRM setting.
The frequencies chosen are reported in Fig.~\ref{fig:DC_freq} with reference to the dispersion curves.

The wavelength for which a wave propagates inside the metamaterial at the considered frequencies is summarized in Table~\ref{tab:frequencies_wavelengths}.

\begin{table}[h!]
    \centering
    \renewcommand{\arraystretch}{1.35}
    \begin{tabular}{|c|c|c|}
        \hline
        \textbf{Frequency [Hz]} & \textbf{Wavelength [cm]} & \rev{\textbf{Ratio }$\frac{\boldsymbol\lambda}{\boldsymbol L_{\rm \boldsymbol s}}$}\\ 
        \hline
        80 & 91 & \rev{6.07}\\ 
        \hline
        140 & 51 & \rev{3.4}\\ 
        \hline
        340 & 18.5 & \rev{1.23}\\ 
        \hline
        420 & 12 & \rev{0.8}\\ 
        \hline
        460 & 10 & \rev{0.67}\\ 
        \hline
        700 & \multicolumn{2}{c|}{band-gap}\\ 
        \hline
        1500 & \multicolumn{2}{c|}{no propagation} \\ 
        \hline
    \end{tabular}
    \caption{Wavelengths \rev{$\lambda$} at which a pressure wave propagates in the infinite metamaterial of Fig.~\ref{fig:4_cuts} at the considered frequencies. The size of the specimen \rev{$L_{\rm s}$} (without the Cauchy plates) is 15 cm.}
    \label{tab:frequencies_wavelengths}
\end{table}
The values of the coefficients $\alpha_{i_{n}}$ and the forces $\beta_{i_{n}}$ are calculated by direct inspection and comparison of the RRM and corresponding microstructured displacement fields.
\rev{The RRMM, being an enriched continuum used in a homogenized framework, captures the response of the metamaterial in an average sense, meaning that not every little detail of the displacement of the microstructural components can be captured, but rather, the overall macroscopic displacement field in an average sense.
This implies that we can change the interface force appropriately by direct inspection of the displacement field and deformed shape, until the response adequately represents the average displacement field.
Moreover, this interface force that produced the adequate response for the enriched continuum, should also bring the traction on the ``homogenized" interfaces close to an average of the microstructured traction on the same interface, since it is a function of the displacement.
This has been used in the opposite way in \cite{ramirez2024effective}, where the calculation of the expression of the interface force on a given boundary was done purely by ``average" traction considerations.
This implies that one could arrive at the correct interface force by trial and error, so that the traction on the interface in the case of modeling the metamaterial using an enriched continuum, becomes an average of the traction on the same interface in the corresponding microstructured simulation.
This particular way of calculating the interface force is more meaningful when the metamaterial specimens are bigger and hence the traction average is more representative.
Therefore, in this work where the specimens are of very small size, the calculation of the coefficients $\alpha_{i_{n}}$ and the forces $\beta_{i_{n}}$, is done through average displacement considerations.}
We first run a parametric sweep for the $\alpha_{i_{x}}$ coefficients, compare the displacement field with that of the microstructured metamaterial and check which values make the response more accurate.
Then, if the Cauchy plates on the left and right hand side of the metamaterial in the microstructured simulation experience bending, also the $\alpha_{i_{y}}$ coefficients must be taken into account.
We apply the same procedure (running a parametric sweep and inspecting the displacement field) until the response is accurate.
Furthermore, forces $\beta_{i_{x}}$ and $\beta_{i_{y}}$ can be used as an extra tool if the response is hard to capture.
Moreover,if boundary effects appear on the top boundary of the microstructured simulation, an interface force must be activated on the top ``free" RRM boundary in the corresponding RRM simulation.

In the following, we also look at the tractions on the Cauchy/RRM interfaces, that could potentially be used as an alternative method to calculate coefficients $\alpha_{i_{n}}$ and forces $\beta_{i_{n}}$ that lead to the correct response.
A method based on the ``average" of the tractions, as discussed previously, is indeed a promising method for bigger specimens where the calculation of ``average" traction is more meaningful.

\rev{Regarding the size of the mesh used in each simulation we indicate the following:\\ for the cases of modeling the metamaterial using the full microstructure, the mesh of the metastructure (homogeneous Cauchy plates and microstructured metamaterial) was composed of approximately (we have four different versions, one for each cut) 46700 triangular quintic Lagrange elements. For the case of modeling the metamaterial using a homogeneous model (either the RRMM or the macro-Cauchy) the mesh was composed of 560 triangular quadratic Lagrange elements. All meshes are the result of individual mesh convergence studies using the adaptive mesh refinement node of \comsol.}
\subsubsection{Frequency: 80 Hz}
We start analyzing the homogenized Cauchy and RRM simulations and the corresponding comparison to the microstructured ones for the frequency of 80 Hz. 
This frequency is relatively low and corresponds to a macro Cauchy-like non-dispersive behavior (see the first point in Fig.~\ref{fig:DC_freq}).

\begin{figure}[h!]
\centering
\begin{tikzpicture}
        \node[anchor=south west,inner sep=0] (image) at (0,0) {\includegraphics[width=0.9\textwidth]{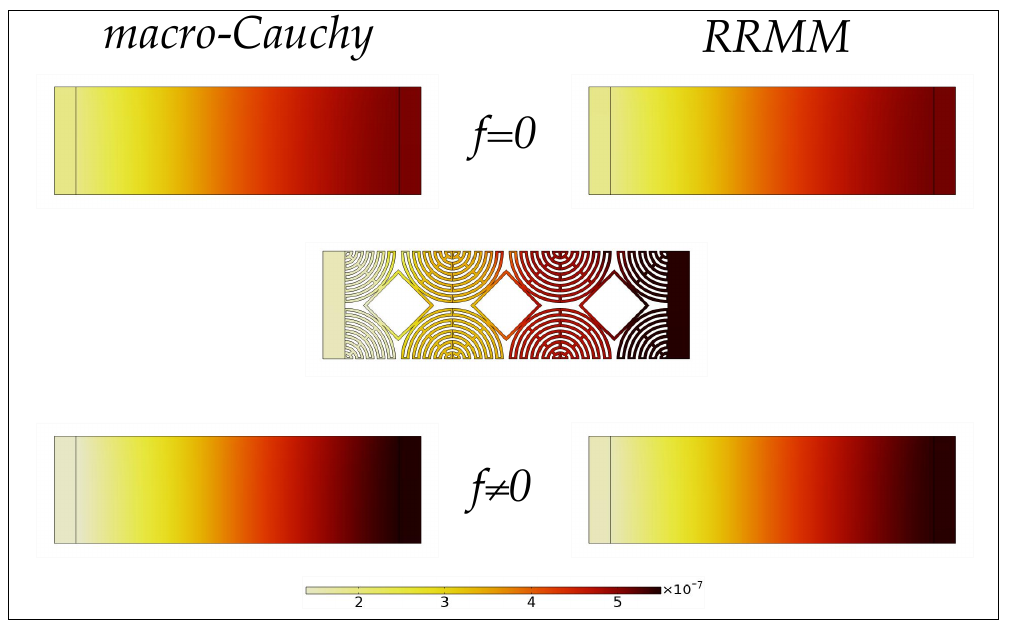}};
        \node[anchor=south] at ($(image.north west)!0.5!(image.north east)$) [yshift=0.1em] {\huge \textbf{80 Hz}};
    \end{tikzpicture}
    \caption{Comparison of the displacement field of the metamaterial specimen $\Beta$ with the macro-Cauchy and the RRMM when $f=0$ and $f\neq 0$ at 80 Hz. When $f\neq 0$ for the macro RRMM, we have: $\alpha_{L_x}=0.9$, $\beta_{L_x}=0$, $\alpha_{L_y}=1$, $\beta_{L_y}=0$, $\alpha_{R_x}=0.6$, $\beta_{R_x}=0$, $\alpha_{R_y}=1$ and $\beta_{R_y}=0$, while for the macro Cauchy: $\alpha_{L_x}=0.87$, $\beta_{L_x}=0$, $\alpha_{L_y}=1$, $\beta_{L_y}=0$, $\alpha_{R_x}=0.5$, $\beta_{R_x}=0$, $\alpha_{R_y}=1$ and $\beta_{R_y}=0$.}
\label{fig:disp80_beta}
\end{figure}
\rev{Figure~\ref{fig:disp80_beta} shows that at the frequency of 80 Hz the RRM solution is almost coinciding with the long-wavelength limit Cauchy solution. We did not include the other unit cell cuts as there were already in perfect agreement without the need of interface forces.}
This is related to the fact that the behavior of the dispersion curves is still linear at the considered lower frequency (see Fig.~\ref{fig:DC_freq}).
However, it is evident that, even at this lower frequency, boundary effects may play a non-negligible role. 
Indeed, while for the ``cuts" $\Alpha$, $\Delta$ and $\Gamma$ there is no need to introduce non-vanishing interface forces, ``cut" $\Beta$ needs a non-vanishing interface force to recover the correct solution (see Fig.~\ref{fig:disp80_beta}). 
This might be related to the fact that ``cut" $\Gamma$ provides a better solid/solid connection between the metamaterial and the Cauchy plates, so that the applied load is better transmitted through the metamaterial's specimen.
This enhanced boundary-driven transmission can be recovered in the RRM setting through suitable non-vanishing interface forces. 
However, also for the $\Beta$ case for which interface forces are necessary to recover the correct solution, it can be recognized that the correction brought by triggering $f^{\rm interface}\neq 0$ is quite small when compared to the corrections which are found for higher frequencies (see Fig.~\ref{fig:disp80_beta} with $f=0$ and $f\neq 0$).
The fact that the interface force is bringing either a small correction or no correction at all is related to the fact that at this low frequency the wavelength is quite larger than the size of the specimen (see Table~\ref{tab:frequencies_wavelengths}).
This implies that for such large wavelength boundary effects are overall quite small or mostly negligible.

\begin{figure}[h!]
\centering
\includegraphics[width=1\textwidth]{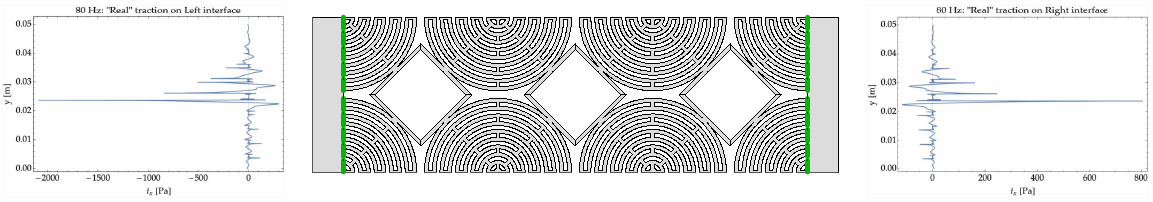}
\caption{Sketch of the tractions on the Cauchy side of the Cauchy plate/metamaterial interfaces at 80 Hz.
The two plots on the two sides of the specimen represent the traction fields on the Cauchy-plate sides along the green lines highlighted in the picture.
Similar patters can be observed for all other ``cuts" and for other frequencies.}
\label{fig:real_tractions}
\end{figure}

We have seen that at the considered frequency of 80 Hz both the RRMM and the macro-Cauchy model give good results when suitable interface forces are introduced. 
This can also be seen when looking at the tractions that arise at the Cauchy-plates/metamaterial interfaces, as shown in Fig.~\ref{fig:real_tractions}.

\begin{figure}[h!]
    \centering
    \includegraphics[width=0.45\textwidth]{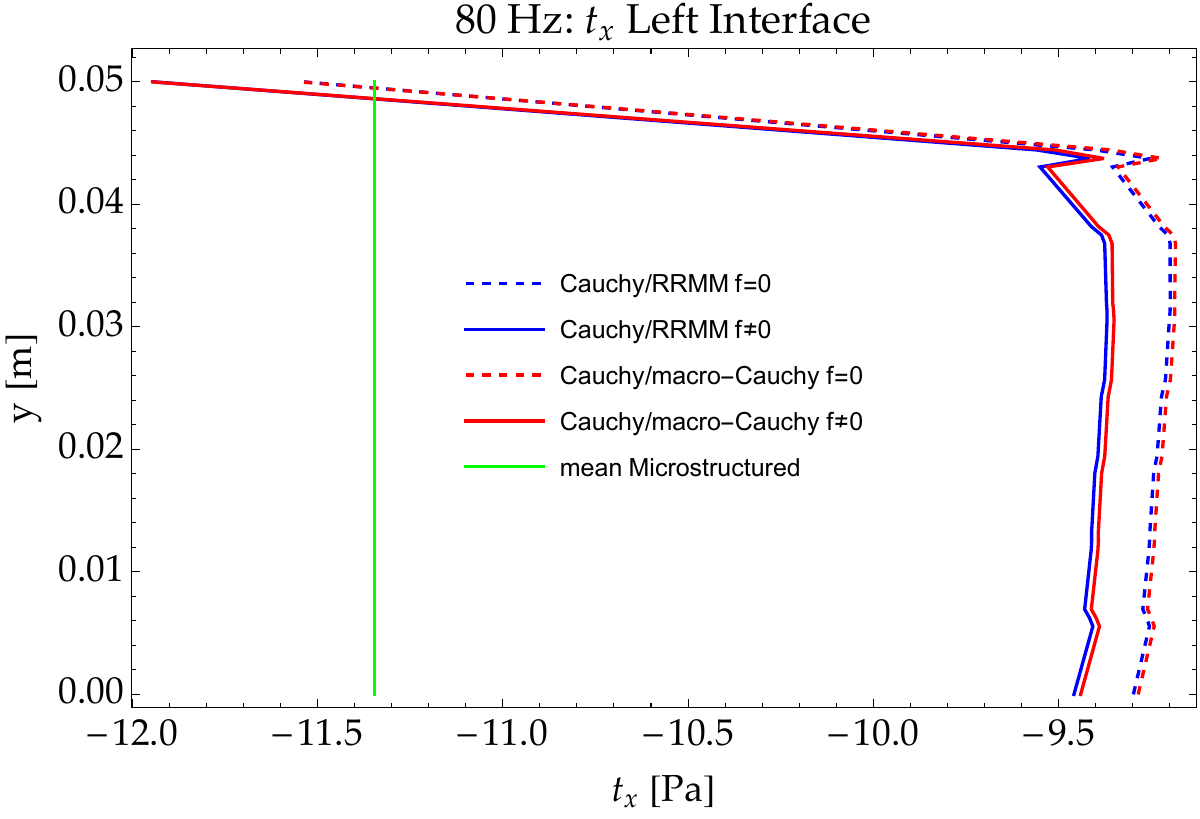}
    \vspace{2mm}
    \includegraphics[width=0.45\textwidth]{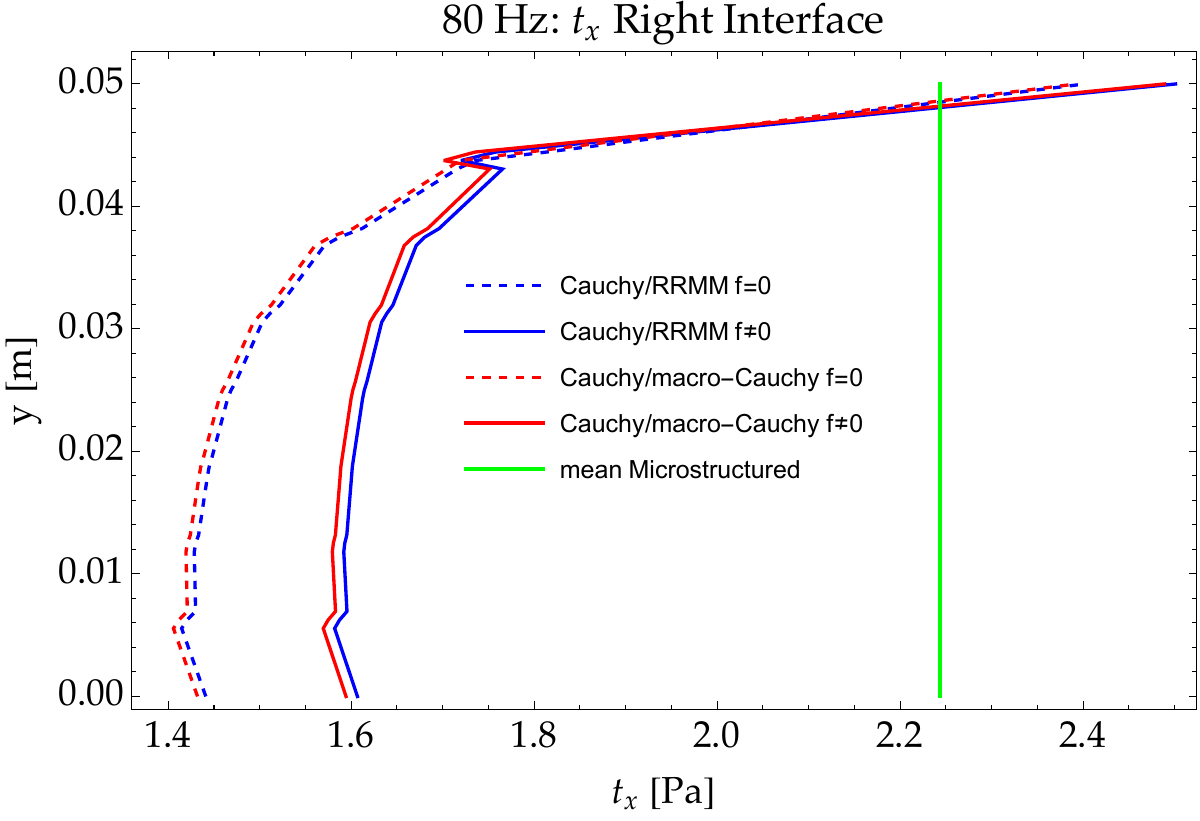}
    \caption{Tractions on the Cauchy side of the Cauchy plate/metamaterial interfaces (left and right) for the RRMM and for the macro-Cauchy when $f=0$ and $f\neq 0$. The tractions shown here are those relative to ``cut" $\Beta$. Analogous reasoning holds true for all other ``cuts".}
\label{fig:rrmm_macro_tractions_80}
\end{figure}

Fig.~\ref{fig:real_tractions} shows the microstructured tractions arising at the Cauchy-plates/metamaterial interfaces (on the Cauchy side), for  ``cut" $\Beta$.
Completely analogous considerations are valid for all other ``cuts".
As a first rough measure, we can compute the ``mean" value of this traction\footnote{The ``mean" is calculated with a standard procedure: The traction $t_{x}$ has a specific value on every point on the Cauchy-plate/metamaterial interface. We sum all these values together and we divide by the number of points. This gives us a number that represents the ``average" value of the traction $t_{x}$ on the interface.} and compare it to the corresponding RRM and macro-Cauchy tractions. 
This comparison is shown in Fig.~\ref{fig:rrmm_macro_tractions_80}, where we can see that there is practically no difference between the RRM and macro-Cauchy tractions both on the left and right interface.

It can also be inferred that the homogenized (both RRMM and macro-Cauchy) tractions become ``closer" to the mean traction calculated starting from the microstructured solution as soon as ``interface" forces are triggered.
This is a good indication that the homogenized framework including the concept of interface forces is a mandatory step if homogenized models have to be used for finite-size metamaterials modeling. 
However, we can see that the homogenized (both RRMM and macro-Cauchy) traction does not exactly coincide with the ``mean" of the microstructured traction which has been evaluated for comparison. 
This is due to the fact that the ``averaging" of the interface microstructured traction (the one shown in Fig.~\ref{fig:real_tractions}) is an operation whose exact definition is an open challenge in the homogenization community.
Practically no work exists which try to incorporate the effect of microscopic heterogeneous boundaries into the existing homogenization procedures. 
Given the complexity of the considered metamaterials and interfaces, as well as the simplifying hypotheses always present in all upscaling techniques, a rigorous definition of such average interface microstructured tractions could be impossible to be fully achieved.

Our approach, being purely macroscopical, refrains from calculating microscopic averages, but shows which form should take the introduced macroscopic interface forces to bring the homogenized solution close to the microstructured one.

While Fig.~\ref{fig:rrmm_macro_tractions_80} shows the x-component of the tractions at the Cauchy-plate/metamaterial interface, completely analogous conclusions hold for the y-components, when they are non-vanishing.
We do not present graphics of the y-component of the tractions for the sake of compactness.
\FloatBarrier

\subsubsection{Frequency: 140 Hz}
We continue analyzing the homogenized Cauchy and RRM simulations and the corresponding comparison to the microstructured ones for the frequency of 140 Hz. 
This frequency is relatively low and still corresponds to a macro Cauchy-like non-dispersive behavior (see the second point in Fig.~\ref{fig:DC_freq}).

\begin{figure}[h!]
\centering
\begin{tikzpicture}
        \node[anchor=south west,inner sep=0] (image) at (0,0) {\includegraphics[width=0.9\textwidth]{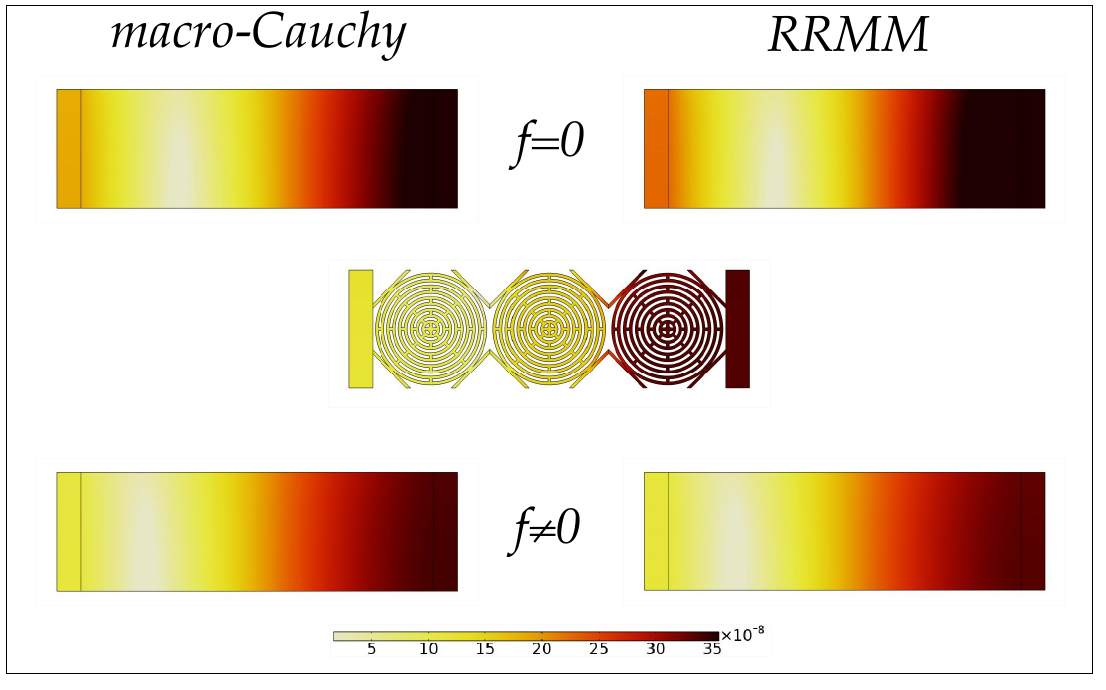}};
        \node[anchor=south] at ($(image.north west)!0.5!(image.north east)$) [yshift=0.1em] {\huge \textbf{140 Hz}};
    \end{tikzpicture}
\caption{Comparison of the displacement field of the metamaterial specimen $\Alpha$ with the macro-Cauchy and the RRMM when $f=0$ and $f\neq 0$ at 140 Hz. When $f\neq 0$ for the RRMM, we have: $\alpha_{L_x}=1.05$, $\beta_{L_x}=0$, $\alpha_{L_y}=1$, $\beta_{L_y}=0$, $\alpha_{R_x}=5$, $\beta_{R_x}=0$, $\alpha_{R_y}=1$ and $\beta_{R_y}=0$, while for the macro Cauchy: $\alpha_{L_x}=0.99$, $\beta_{L_x}=0$, $\alpha_{L_y}=1$, $\beta_{L_y}=0$, $\alpha_{R_x}=3$, $\beta_{R_x}=0$, $\alpha_{R_y}=1$ and $\beta_{R_y}=0$.}
\label{fig:disp140_alpha}
\end{figure}

\begin{figure}[h!]
\centering
\begin{tikzpicture}
        \node[anchor=south west,inner sep=0] (image) at (0,0) {\includegraphics[width=0.9\textwidth]{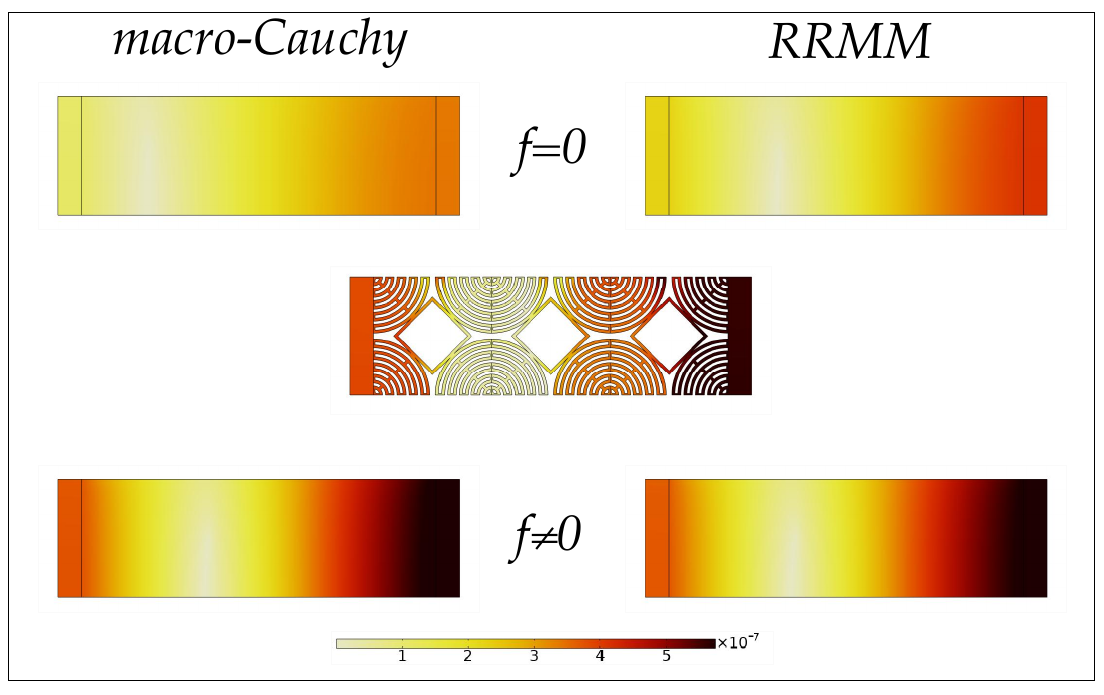}};
        \node[anchor=south] at ($(image.north west)!0.5!(image.north east)$) [yshift=0.1em] {\huge \textbf{140 Hz}};
    \end{tikzpicture}
\caption{Comparison of the displacement field of the metamaterial specimen $\Beta$ with the macro-Cauchy and the RRMM when $f=0$ and $f\neq 0$ at 140 Hz. When $f\neq 0$ for the RRMM, we have: $\alpha_{L_x}=0.8$, $\beta_{L_x}=0$, $\alpha_{L_y}=1$, $\beta_{L_y}=0$, $\alpha_{R_x}=0.75$, $\beta_{R_x}=0$, $\alpha_{R_y}=1$ and $\beta_{R_y}=0$, while for the macro Cauchy: $\alpha_{L_x}=0.75$, $\beta_{L_x}=0$, $\alpha_{L_y}=1$, $\beta_{L_y}=0$, $\alpha_{R_x}=0.65$, $\beta_{R_x}=0$, $\alpha_{R_y}=1$ and $\beta_{R_y}=0$.}
\label{fig:disp140_beta}
\end{figure}
\begin{figure}[h!]
\centering
\begin{tikzpicture}
        \node[anchor=south west,inner sep=0] (image) at (0,0) {\includegraphics[width=0.9\textwidth]{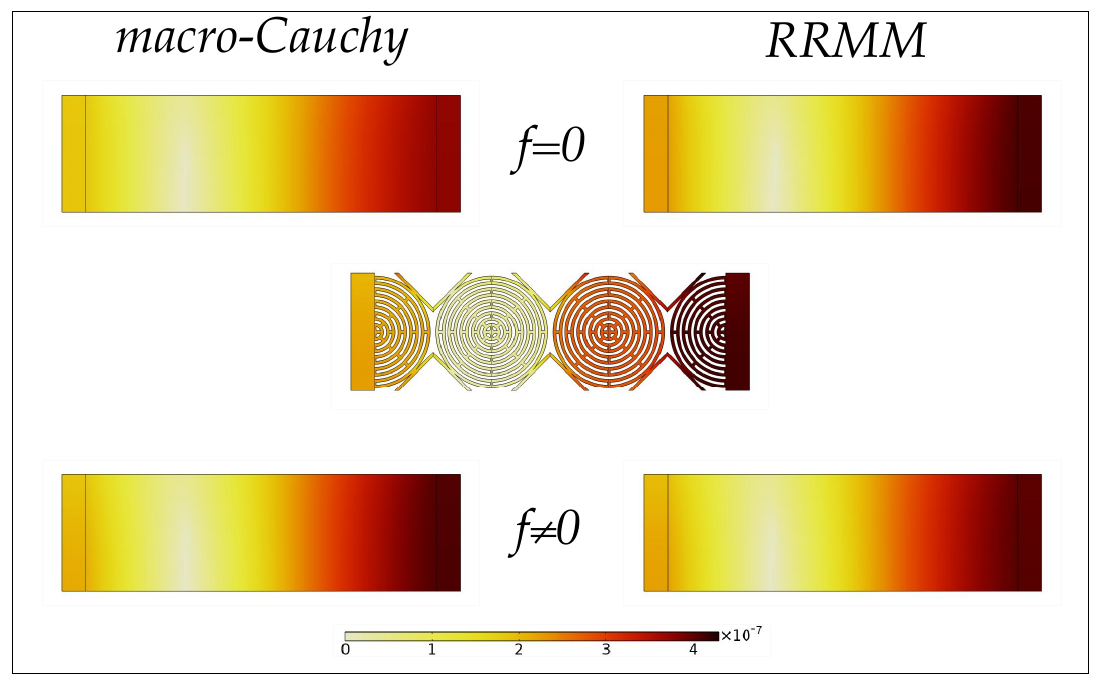}};
        \node[anchor=south] at ($(image.north west)!0.5!(image.north east)$) [yshift=0.1em] {\huge \textbf{140 Hz}};
    \end{tikzpicture}
\caption{Comparison of the displacement field of the metamaterial specimen $\Gamma$ with the macro-Cauchy and the RRMM when $f=0$ and $f\neq 0$ at 140 Hz. When $f\neq 0$ for the RRMM, we have: $\alpha_{L_x}=1$, $\beta_{L_x}=0$, $\alpha_{L_y}=1$, $\beta_{L_y}=-18$, $\alpha_{R_x}=1.1$, $\beta_{R_x}=0$, $\alpha_{R_y}=1$ and $\beta_{R_y}=0$, while for the macro Cauchy: $\alpha_{L_x}=0.93$, $\beta_{L_x}=0$, $\alpha_{L_y}=1$, $\beta_{L_y}=-18$, $\alpha_{R_x}=0.5$, $\beta_{R_x}=0$, $\alpha_{R_y}=1$ and $\beta_{R_y}=0$.}
\label{fig:disp140_gamma}
\end{figure}
\begin{figure}[h!]
\centering
\begin{tikzpicture}
        \node[anchor=south west,inner sep=0] (image) at (0,0) {\includegraphics[width=0.9\textwidth]{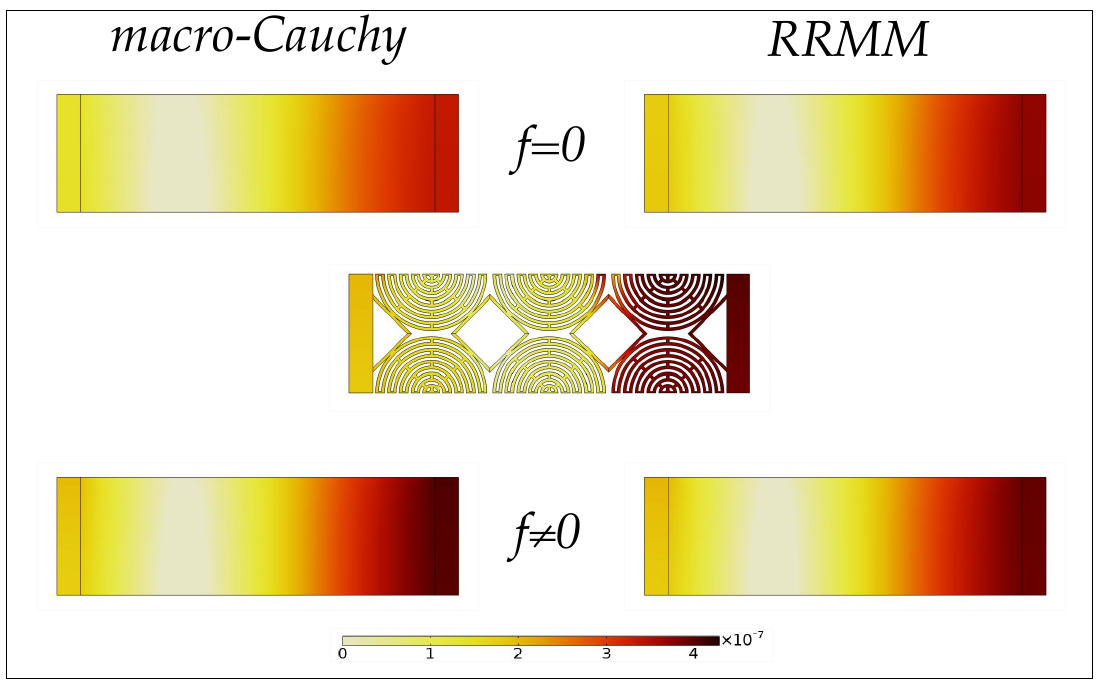}};
        \node[anchor=south] at ($(image.north west)!0.5!(image.north east)$) [yshift=0.1em] {\huge \textbf{140 Hz}};
    \end{tikzpicture}
\caption{Comparison of the displacement field of the metamaterial specimen $\Delta$ with the macro-Cauchy and the RRMM when $f=0$ and $f\neq 0$ at 140 Hz. When $f\neq 0$ for the RRMM, we have: $\alpha_{L_x}=0.96$, $\beta_{L_x}=0$, $\alpha_{L_y}=1$, $\beta_{L_y}=20$, $\alpha_{R_x}=1.01$, $\beta_{R_x}=0$, $\alpha_{R_y}=1$ and $\beta_{R_y}=5$, while for the macro Cauchy: $\alpha_{L_x}=0.89$, $\beta_{L_x}=0$, $\alpha_{L_y}=1$, $\beta_{L_y}=20$, $\alpha_{R_x}=0.9$, $\beta_{R_x}=0$, $\alpha_{R_y}=1$ and $\beta_{R_y}=5$.}
\label{fig:disp140_delta}
\end{figure}
Figures~\ref{fig:disp140_alpha},~\ref{fig:disp140_beta},~\ref{fig:disp140_gamma} and~\ref{fig:disp140_delta} show that at the frequency of 140 Hz both the RRM and the long-wavelength limit (macro-Cauchy) model can recover well the microstructured solution as far as suitable interface forces to discriminate between the 4 different cuts are calibrated.
Indeed boundary effects are clearly more important here than for the frequency of 80 Hz, since the solution is very different for the 4 considered cuts, while this strong difference between the different cuts was not present at lower frequencies.
It is clear that at this higher frequency the wavelength decreases to an extent that the different connections between the Cauchy plate and the metamaterial start to macroscopically affect the travelling wave.
This means that interface effects are more important than the bulk behavior at the considered frequency and specimen's size.
This must be necessarily accounted for via the introduction of suitable interface forces when considering the homogenized modeling of the considered benchmark test.

\begin{figure}[h!]
    \centering
    \includegraphics[width=0.45\textwidth]{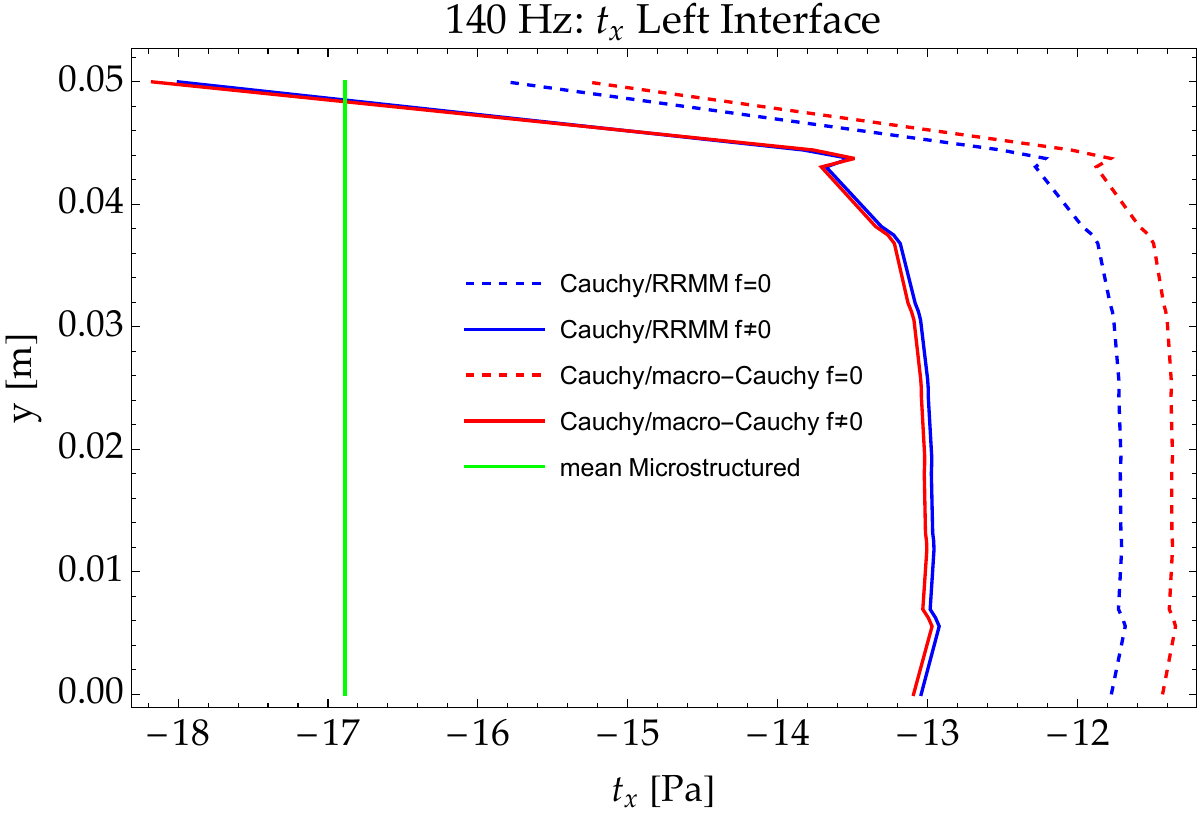}
    \vspace{2mm}
    \includegraphics[width=0.45\textwidth]{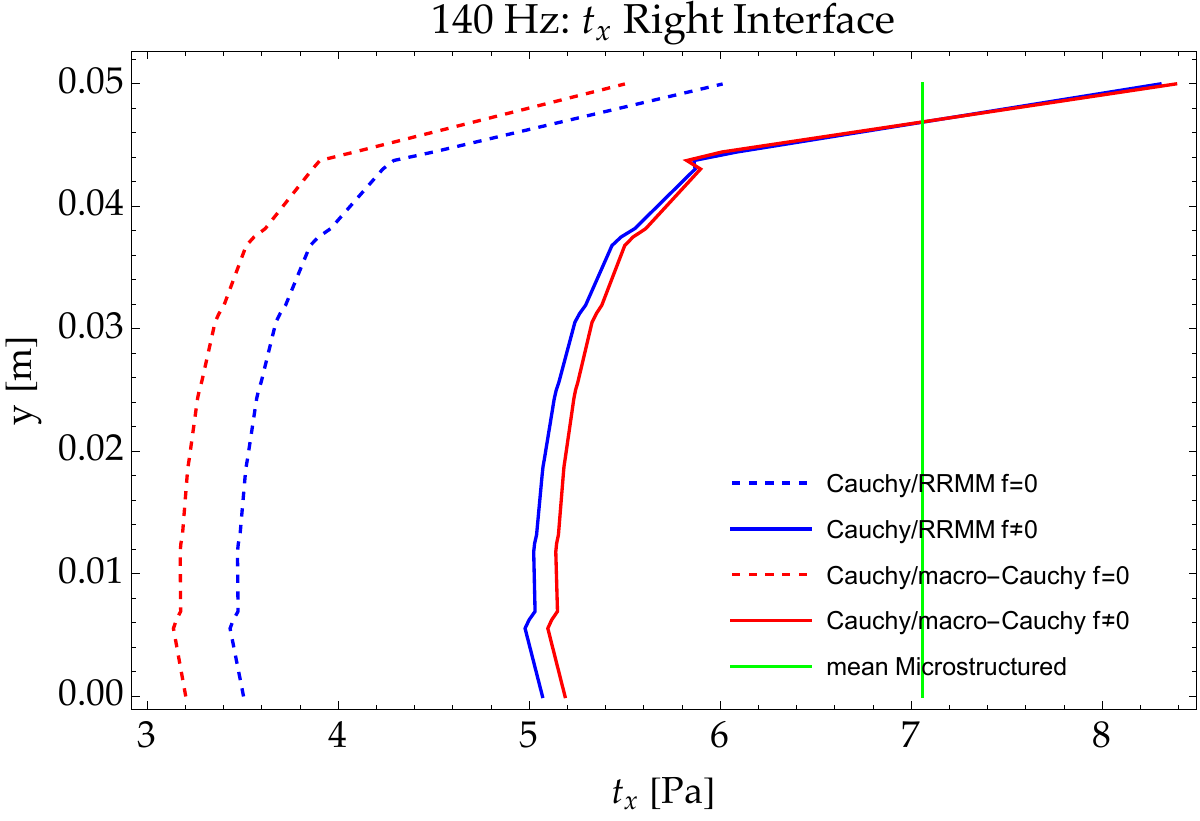}
    \caption{Tractions on the Cauchy side of the Cauchy plate/metamaterial interfaces (left and right) for the RRMM and for the macro-Cauchy when $f=0$ and $f\neq 0$. The tractions shown here are those relative to ``cut" $\Beta$. Analogous reasoning holds true for all other ``cuts".}
\label{fig:rrmm_macro_tractions_140}
\end{figure}
Completely analogous conclusions can be drawn here for the interface forces as those presented for a frequency of 80 Hz.
This means that both the RRMM and the macro-Cauchy interface forces get ``closer" to the microstructured traction's average as soon as triggering interface forces.
At the current frequency of 140 Hz (see Fig.~\ref{fig:rrmm_macro_tractions_140})\footnote{The traction comparison is again done for the case of ``cut" $\Beta$ as is done for every other case in following sections. The reason for choosing this ``cut" is that it is one of the ``cuts" with the most surface area on the Cauchy plate/metamaterial interface and therefore the mean traction calculation is more meaningful.} the two solutions are equally good after the introduction of suitable interface forces since the dispersion is still small in this low frequency.
\FloatBarrier

%
%
%
\subsubsection{Frequency: 340 Hz}
\rev{At this frequency and also for values slightly lower or higher in the interval of 200 Hz to 400 Hz, the four cuts have massive differences. This is because the wavelength gets smaller (around 1-2 times the size of the specimen, interacts with the microstructure and these boundary effects are now governing the whole response of the $3\times2$ specimen. 
We expect that in this frequency interval, structural resonances of the entire specimen can also be easily triggered.
As a consequence, the interface forces arising at the Cauchy-plates metamaterial's interfaces could experience strong oscillations making expression~(\ref{eq:interface_force_a-1}) for the macroscopic interface force less capable to capture such complex interface behavior.
In this frequency region the pressure mode is very dispersive, so we avoid using the macro-Cauchy, since it does not have the ability to capture the dispersion (see the third point in Fig.~\ref{fig:DC_freq}).
}

\begin{figure}[h!]
    \centering
     \begin{tikzpicture}
        \node[anchor=south west,inner sep=0] (image) at (0,0) {\includegraphics[width=0.8\textwidth]{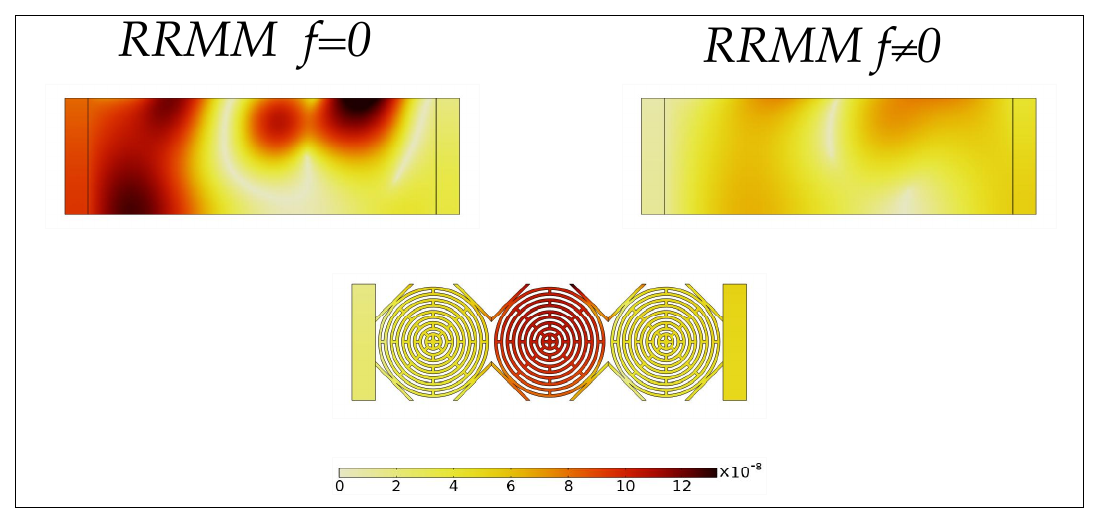}};
        \node[anchor=south] at ($(image.north west)!0.5!(image.north east)$) [yshift=0.1em] {\huge \textbf{340 Hz}};
    \end{tikzpicture}
    \caption{Comparison of the displacement field of the metamaterial specimen $\Alpha$ with the RRMM when $f=0$ and $f\neq 0$ at 340 Hz. For $f\neq 0$, we have: $\alpha_{L_x}=2$, $\beta_{L_x}=0$, $\alpha_{L_y}=1$, $\beta_{L_y}=0$, $\alpha_{R_x}=-2$, $\beta_{R_x}=0$, $\alpha_{R_y}=1$, $\beta_{R_y}=0$.}
    \label{fig:disp340_alpha}
\end{figure}

\begin{figure}[h!]
    \centering
     \begin{tikzpicture}
        \node[anchor=south west,inner sep=0] (image) at (0,0) {\includegraphics[width=0.8\textwidth]{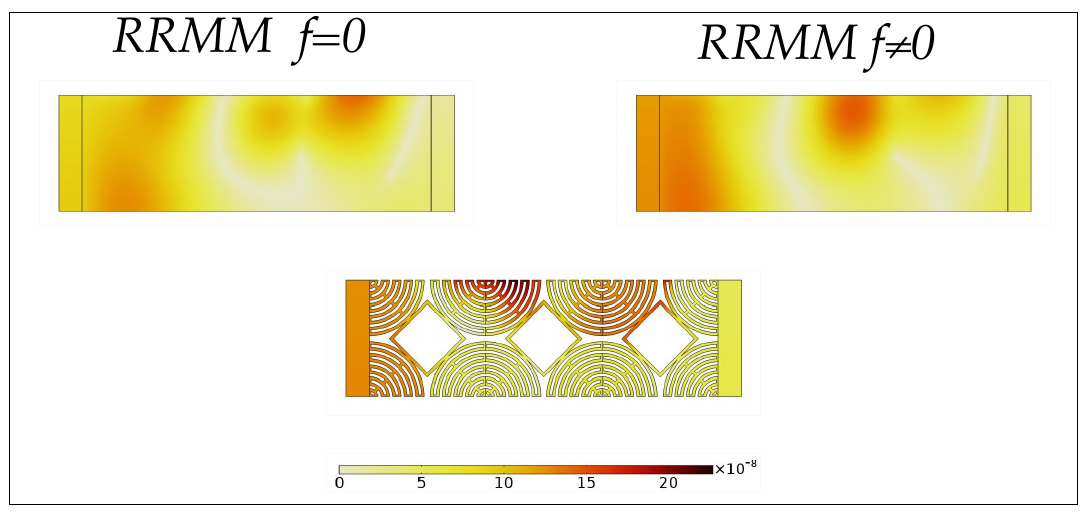}};
        \node[anchor=south] at ($(image.north west)!0.5!(image.north east)$) [yshift=0.1em] {\huge \textbf{340 Hz}};
    \end{tikzpicture}
    \caption{Comparison of the displacement field of the metamaterial specimen $\Beta$ with the RRMM when $f=0$ and $f\neq 0$ at 340 Hz. For $f\neq 0$, we have: $\alpha_{L_x}=0.8$, $\beta_{L_x}=0$, $\alpha_{L_y}=1$, $\beta_{L_y}=0$, $\alpha_{R_x}=0.6$, $\beta_{R_x}=0$, $\alpha_{R_y}=1$, $\beta_{R_y}=0$.}
    \label{fig:disp340_beta}
\end{figure}

\begin{figure}[h!]
    \centering
     \begin{tikzpicture}
        \node[anchor=south west,inner sep=0] (image) at (0,0) {\includegraphics[width=0.8\textwidth]{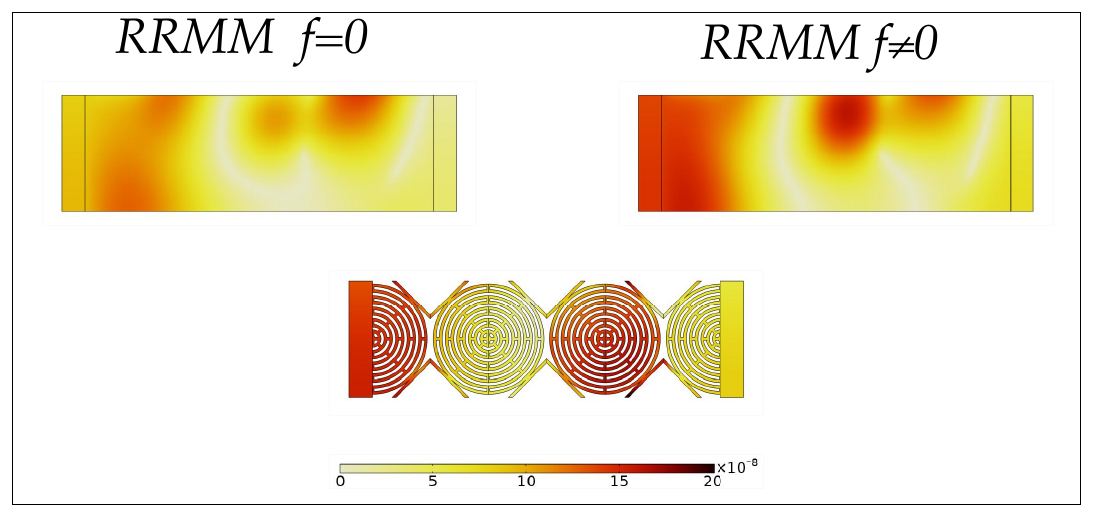}};
        \node[anchor=south] at ($(image.north west)!0.5!(image.north east)$) [yshift=0.1em] {\huge \textbf{340 Hz}};
    \end{tikzpicture}
    \caption{Comparison of the displacement field of the metamaterial specimen $\Gamma$ with the RRMM when $f=0$ and $f\neq 0$ at 340 Hz. For $f\neq 0$, we have: $\alpha_{L_x}=0.6$, $\beta_{L_x}=0$, $\alpha_{L_y}=1$, $\beta_{L_y}=0$, $\alpha_{R_x}=1$, $\beta_{R_x}=0$, $\alpha_{R_y}=1$, $\beta_{R_y}=0$.}
    \label{fig:disp340_gamma}
\end{figure}

\begin{figure}[H]
    \centering
     \begin{tikzpicture}
        \node[anchor=south west,inner sep=0] (image) at (0,0) {\includegraphics[width=0.8\textwidth]{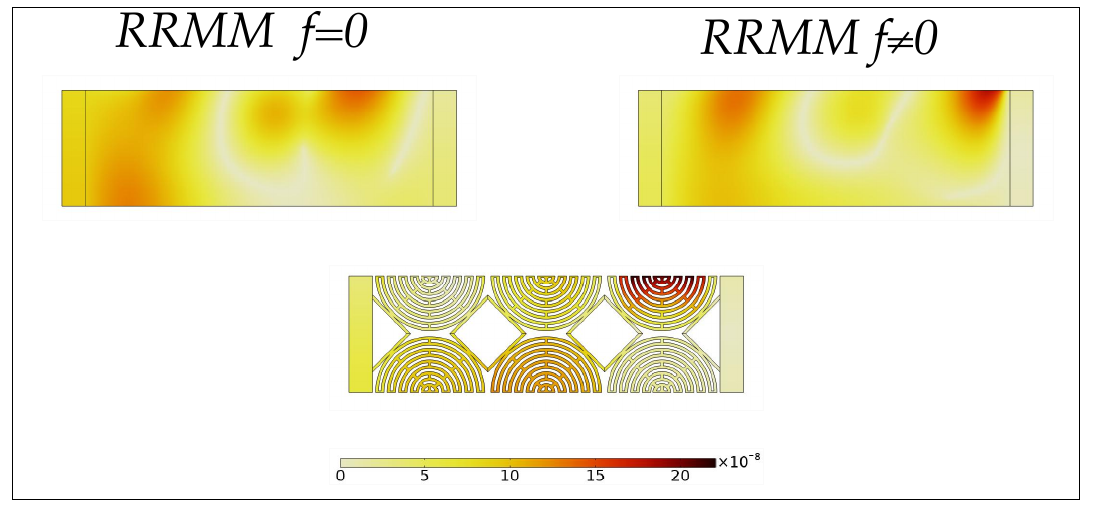}};
        \node[anchor=south] at ($(image.north west)!0.5!(image.north east)$) [yshift=0.1em] {\huge \textbf{340 Hz}};
    \end{tikzpicture}
    \caption{Comparison of the displacement field of the metamaterial specimen $\Delta$ with the RRMM when $f=0$ and $f\neq 0$ at 340 Hz. For $f\neq 0$, we have: $\alpha_{L_x}=1.3$, $\beta_{L_x}=0$, $\alpha_{L_y}=1$, $\beta_{L_y}=0$, $\alpha_{R_x}=1$, $\beta_{R_x}=0$, $\alpha_{R_y}=1$, $\beta_{R_y}=0$.
     An interface force was used on the top ``free" RRMM boundary with the expression: $f^{int}=-50\exp(\frac{-(x-0.08)^2}{0.0002})$.}
    \label{fig:disp340_delta}
\end{figure}

Figs.~\ref{fig:disp340_alpha}~-~\ref{fig:disp340_delta} show that in this frequency interval the wavelength is comparable to the size of the specimen, therefore more complex expressions of $f^{\rm interface}$ would be needed compared to linear ones (\ref{eq:interface_force_a-1}).
Indeed, it is apparent that interface forces of the form~(\ref{eq:interface_force_a-1}) improve the solutions for all ``cuts", but sometimes some details of the microstructured (``real") solution are not entirely captured.
When the frequency increases again, approaching the band-gap region, expression~(\ref{eq:interface_force_a-1}) is again sufficient to catch the overall solution.
These findings point out to the general result that macroscopic interface effects can be predominant as soon as the wavelength is comparable to the specimen's size or smaller.


Particular attention should be paid to cases where one wants to use homogenized models specifically for frequencies with corresponding wavelengths comparable to or smaller than the specimen's size, when these frequencies are not ``close enough" to a band-gap region in order for the metamaterial’s attenuation mechanisms to start being activated.

%
%
%
\subsubsection{Frequency: 420 Hz}
We continue analyzing the homogenized Cauchy and RRM simulations and the corresponding comparison to the microstructured ones for the frequency of 420 Hz.
\rev{Here, the wavelength is smaller than the one at 340 Hz (see Table~\ref{tab:frequencies_wavelengths}) which would indicate more pronounced boundary effects,  but because we are now close to a band-gap region, the attenuation mechanisms of the metamaterial are being activated, causing some destructive interference of the waves instead of more pronounced boundary effects. Hence, expression~(\ref{eq:interface_force_a-1}) is again sufficient to capture the overall response.}
\begin{figure}[h!]
    \centering
    \begin{tikzpicture}
        \node[anchor=south west,inner sep=0] (image) at (0,0) {\includegraphics[width=0.9\textwidth]{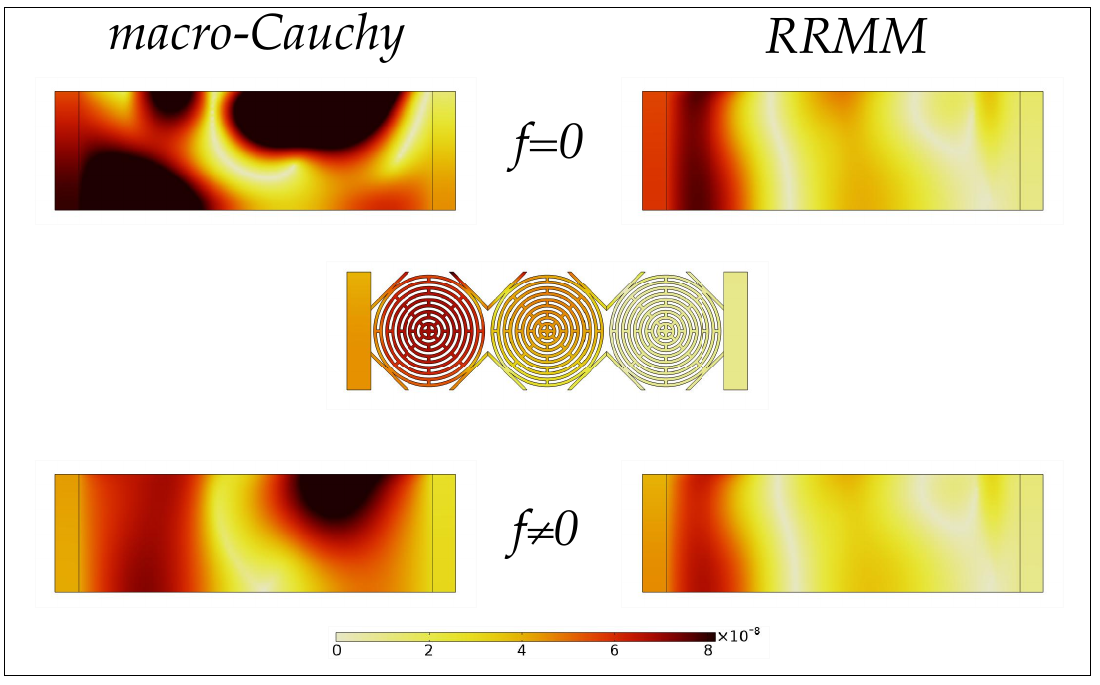}};
        \node[anchor=south] at ($(image.north west)!0.5!(image.north east)$) [yshift=0.1em] {\huge \textbf{420 Hz}};
    \end{tikzpicture}
    \caption{Comparison of the displacement field of the metamaterial specimen $\Alpha$ with the macro-Cauchy and the RRMM when $f=0$ and $f\neq 0$ at 420 Hz. When $f\neq 0$ for the RRMM, we have: $\alpha_{L_x}=1.3$, $\beta_{L_x}=0$, $\alpha_{L_y}=1$, $\beta_{L_y}=2$, $\alpha_{R_x}=0.8$, $\beta_{R_x}=0$, $\alpha_{R_y}=1$ and $\beta_{R_y}=0$, while for the macro Cauchy: $\alpha_{L_x}=1$, $\beta_{L_x}=0$, $\alpha_{L_y}=1$, $\beta_{L_y}=0$, $\alpha_{R_x}=-0.4$, $\beta_{R_x}=0$, $\alpha_{R_y}=1$ and $\beta_{R_y}=0$.}
    \label{fig:disp420_alpha}
\end{figure}
\begin{figure}[h!]
    \centering
    \begin{tikzpicture}
        \node[anchor=south west,inner sep=0] (image) at (0,0) {\includegraphics[width=0.9\textwidth]{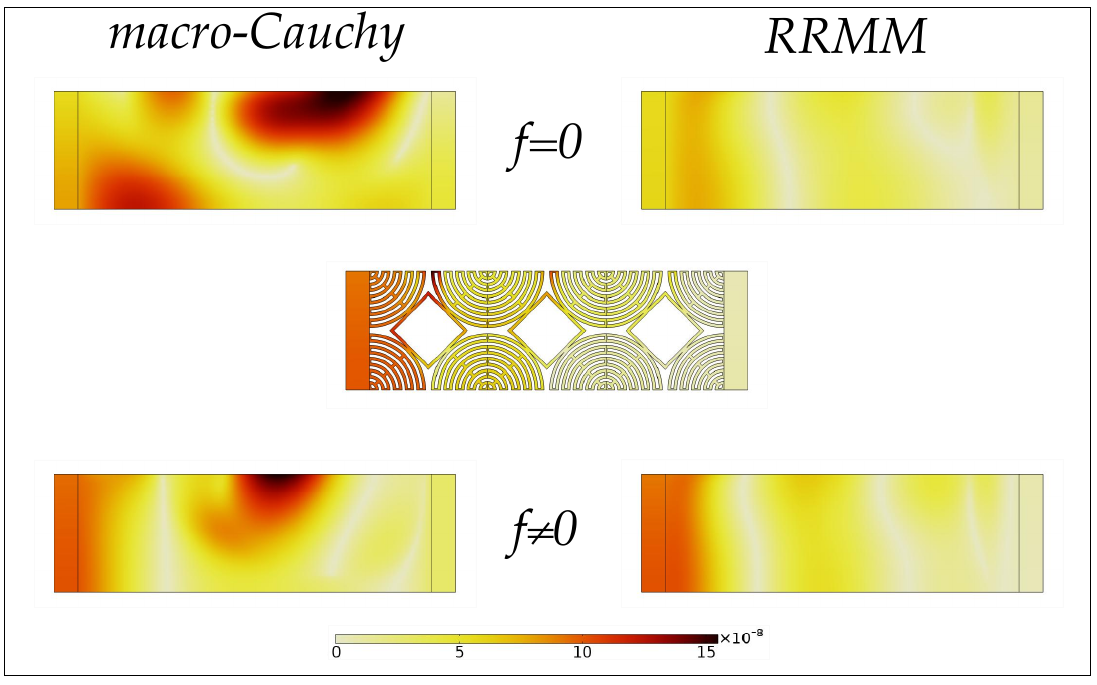}};
        \node[anchor=south] at ($(image.north west)!0.5!(image.north east)$) [yshift=0.1em] {\huge \textbf{420 Hz}};
    \end{tikzpicture}
    \caption{Comparison of the displacement field of the metamaterial specimen $\Beta$ with the macro-Cauchy and the RRMM when $f=0$ and $f\neq 0$ at 420 Hz. When $f\neq 0$ for the RRMM, we have: $\alpha_{L_x}=0.5$, $\beta_{L_x}=0$, $\alpha_{L_y}=1$, $\beta_{L_y}=5$, $\alpha_{R_x}=1.1$, $\beta_{R_x}=0$, $\alpha_{R_y}=1$ and $\beta_{R_y}=-5$, while for the macro Cauchy: $\alpha_{L_x}=-1$, $\beta_{L_x}=0$, $\alpha_{L_y}=1$, $\beta_{L_y}=0$, $\alpha_{R_x}=0.3$, $\beta_{R_x}=0$, $\alpha_{R_y}=1$ and $\beta_{R_y}=0$.}
\label{fig:disp420_beta}
\end{figure}
\begin{figure}[h!]
    \centering
    \begin{tikzpicture}
        \node[anchor=south west,inner sep=0] (image) at (0,0) {\includegraphics[width=0.9\textwidth]{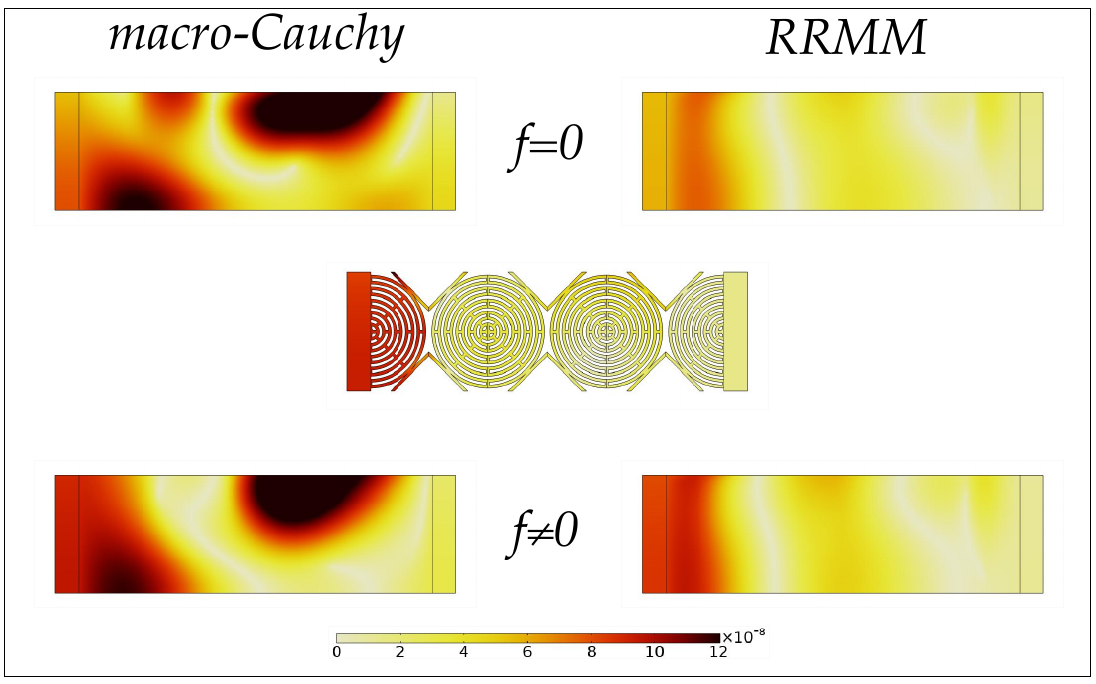}};
        \node[anchor=south] at ($(image.north west)!0.5!(image.north east)$) [yshift=0.1em] {\huge \textbf{420 Hz}};
    \end{tikzpicture}
    \caption{Comparison of the displacement field of the metamaterial specimen $\Gamma$ with the macro-Cauchy and the RRMM when $f=0$ and $f\neq 0$ at 420 Hz. When $f\neq 0$ for the RRMM, we have: $\alpha_{L_x}=0.6$, $\beta_{L_x}=0$, $\alpha_{L_y}=1$, $\beta_{L_y}=2$, $\alpha_{R_x}=1.4$, $\beta_{R_x}=0$, $\alpha_{R_y}=1$ and $\beta_{R_y}=-5$, while for the macro Cauchy: $\alpha_{L_x}=0.5$, $\beta_{L_x}=0$, $\alpha_{L_y}=1$, $\beta_{L_y}=0$, $\alpha_{R_x}=0.3$, $\beta_{R_x}=0$, $\alpha_{R_y}=1$ and $\beta_{R_y}=0$.}
\label{fig:disp420_gamma}
\end{figure}
\begin{figure}[h!]
    \centering
    \begin{tikzpicture}
        \node[anchor=south west,inner sep=0] (image) at (0,0) {\includegraphics[width=0.9\textwidth]{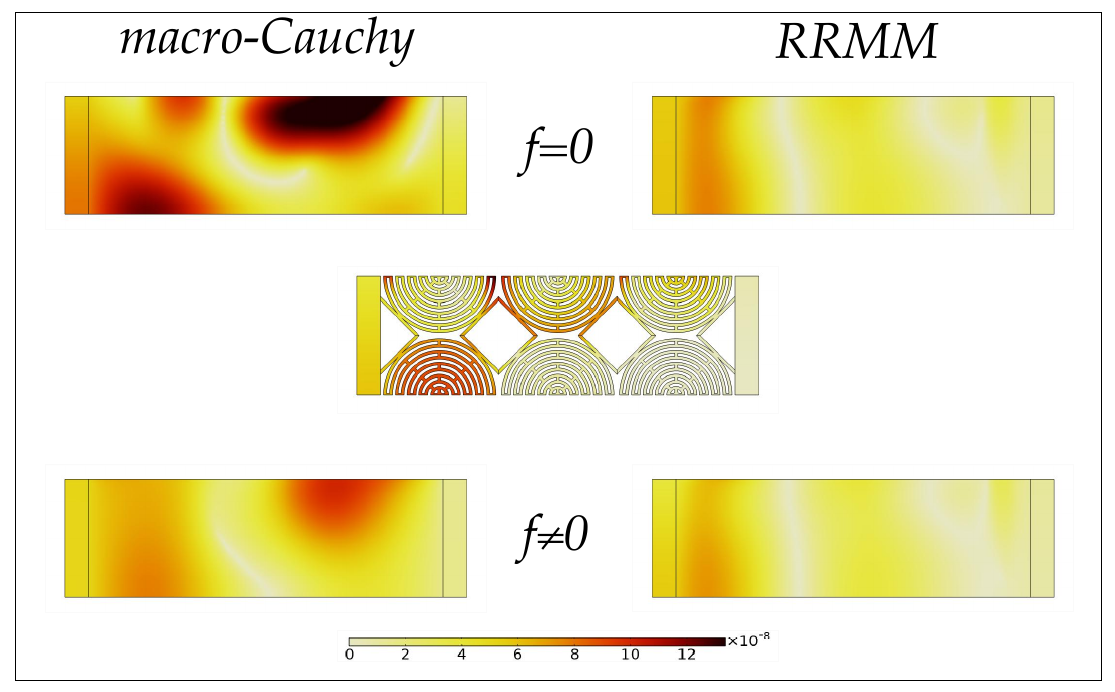}};
        \node[anchor=south] at ($(image.north west)!0.5!(image.north east)$) [yshift=0.1em] {\huge \textbf{420 Hz}};
    \end{tikzpicture}
    \caption{Comparison of the displacement field of the metamaterial specimen $\Delta$ with the macro-Cauchy and the RRMM when $f=0$ and $f\neq 0$ at 420 Hz. When $f\neq 0$ for the RRMM, we have: $\alpha_{L_x}=1.2$, $\beta_{L_x}=0$, $\alpha_{L_y}=1$, $\beta_{L_y}=10$, $\alpha_{R_x}=0.8$, $\beta_{R_x}=0$, $\alpha_{R_y}=1$ and $\beta_{R_y}=0$, while for the macro Cauchy: $\alpha_{L_x}=1$, $\beta_{L_x}=0$, $\alpha_{L_y}=1$, $\beta_{L_y}=0$, $\alpha_{R_x}=-0.2$, $\beta_{R_x}=0$, $\alpha_{R_y}=1$ and $\beta_{R_y}=0$.}
    \label{fig:disp420_delta}
\end{figure}
Moreover, the dispersion curves start showing a strong dispersive behavior (see Fig.~\ref{fig:DC_freq}).
This implies that, while the RRMM gives good results when introducing the suitable interface forces, the long-wavelength limit Cauchy model is not able anymore to recover the correct behavior even when interface forces are triggered.
The long-wavelength limit Cauchy model gives rise to unphysical responses which are due to the fact that dispersion cannot be described in the framework of Cauchy linear-elasticity.

\begin{figure}[h!]
    \centering
    \includegraphics[width=0.45\textwidth]{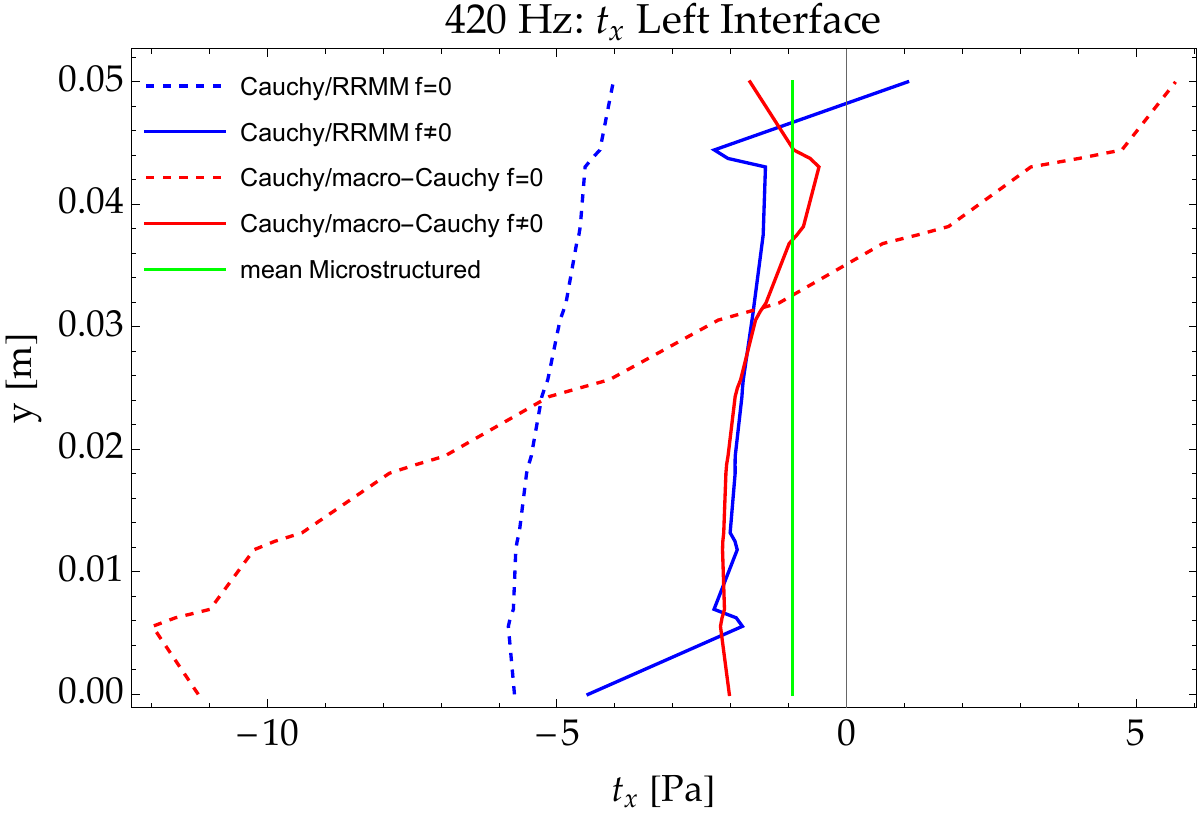}
    \vspace{2mm}
    \includegraphics[width=0.45\textwidth]{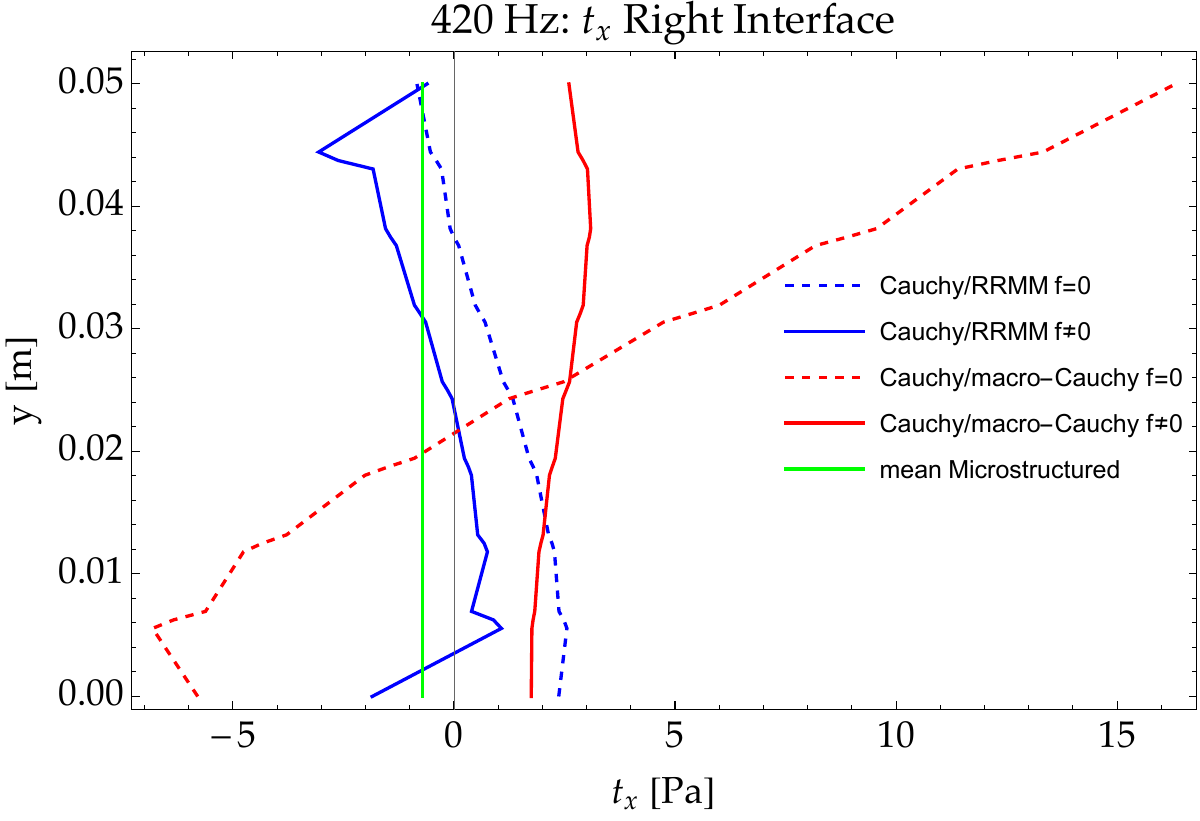}
    \caption{Tractions on the Cauchy side of the Cauchy plate/metamaterial interfaces (left and right) for the RRMM and for the macro-Cauchy when $f=0$ and $f\neq 0$. The tractions shown here are those relative to ``cut" $\Beta$. Analogous reasoning holds true for all other ``cuts".}
\label{fig:rrmm_macro_tractions_420}
\end{figure}
Here, we can observe that the initial (without interface forces correction) tractions for the Cauchy model are not only quantitatively less close to the mean microstructured traction than the RRM one, but we can also start seeing a qualitative deviation of the macro-Cauchy initial traction. 
This is because a correct description of the bulk response becomes more and more important at higher frequencies where dispersion becomes higher. 
Moreover, we can see that while the corrected RRM and macro-Cauchy tractions with interface forces are comparable on the left interface, the RRM one is by far better on the right interface.
We thus see how the RRM response starts having better performances as soon as frequency increases.
We note that according to Table~\ref{tab:frequencies_wavelengths}, the wavelength is already smaller than twice the size of the structure and therefore comparable to the size of the unit cell.
The overall response starts deviating from a plane-wave-like propagation and the so-called separation of scales doesn't hold in this case ($\lambda\gg L_{c}$).
Nevertheless, the RRMM is still able to recover a good solution as soon as considering suitable interface forces.
This fact points out again the solidity of our inverse (top-down) procedure that unlike classical homogenization procedures (bottom-up) doesn't need to introduce hypotheses like the separation of scales.
\FloatBarrier

%
%
%
\subsubsection{Frequency: 460 Hz}
We continue analyzing the homogenized Cauchy and RRM simulations and the corresponding comparison to the microstructured ones for the frequency of 460 Hz. 
This frequency shows strong dispersive behavior (see the seventh point in Fig.~\ref{fig:DC_freq}) and is directly below the band-gap.

\begin{figure}[h!]
\centering
\begin{tikzpicture}
        \node[anchor=south west,inner sep=0] (image) at (0,0) {\includegraphics[width=0.9\textwidth]{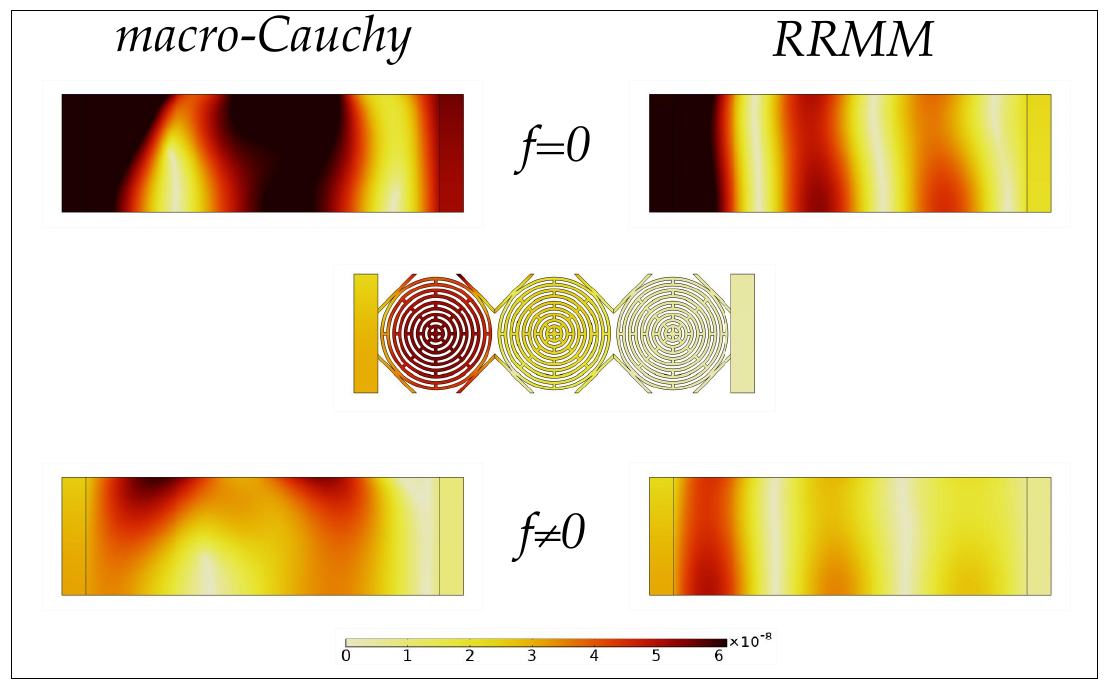}};
        \node[anchor=south] at ($(image.north west)!0.5!(image.north east)$) [yshift=0.1em] {\huge \textbf{460 Hz}};
    \end{tikzpicture}
\caption{Comparison of the displacement field of the metamaterial specimen $\Alpha$ with the macro-Cauchy and the RRMM when $f=0$ and $f\neq 0$ at 460 Hz. When $f\neq 0$ for the RRMM, we have: $\alpha_{L_x}=2$, $\beta_{L_x}=0$, $\alpha_{L_y}=1$, $\beta_{L_y}=4$, $\alpha_{R_x}=-0.4$, $\beta_{R_x}=0$, $\alpha_{R_y}=1$ and $\beta_{R_y}=0$, while for the macro Cauchy: $\alpha_{L_x}=1.9$, $\beta_{L_x}=0$, $\alpha_{L_y}=1$, $\beta_{L_y}=10$, $\alpha_{R_x}=0.2$, $\beta_{R_x}=0$, $\alpha_{R_y}=1$ and $\beta_{R_y}=0$.}
\label{fig:disp460_alpha}
\end{figure}
\begin{figure}[h!]
\centering
\begin{tikzpicture}
        \node[anchor=south west,inner sep=0] (image) at (0,0) {\includegraphics[width=0.9\textwidth]{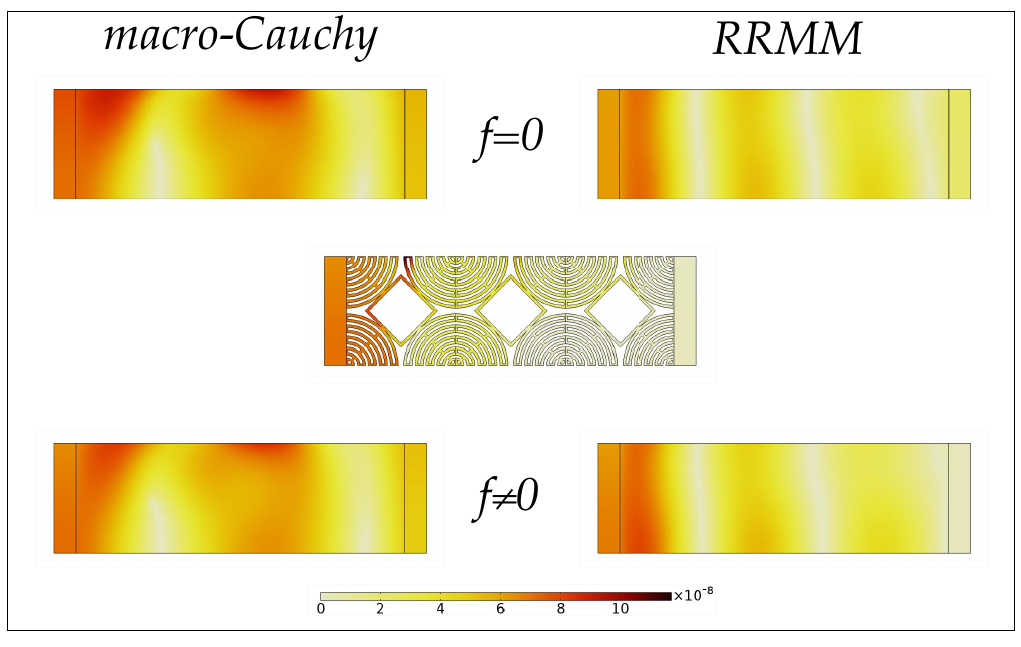}};
        \node[anchor=south] at ($(image.north west)!0.5!(image.north east)$) [yshift=0.1em] {\huge \textbf{460 Hz}};
    \end{tikzpicture}
\caption{Comparison of the displacement field of the metamaterial specimen $\Beta$ with the macro-Cauchy and the RRMM when $f=0$ and $f\neq 0$ at 460 Hz. When $f\neq 0$ for the RRMM, we have: $\alpha_{L_x}=0.8$, $\beta_{L_x}=0$, $\alpha_{L_y}=1$, $\beta_{L_y}=3$, $\alpha_{R_x}=0.1$, $\beta_{R_x}=0$, $\alpha_{R_y}=1$ and $\beta_{R_y}=0$, while for the macro Cauchy: $\alpha_{L_x}=1.2$, $\beta_{L_x}=0$, $\alpha_{L_y}=1$, $\beta_{L_y}=10$, $\alpha_{R_x}=1$, $\beta_{R_x}=0$, $\alpha_{R_y}=1$ and $\beta_{R_y}=0$.}
\label{fig:disp460_beta}
\end{figure}
\begin{figure}[h!]
\centering
\begin{tikzpicture}
        \node[anchor=south west,inner sep=0] (image) at (0,0) {\includegraphics[width=0.9\textwidth]{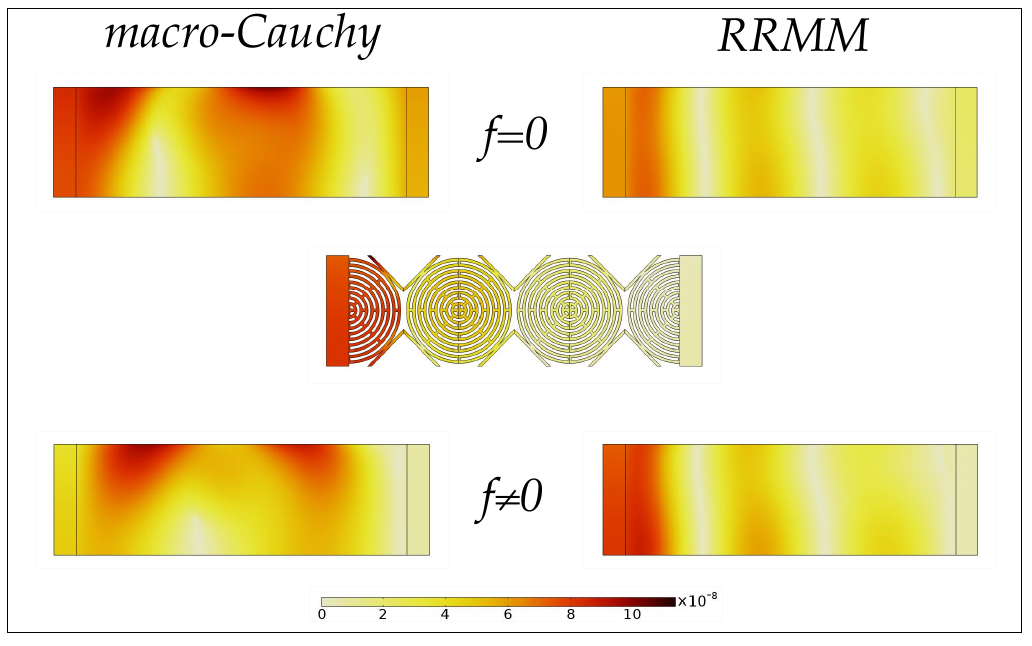}};
        \node[anchor=south] at ($(image.north west)!0.5!(image.north east)$) [yshift=0.1em] {\huge \textbf{460 Hz}};
    \end{tikzpicture}
\caption{Comparison of the displacement field of the metamaterial specimen $\Gamma$ with the macro-Cauchy and the RRMM when $f=0$ and $f\neq 0$ at 460 Hz. When $f\neq 0$ for the RRMM, we have: $\alpha_{L_x}=0.55$, $\beta_{L_x}=0$, $\alpha_{L_y}=1$, $\beta_{L_y}=4$, $\alpha_{R_x}=0.2$, $\beta_{R_x}=0$, $\alpha_{R_y}=1$ and $\beta_{R_y}=0$, while for the macro Cauchy: $\alpha_{L_x}=1.1$, $\beta_{L_x}=0$, $\alpha_{L_y}=1$, $\beta_{L_y}=10$, $\alpha_{R_x}=1$, $\beta_{R_x}=0$, $\alpha_{R_y}=1$ and $\beta_{R_y}=0$.}
\label{fig:disp460_gamma}
\end{figure}
\begin{figure}[h!]
\centering
\begin{tikzpicture}
        \node[anchor=south west,inner sep=0] (image) at (0,0) {\includegraphics[width=0.9\textwidth]{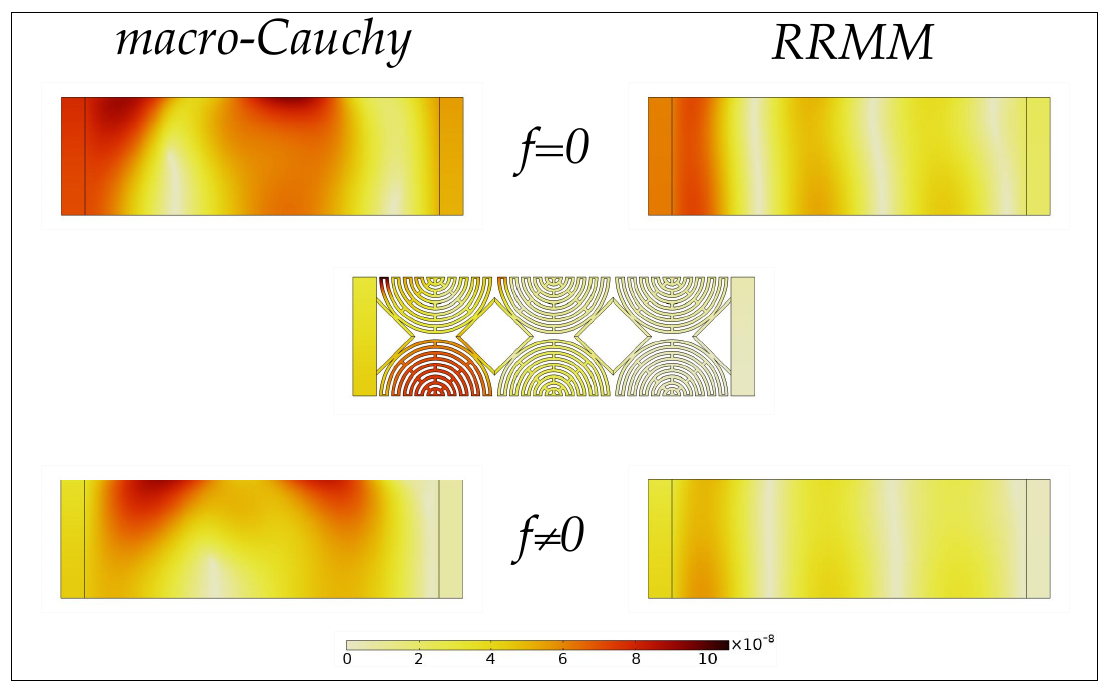}};
        \node[anchor=south] at ($(image.north west)!0.5!(image.north east)$) [yshift=0.1em] {\huge \textbf{460 Hz}};
    \end{tikzpicture}
\caption{Comparison of the displacement field of the metamaterial specimen $\Delta$ with the macro-Cauchy and the RRMM when $f=0$ and $f\neq 0$ at 460 Hz. When $f\neq 0$ for the RRMM, we have: $\alpha_{L_x}=1.7$, $\beta_{L_x}=0$, $\alpha_{L_y}=1$, $\beta_{L_y}=6$, $\alpha_{R_x}=0.1$, $\beta_{R_x}=0$, $\alpha_{R_y}=1$ and $\beta_{R_y}=0$, while for the macro Cauchy: $\alpha_{L_x}=1$, $\beta_{L_x}=0$, $\alpha_{L_y}=1$, $\beta_{L_y}=10$, $\alpha_{R_x}=0.1$, $\beta_{R_x}=0$, $\alpha_{R_y}=1$ and $\beta_{R_y}=-8$.}
\label{fig:disp460_delta}
\end{figure}
For the frequency of 460 Hz considerations analogous to the frequency of 420 Hz hold.

\begin{figure}[h!]
    \centering
    \includegraphics[width=0.45\textwidth]{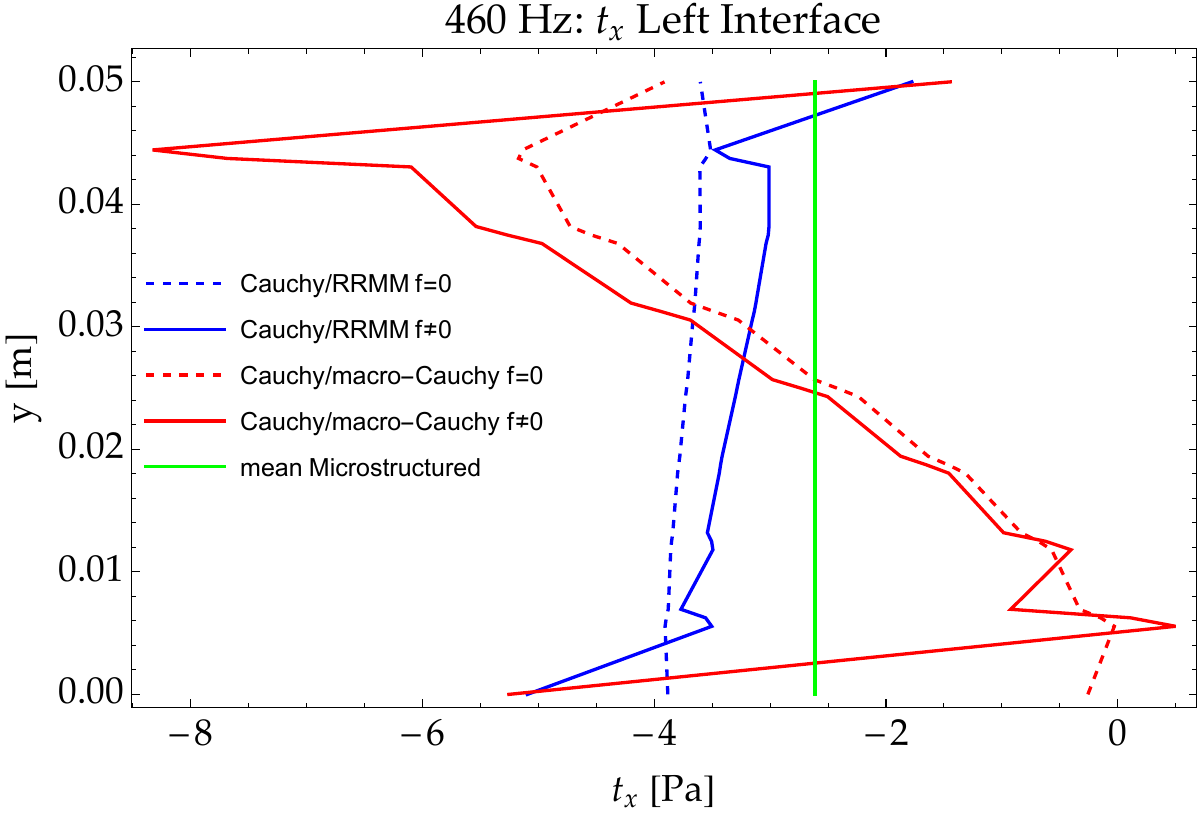}
    \vspace{2mm}
    \includegraphics[width=0.45\textwidth]{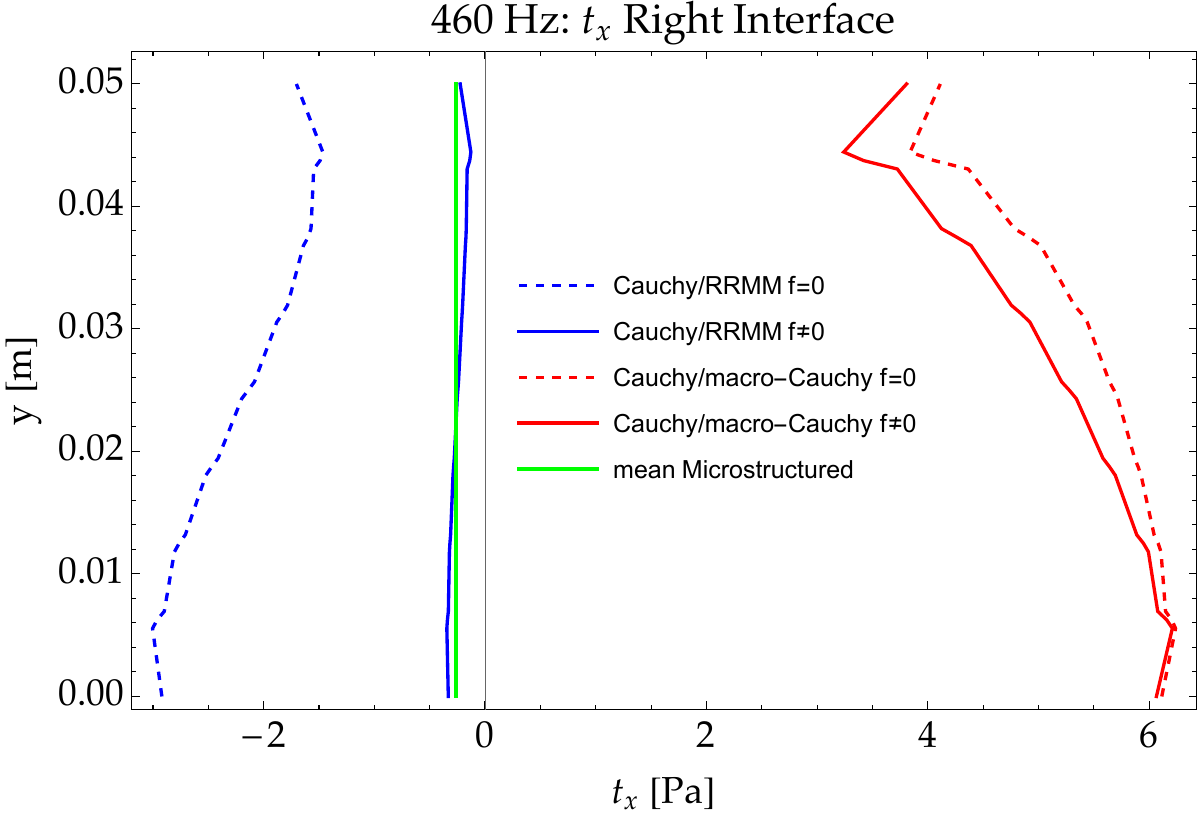}
    \caption{Tractions on the Cauchy side of the Cauchy plate/metamaterial interfaces (left and right) for the RRMM and for the macro-Cauchy when $f=0$ and $f\neq 0$. The tractions shown here are those relative to ``cut" $\Beta$. Analogous reasoning holds true for all other ``cuts".}
\label{fig:rrmm_macro_tractions_460}
\end{figure}
At the present frequency, it becomes more evident how the RRMM outperforms with respect to the macro-Cauchy as soon as suitable interface forces are introduced. All the considerations done for the frequency of 420 Hz also apply here.
The wavelength at the present frequency is closer to the size of the unit cell than the previous case. 
The separation of scales is not holding here, but the RRMM still gives good solutions.
\FloatBarrier

%
%
%
\subsubsection{Frequency: 700 Hz}
We continue analyzing the homogenized Cauchy and RRM simulations and the corresponding comparison to the microstructured ones for the frequency of 700 Hz. 
This frequency is in the lower part of the band-gap (see the eighth point in Fig.~\ref{fig:DC_freq}).

\begin{figure}[h!]
    \centering
    \begin{tikzpicture}
        \node[anchor=south west,inner sep=0] (image) at (0,0) {\includegraphics[width=0.9\textwidth]{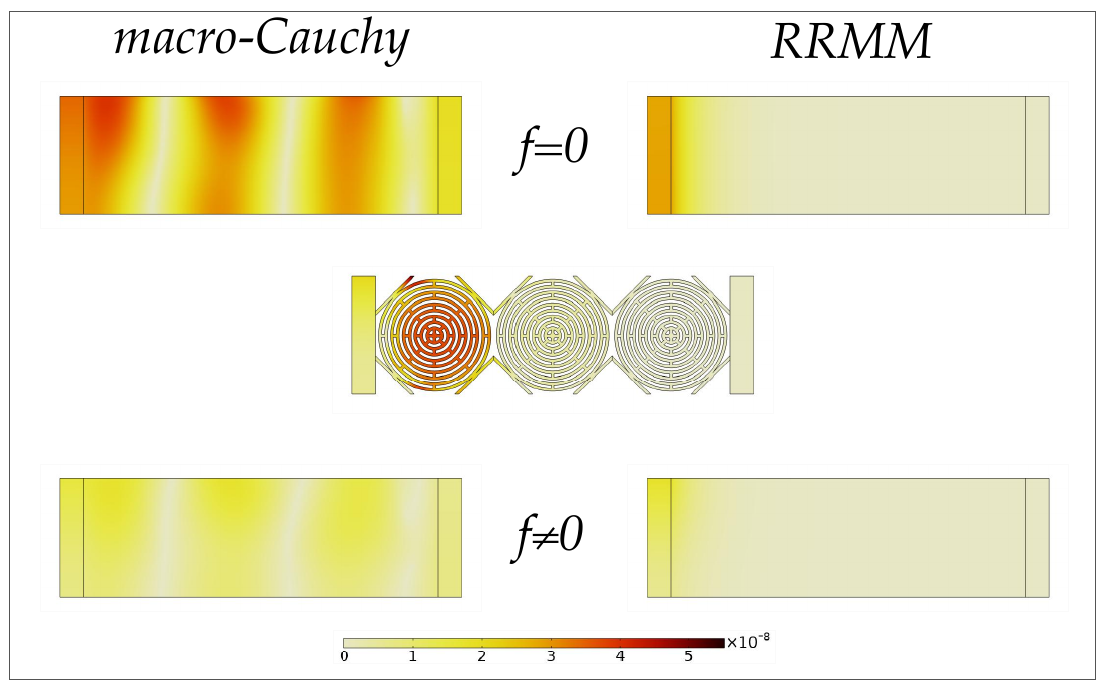}};
        \node[anchor=south] at ($(image.north west)!0.5!(image.north east)$) [yshift=0.1em] {\huge \textbf{700 Hz (lower band-gap)}};
    \end{tikzpicture}
    \caption{Comparison of the displacement field of the metamaterial specimen $\Alpha$ with the macro-Cauchy and the RRMM when $f=0$ and $f\neq 0$ at 700 Hz. When $f\neq 0$ for the RRMM, we have: $\alpha_{L_x}=6$, $\beta_{L_x}=0$, $\alpha_{L_y}=1$, $\beta_{L_y}=5$, $\alpha_{R_x}=1$, $\beta_{R_x}=0$, $\alpha_{R_y}=1$ and $\beta_{R_y}=0$, while for the macro Cauchy: $\alpha_{L_x}=5$, $\beta_{L_x}=0$, $\alpha_{L_y}=1$, $\beta_{L_y}=0$, $\alpha_{R_x}=1$, $\beta_{R_x}=0$, $\alpha_{R_y}=1$ and $\beta_{R_y}=0$.}
\label{fig:disp700_alpha}
\end{figure}
\begin{figure}[h!]
    \centering
    \begin{tikzpicture}
        \node[anchor=south west,inner sep=0] (image) at (0,0) {\includegraphics[width=0.9\textwidth]{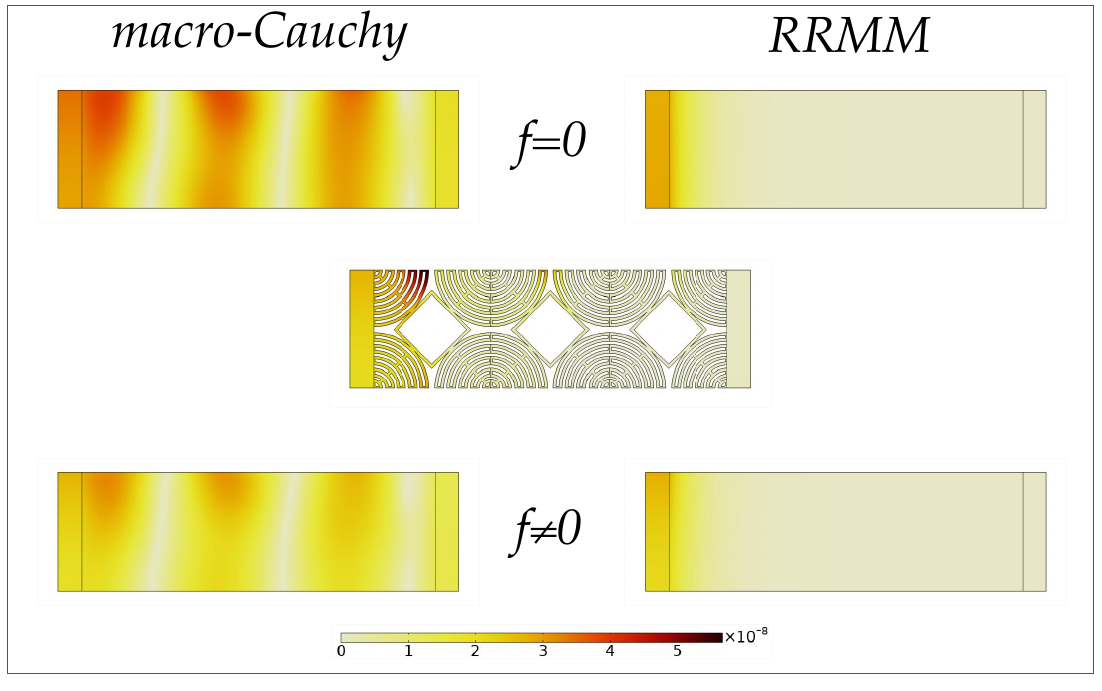}};
        \node[anchor=south] at ($(image.north west)!0.5!(image.north east)$) [yshift=0.1em] {\huge \textbf{700 Hz (lower band-gap)}};
    \end{tikzpicture}
    \caption{Comparison of the displacement field of the metamaterial specimen $\Beta$ with the macro-Cauchy and the RRMM when $f=0$ and $f\neq 0$ at 700 Hz. When $f\neq 0$ for the RRMM, we have: $\alpha_{L_x}=1.6$, $\beta_{L_x}=0$, $\alpha_{L_y}=1$, $\beta_{L_y}=-2$, $\alpha_{R_x}=1$, $\beta_{R_x}=0$, $\alpha_{R_y}=1$ and $\beta_{R_y}=0$, while for the macro Cauchy: $\alpha_{L_x}=2$, $\beta_{L_x}=0$, $\alpha_{L_y}=1$, $\beta_{L_y}=1$, $\alpha_{R_x}=1$, $\beta_{R_x}=0$, $\alpha_{R_y}=1$ and $\beta_{R_y}=0$.}
    \label{fig:disp700_beta}
\end{figure}
    \begin{figure}[h!]
    \centering
    \begin{tikzpicture}
        \node[anchor=south west,inner sep=0] (image) at (0,0) {\includegraphics[width=0.9\textwidth]{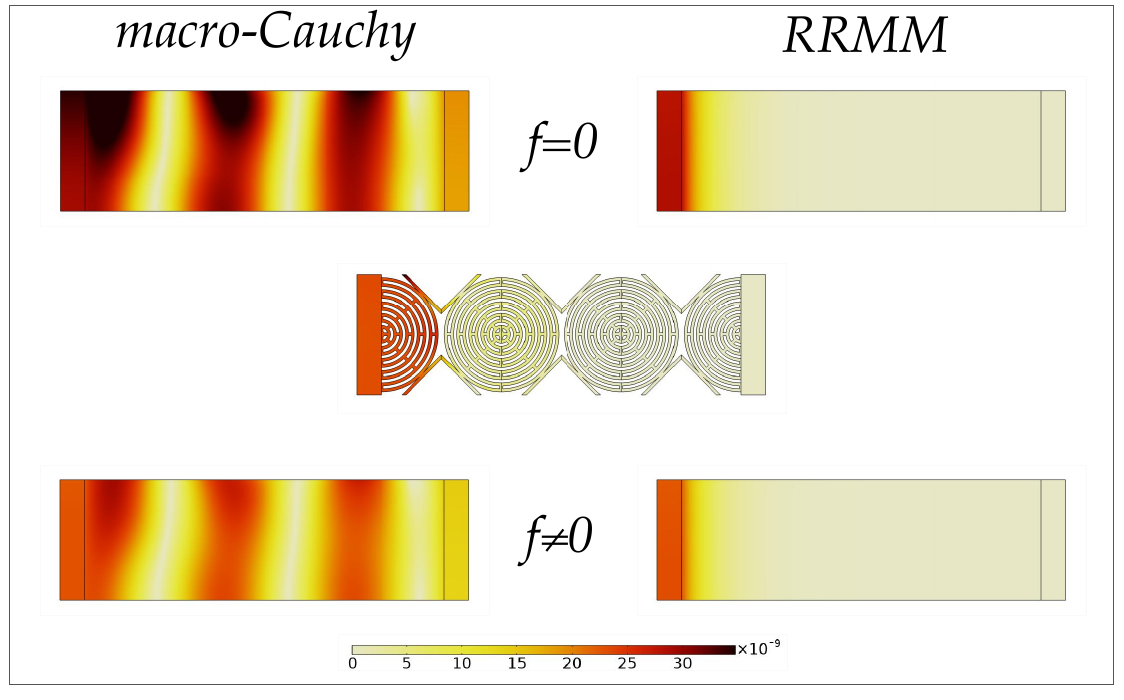}};
        \node[anchor=south] at ($(image.north west)!0.5!(image.north east)$) [yshift=0.1em] {\huge \textbf{700 Hz (lower band-gap)}};
    \end{tikzpicture}
    \caption{Comparison of the displacement field of the metamaterial specimen $\Gamma$ with the macro-Cauchy and the RRMM when $f=0$ and $f\neq 0$ at 700 Hz. When $f\neq 0$ for the RRMM, we have: $\alpha_{L_x}=1.7$, $\beta_{L_x}=0$, $\alpha_{L_y}=1$, $\beta_{L_y}=0$, $\alpha_{R_x}=1$, $\beta_{R_x}=0$, $\alpha_{R_y}=1$ and $\beta_{R_y}=0$, while for the macro Cauchy: $\alpha_{L_x}=2$, $\beta_{L_x}=0$, $\alpha_{L_y}=1$, $\beta_{L_y}=5$, $\alpha_{R_x}=1$, $\beta_{R_x}=0$, $\alpha_{R_y}=1$ and $\beta_{R_y}=0$.}
    \label{fig:disp700_gamma}
\end{figure}
\begin{figure}[h!]
    \centering
    \begin{tikzpicture}
        \node[anchor=south west,inner sep=0] (image) at (0,0) {\includegraphics[width=0.9\textwidth]{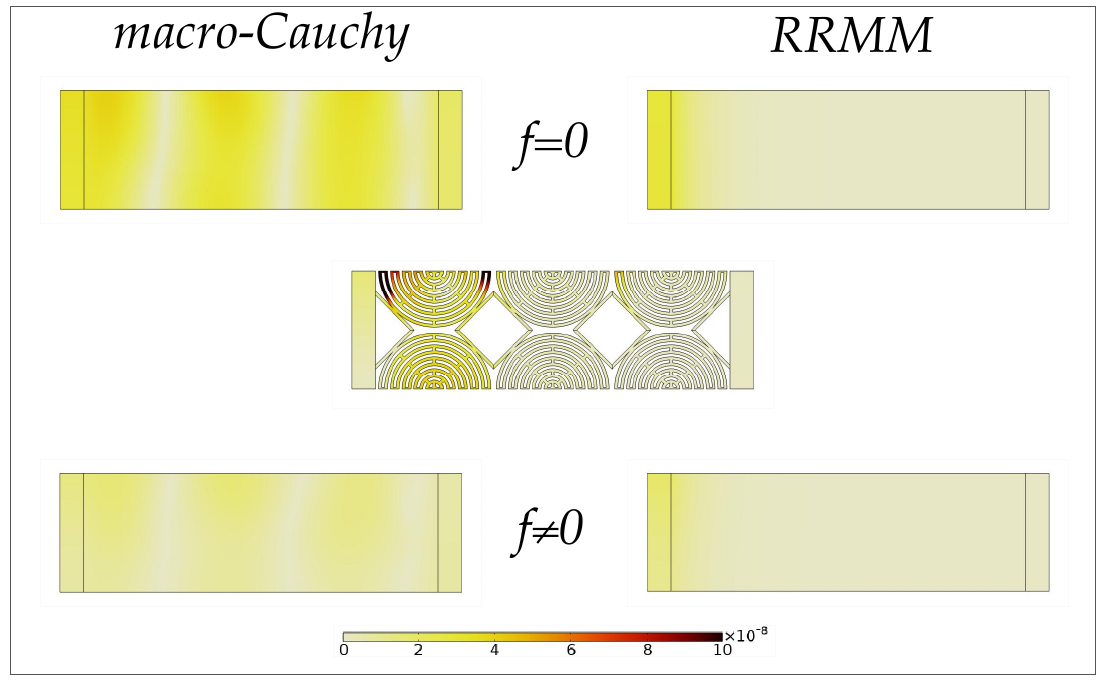}};
        \node[anchor=south] at ($(image.north west)!0.5!(image.north east)$) [yshift=0.1em] {\huge \textbf{700 Hz (lower band-gap)}};
    \end{tikzpicture}
    \caption{Comparison of the displacement field of the metamaterial specimen $\Delta$ with the macro-Cauchy and the RRMM when $f=0$ and $f\neq 0$ at 700 Hz. When $f\neq 0$ for the RRMM, we have: $\alpha_{L_x}=4$, $\beta_{L_x}=0$, $\alpha_{L_y}=1$, $\beta_{L_y}=2$, $\alpha_{R_x}=1$, $\beta_{R_x}=0$, $\alpha_{R_y}=1$ and $\beta_{R_y}=0$, while for the macro Cauchy: $\alpha_{L_x}=5.5$, $\beta_{L_x}=0$, $\alpha_{L_y}=1$, $\beta_{L_y}=0$, $\alpha_{R_x}=1$, $\beta_{R_x}=0$, $\alpha_{R_y}=1$ and $\beta_{R_y}=0$.}
    \label{fig:disp700_delta}
\end{figure}

The frequency of 700 Hz is located in the band-gap region of the considered metamaterial.
Boundary effects are limited to the interface close to the surface where the external load is applied (left interface).
Since wave propagation is not allowed through the metamaterial's bulk, the deformation is concentrated at the first unit cell.
Depending on the unit cell's ``cut", the solution is quite different and macroscopic bending of the left Cauchy plate can occur as a consequence of a heterogeneous micro-deformation field of the unit cells belonging to the first cell's layer close to the left interface.

\FloatBarrier

The RRMM can correctly describe the band-gap behavior of the considered specimen for all 4 ``cuts", as soon as suitable interface forces are introduced. 
On the other hand the long-wavelength limit Cauchy model fails to recover the correct solution also when triggering interface forces. 
This is due to the fact that Cauchy models are not able to describe band-gap behaviors.

\begin{figure}[h!]
    \centering
    \includegraphics[width=0.45\textwidth]{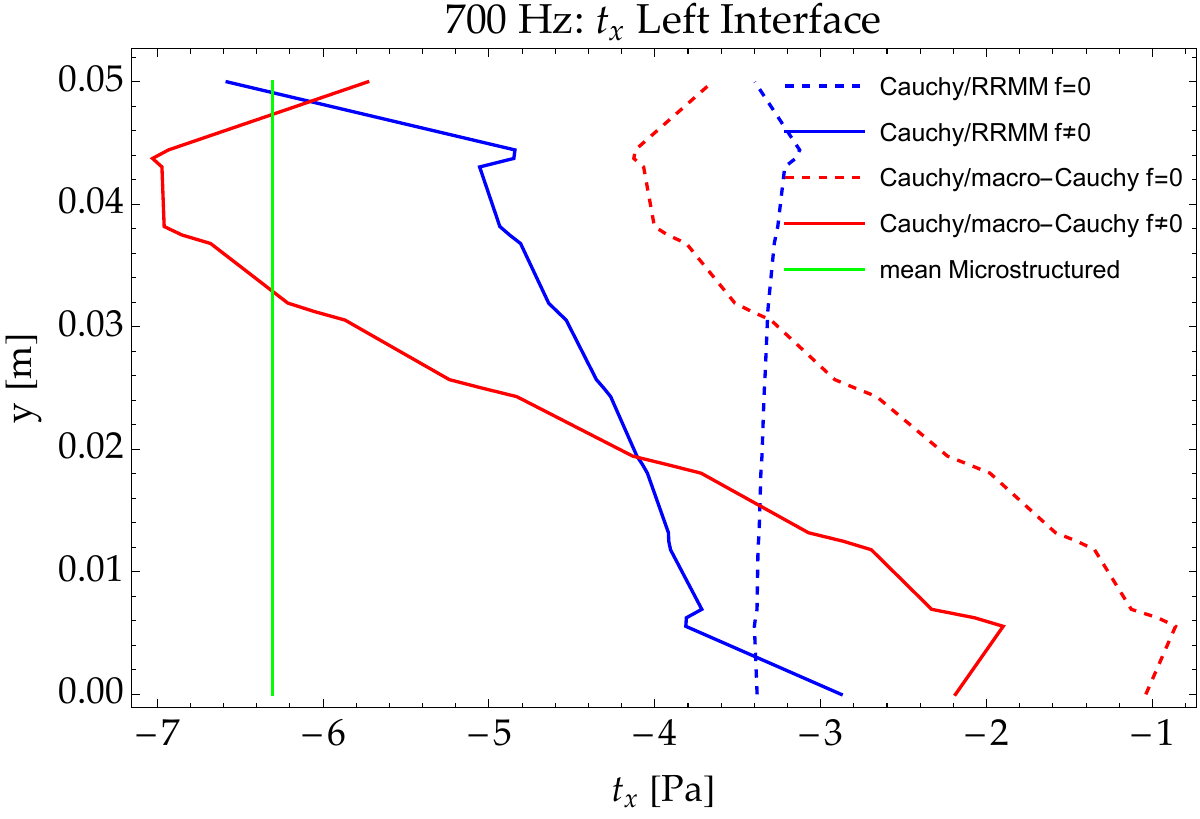}
    \vspace{2mm}
    \includegraphics[width=0.45\textwidth]{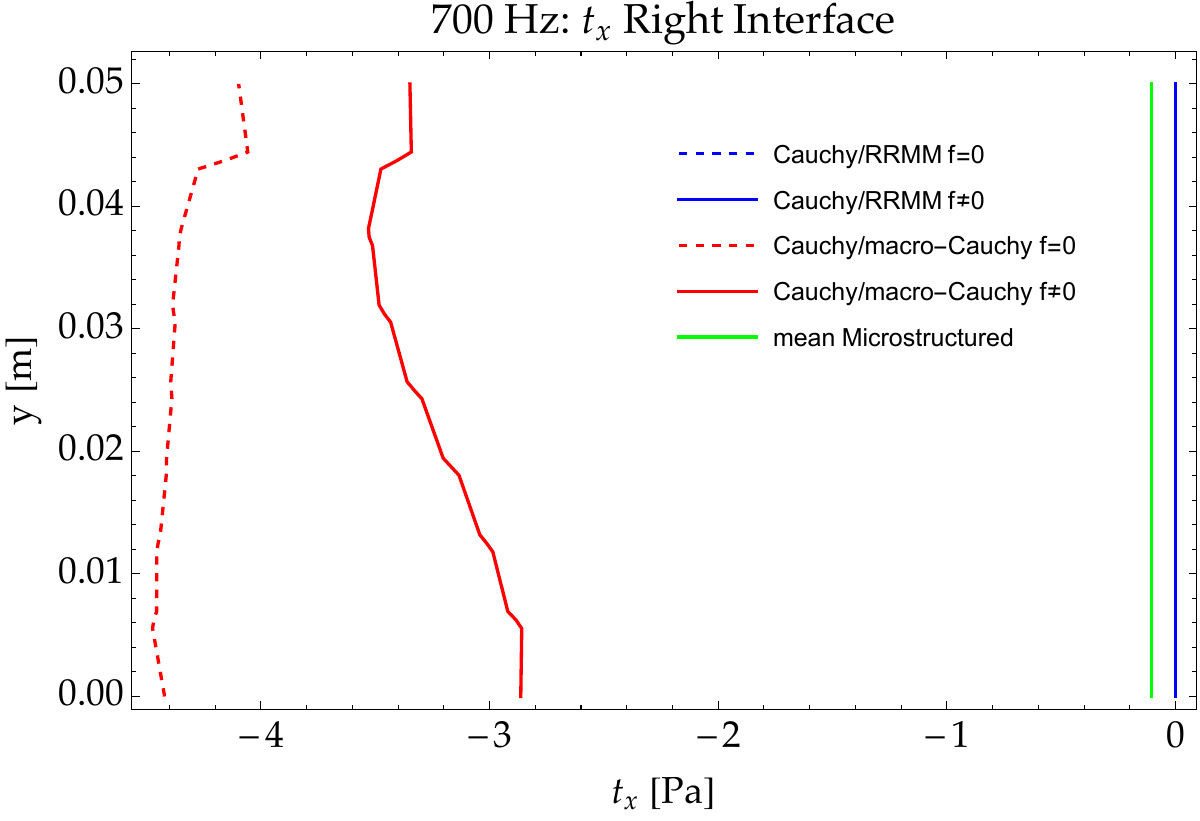}
    \caption{Tractions on the Cauchy side of the Cauchy plate/metamaterial interfaces (left and right) for the RRMM and for the macro-Cauchy when $f=0$ and $f\neq 0$. The tractions shown here are those relative to ``cut" $\Beta$. Analogous reasoning holds true for all other ``cuts".}
    \label{fig:rrmm_macro_tractions_700}
\end{figure}

%
%
%
\subsubsection{Frequency: 1500 Hz}
We continue analyzing the homogenized Cauchy and RRM simulations and the corresponding comparison to the microstructured ones for the frequency of 1500 Hz. 
This frequency is in the upper limit of the band-gap (see the tenth point in Fig.~\ref{fig:DC_freq}).
Considerations similar to the frequency of 700 Hz and 1100 Hz hold here.

\begin{figure}[h!]
    \centering
     \begin{tikzpicture}
        \node[anchor=south west,inner sep=0] (image) at (0,0) {\includegraphics[width=0.9\textwidth]{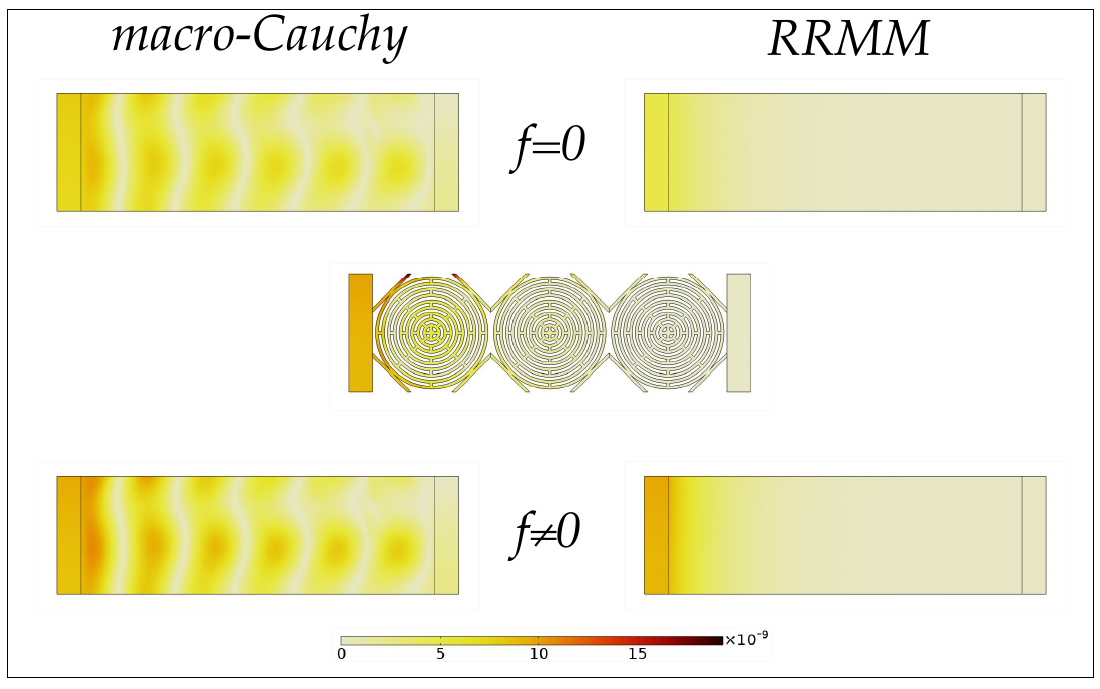}};
        \node[anchor=south] at ($(image.north west)!0.5!(image.north east)$) [yshift=0.1em] {\huge \textbf{1500 Hz (upper band-gap limit)}};
    \end{tikzpicture}
    \caption{Comparison of the displacement field of the metamaterial specimen $\Alpha$ with the macro-Cauchy and the RRMM when $f=0$ and $f\neq 0$ at 1500 Hz. When $f\neq 0$ for the RRMM, we have: $\alpha_{L_x}=-0.01$, $\beta_{L_x}=0$, $\alpha_{L_y}=1$, $\beta_{L_y}=0.5$, $\alpha_{R_x}=1$, $\beta_{R_x}=0$, $\alpha_{R_y}=1$ and $\beta_{R_y}=0$, while for the macro Cauchy: $\alpha_{L_x}=1$, $\beta_{L_x}=-2$, $\alpha_{L_y}=1$, $\beta_{L_y}=0$, $\alpha_{R_x}=1$, $\beta_{R_x}=0$, $\alpha_{R_y}=1$ and $\beta_{R_y}=0$.}
    \label{fig:disp1500_alpha}
\end{figure}
\begin{figure}[h!]
    \centering
    \begin{tikzpicture}
        \node[anchor=south west,inner sep=0] (image) at (0,0) {\includegraphics[width=0.9\textwidth]{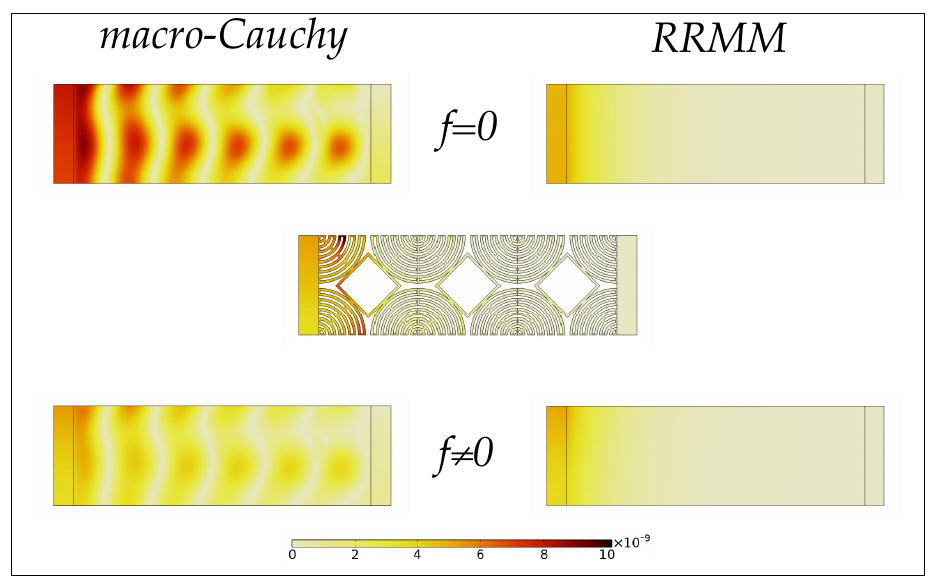}};
        \node[anchor=south] at ($(image.north west)!0.5!(image.north east)$) [yshift=0.1em] {\huge \textbf{1500 Hz (upper band-gap limit)}};
    \end{tikzpicture}
    \caption{Comparison of the displacement field of the metamaterial specimen $\Beta$ with the macro-Cauchy and the RRMM when $f=0$ and $f\neq 0$ at 1500 Hz. When $f\neq 0$ for the RRMM, we have: $\alpha_{L_x}=1.3$, $\beta_{L_x}=0$, $\alpha_{L_y}=1$, $\beta_{L_y}=3$, $\alpha_{R_x}=1$, $\beta_{R_x}=0$, $\alpha_{R_y}=1$ and $\beta_{R_y}=0$, while for the macro Cauchy: $\alpha_{L_x}=1$, $\beta_{L_x}=4.2$, $\alpha_{L_y}=1$, $\beta_{L_y}=1.1$, $\alpha_{R_x}=1$, $\beta_{R_x}=0$, $\alpha_{R_y}=1$ and $\beta_{R_y}=0$.}
    \label{fig:disp1500_beta}
\end{figure}
\begin{figure}[h!]
    \centering
    \begin{tikzpicture}
        \node[anchor=south west,inner sep=0] (image) at (0,0) {\includegraphics[width=0.9\textwidth]{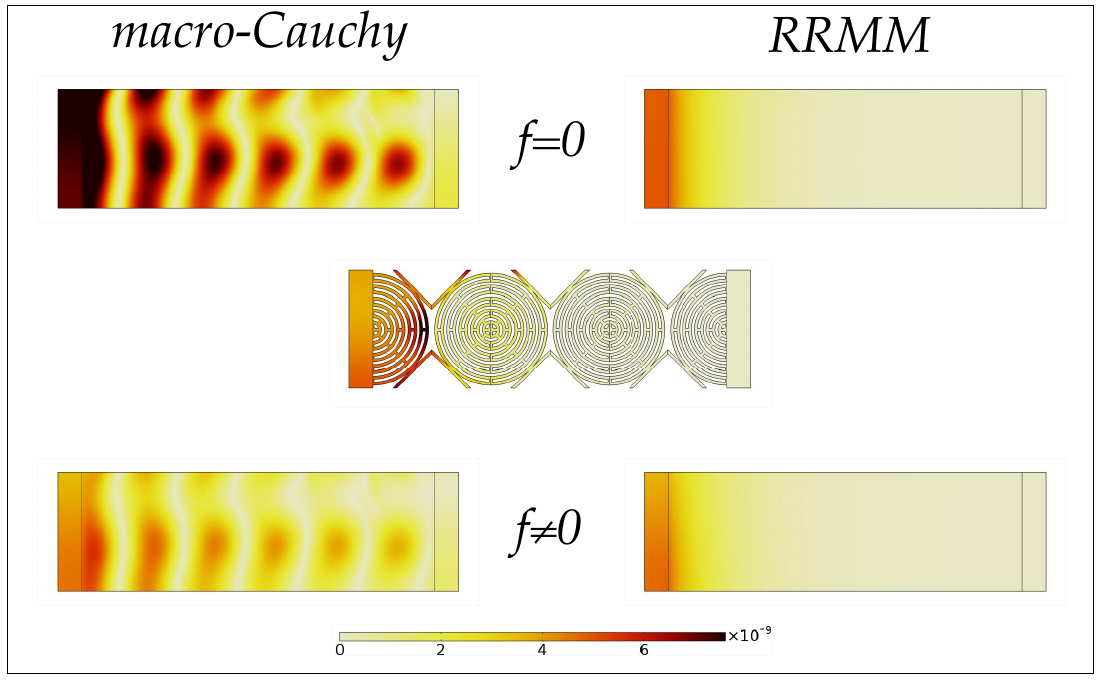}};
        \node[anchor=south] at ($(image.north west)!0.5!(image.north east)$) [yshift=0.1em] {\huge \textbf{1500 Hz (upper band-gap limit)}};
    \end{tikzpicture}
    \caption{Comparison of the displacement field of the metamaterial specimen $\Gamma$ with the macro-Cauchy and the RRMM when $f=0$ and $f\neq 0$ at 1500 Hz. When $f\neq 0$ for the RRMM, we have: $\alpha_{L_x}=1.3$, $\beta_{L_x}=0$, $\alpha_{L_y}=1$, $\beta_{L_y}=-2$, $\alpha_{R_x}=1$, $\beta_{R_x}=0$, $\alpha_{R_y}=1$ and $\beta_{R_y}=0$, while for the macro Cauchy: $\alpha_{L_x}=1$, $\beta_{L_x}=4.4$, $\alpha_{L_y}=1$, $\beta_{L_y}=-1.4$, $\alpha_{R_x}=1$, $\beta_{R_x}=0$, $\alpha_{R_y}=1$ and $\beta_{R_y}=0$.}
    \label{fig:disp1500_gamma}
\end{figure}
\begin{figure}[h!]
    \centering
    \begin{tikzpicture}
        \node[anchor=south west,inner sep=0] (image) at (0,0) {\includegraphics[width=0.9\textwidth]{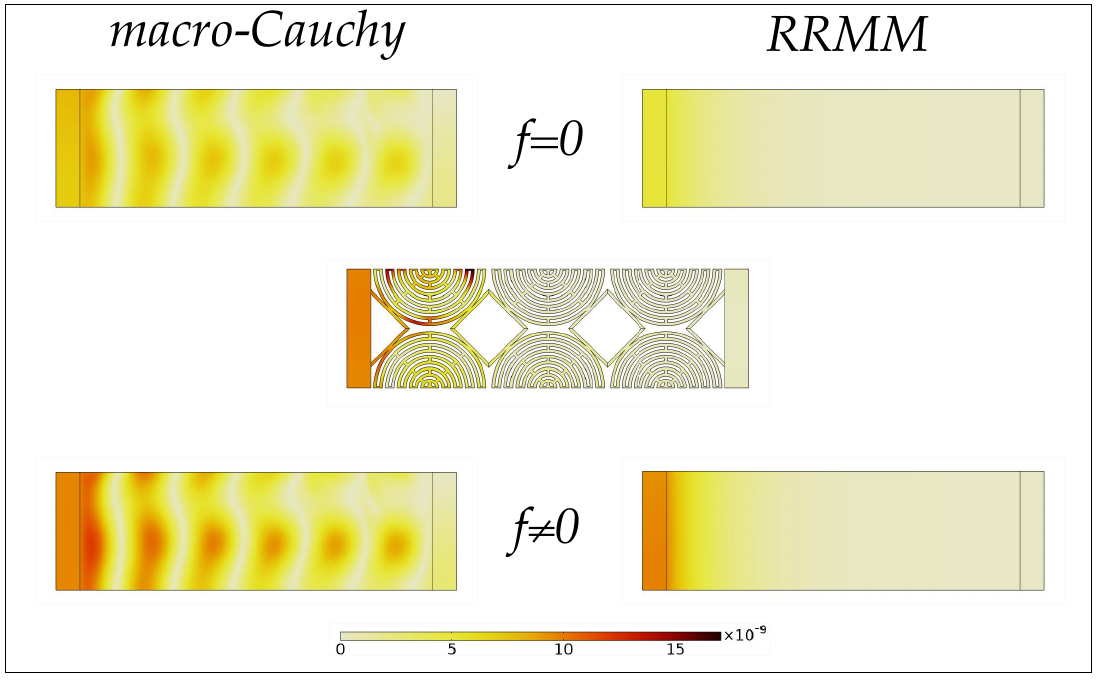}};
        \node[anchor=south] at ($(image.north west)!0.5!(image.north east)$) [yshift=0.1em] {\huge \textbf{1500 Hz (upper band-gap limit)}};
    \end{tikzpicture}
    \caption{Comparison of the displacement field of the metamaterial specimen $\Delta$ with the macro-Cauchy and the RRMM when $f=0$ and $f\neq 0$ at 1500 Hz. When $f\neq 0$ for the RRMM, we have: $\alpha_{L_x}=-0.05$, $\beta_{L_x}=0$, $\alpha_{L_y}=1$, $\beta_{L_y}=0$, $\alpha_{R_x}=1$, $\beta_{R_x}=0$, $\alpha_{R_y}=1$ and $\beta_{R_y}=0$, while for the macro Cauchy: $\alpha_{L_x}=1$, $\beta_{L_x}=-3$, $\alpha_{L_y}=1$, $\beta_{L_y}=-1$, $\alpha_{R_x}=1$, $\beta_{R_x}=0$, $\alpha_{R_y}=1$ and $\beta_{R_y}=0$.}
    \label{fig:disp1500_delta}
\end{figure}
 
\begin{figure}[h!]
    \centering
    \includegraphics[width=0.45\textwidth]{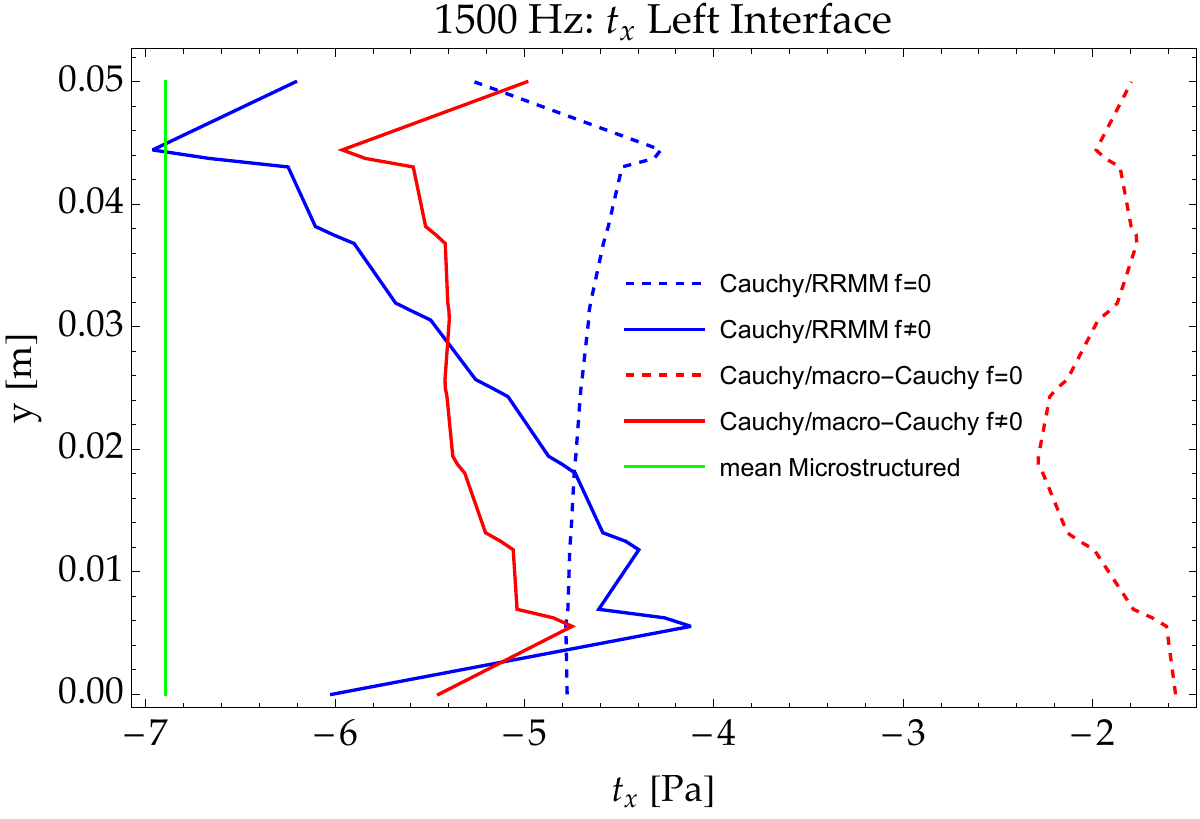}
    \vspace{2mm}
    \includegraphics[width=0.45\textwidth]{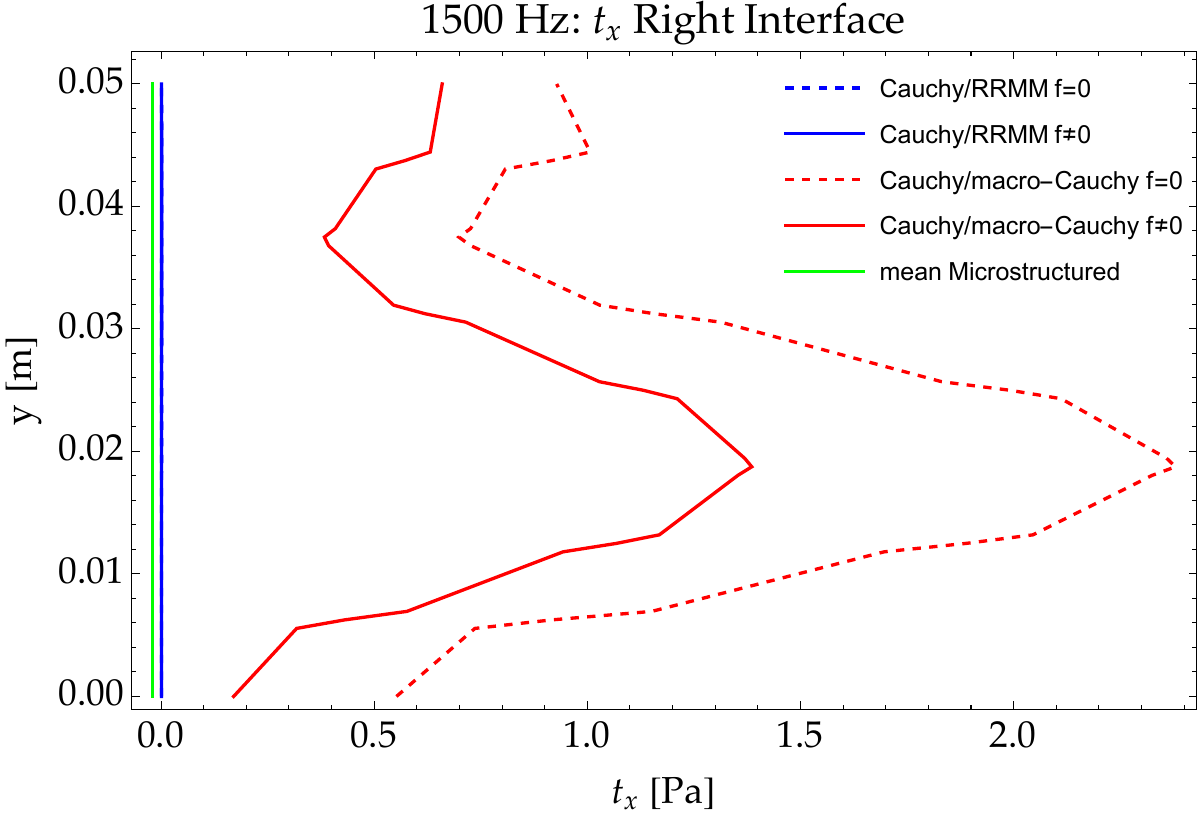}
    \caption{Tractions on the Cauchy side of the Cauchy plate/metamaterial interfaces (left and right) for the RRMM and for the macro-Cauchy when $f=0$ and $f\neq 0$. The tractions shown here are those relative to ``cut" $\Beta$. Analogous reasoning holds true for all other ``cuts".}
    \label{fig:rrmm_macro_tractions_1500}
\end{figure}

%
%
%
\section{``Cut" dependent effectiveness of the metamaterial}\label{sec:transm}
In this section, we will discuss the implications of choosing a different unit cell ``cut" on the microstructured metamaterial's behavior with focus on transmissibility.
In most cases, the frequency region of interest on a transmissibility plot is the band-gap region, where we witness very low values of wave transmission, and therefore this region can be used for vibration isolation.
In the following, we also show that the possibility of constructing the finite-size metamaterial from different unit cell ``cuts" could potentially reveal new frequency regions of interest.
\subsection{Transmissibility}
We will now discuss the transmissibility of the four finite-size specimens.
Transmissibility is defined as the ratio of output to input acceleration (or equivalently displacement, velocity or force) for a given vibration test.
\rev{In our case, we choose the absolute value of the acceleration at each point on the right Cauchy plate as the measure of output, and similarly the input is the absolute value of the acceleration at each point on the left Cauchy plate where the excitation force is applied. 
To have a single value of transmissibility for each frequency (instead of a point-wise comparison), we take their averages on each plate and call Transmissibility the ratio}
\begin{equation}
    \text{Transmissibility}=\rev{\frac{\avg\abs{\ddot{u}_{\rm{output}}}}{\avg\abs{\ddot{u}_{\rm{input}}}}}.
\end{equation}
where $\avg$ is an average operator that calculates an average on the right plate for the output and on the left plate for the input.

\begin{figure}[h!]
\centering
\includegraphics[width=0.60\textwidth]{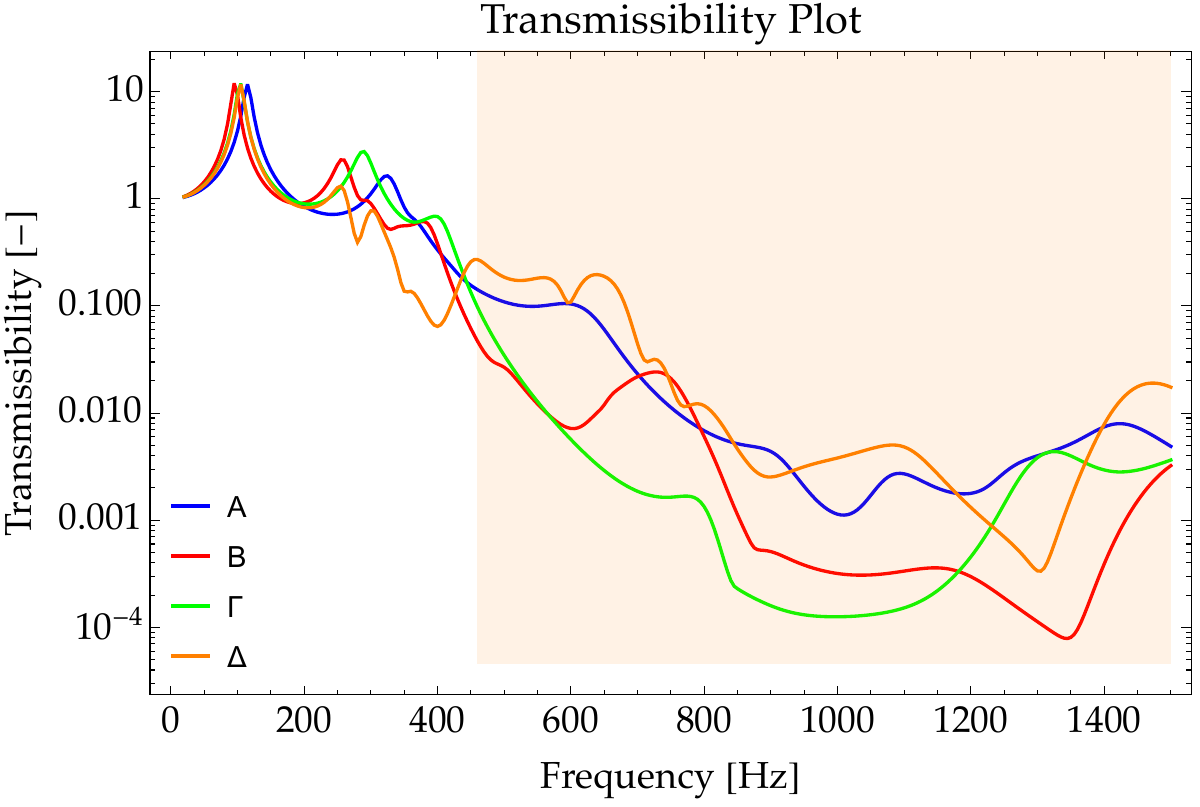}
\caption{Transmissibility plot for the four microstructured specimens. The $y-$axis is presented in a logarithmic scale and the band-gap range is indicated with light orange color.}
\label{fig:transm_plot}
\end{figure}
The transmissibility plot of all the four microstructured metamaterials, for the pressure test presented in section~\ref{Sec:FE_S} can be seen in Fig.~\ref{fig:transm_plot} where we can observe vast differences in the transmissibility values in the band-gap region between the four specimens. 
This emphasises again that even if the four ``cuts" produce the same infinitely big metamaterial, given the finite size of real applications, the choice of unit cell ``cut'' becomes of paramount importance.

By choosing the four different ``cuts" for constructing our finite size specimen, we end up with 4 different metamaterial's specimens of finite size. 
One could believe that the only common thing between the four periodic structures, is the fact that that they correspond to the same dispersion curves (in an infinitely big domain) and thus they should have the same ``vibrational characteristics''. 
However, due to the finite size of these structures, we end up with a different geometry on the boundaries of each specimen, which gives us vastly different boundary effects.
Furthermore, different structures even of the same size can possess different eigenfrequencies of vibration, leading to different results for identical frequencies. 
All these factors lead to the difference in the transmissibility of the four specimens in Fig.\ \ref{fig:transm_plot}.


\subsection{Reduced transmissibility in a non-band-gap frequency range}
\begin{figure}[h!]
\centering
\includegraphics[width=0.60\textwidth]{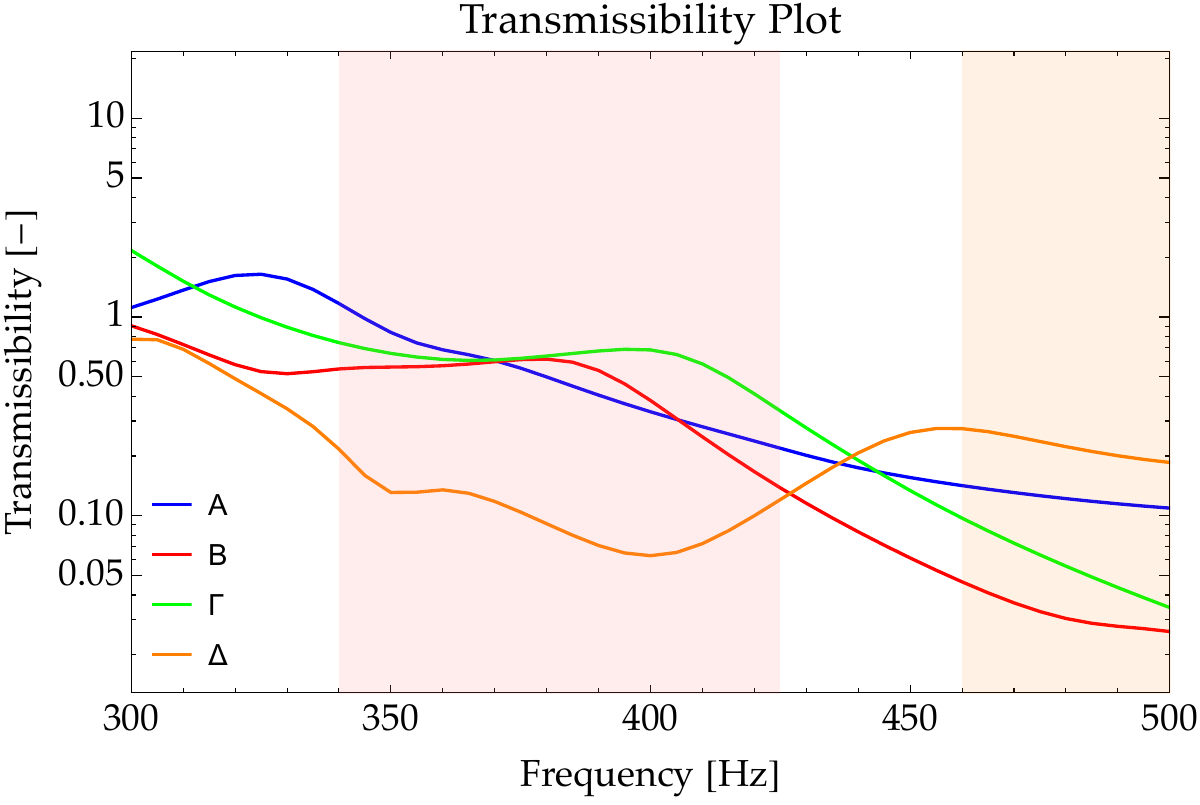}
\caption{Zoom in the Transmissibility plot: the frequency region of interest is indicated with light red color.}
\label{fig:transm_plot_zoom}
\end{figure}

\begin{figure}[h!]
\centering
\includegraphics[width=0.9\textwidth]{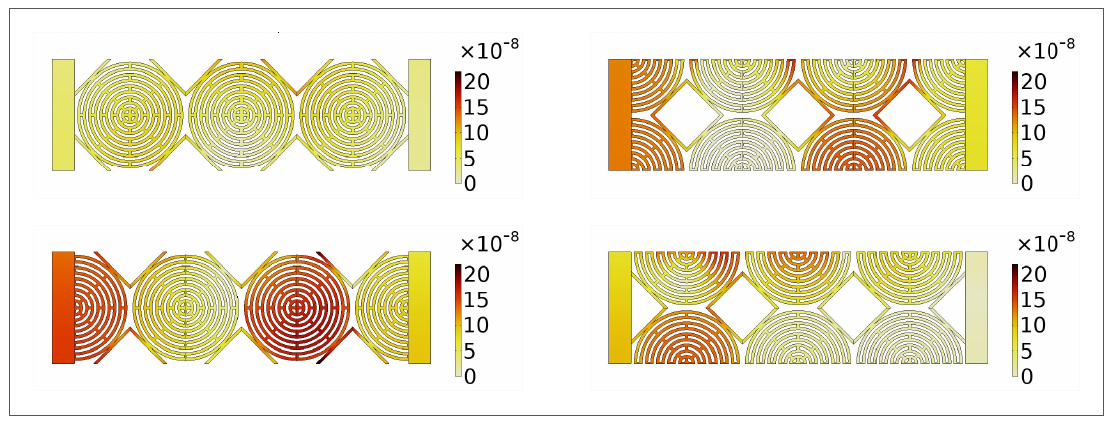}
\caption{Displacement field at 350 Hz for the four specimens. Specimen $\Alpha$ (top left), $\Beta$ (top right), $\Gamma$ (bottom left) and $\Delta$ (bottom right).}
\label{fig:350_disp}
\end{figure}

The transmissibility plot (Fig.~\ref{fig:transm_plot}) reveals a small region (340-425 Hz), for which only cut $\Delta$ has a reduced transmissibility, and this region does not fall in the frequency region of the band-gap.
Specifically, the transmissibility values for cut $\Delta$ in this region are around 10 \%.
These transmissibility values are not comparable to the band-gap, but still the attenuation is big enough to be used for potential shielding applications.
A zoom in the transmissibility plot in Fig.\ \ref{fig:transm_plot_zoom} shows better the discussed absorption property and the displacement field for an indicative frequency of 350 Hz in this region can be seen in Fig.~\ref{fig:350_disp}.

 
\subsection{Independent tests using bigger specimens: limits of Bloch-Floquet analysis}\label{sec:size}
Since the analyzed 3x2 specimens have vastly different displacement field solutions, we will now also increase the size of the structures in steps, to understand how much it needs to be increased in order for the four specimens to have a similar displacement field, i.e.\ what is the size of the specimens to consider boundary effects (and thus the need for interface forces in the RRM setting) as negligible. 
We proceed with this task but we change the excitation to a displacement instead of a force so that we are able to compare the convergence of the 4 different cases towards one big metamaterial sample.
We present here the results for specimens of size 3x2, 9x6, 15x10, 30x20 and 45x30 for the frequency of 200 Hz and all the four ``cuts".
\subsubsection{3x2 cells, $L_{\rm specimen}$=15 cm, wavelength $\lambda$=35 cm}
\begin{figure}[H]
\centering
\includegraphics[width=0.9\textwidth]{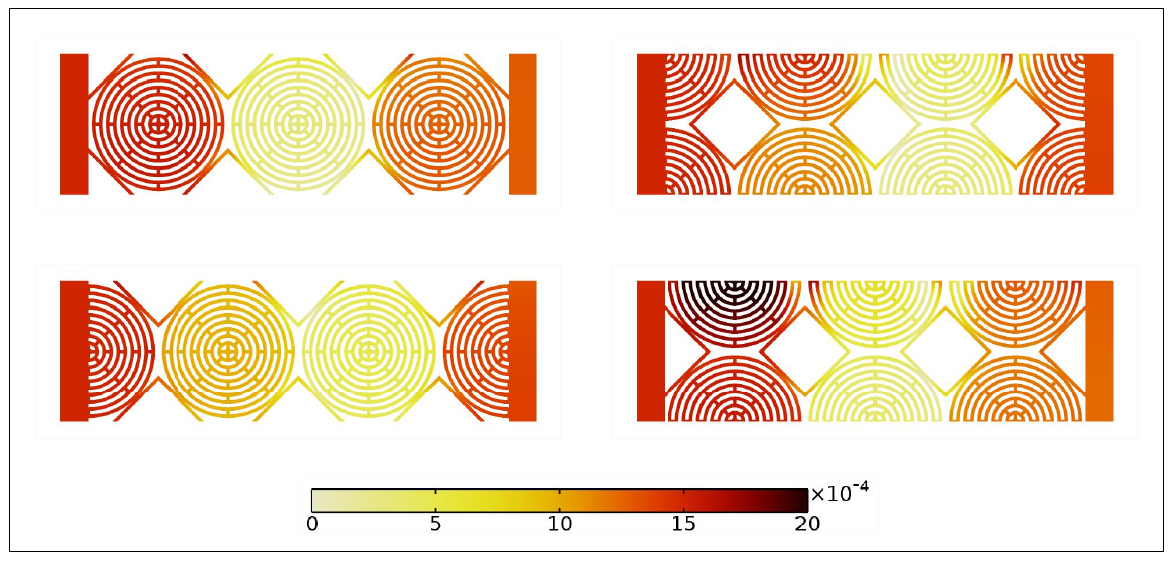}
\caption{Displacement field for the four original 3x2 specimens at the frequency of 200 Hz. Specimen made out of cut 
$\Alpha$ (Top Left), cut $\Beta$ (Top Right), cut $\Gamma$ (Bottom Left) and cut $\Delta$ (Bottom Right).}
\label{fig:3x2_disp_test}
\end{figure}

\subsubsection{15x10 cells, $L_{\rm specimen}$=75 cm, wavelength $\lambda$=35 cm}
\begin{figure}[H]
\centering
\includegraphics[width=0.90\textwidth]{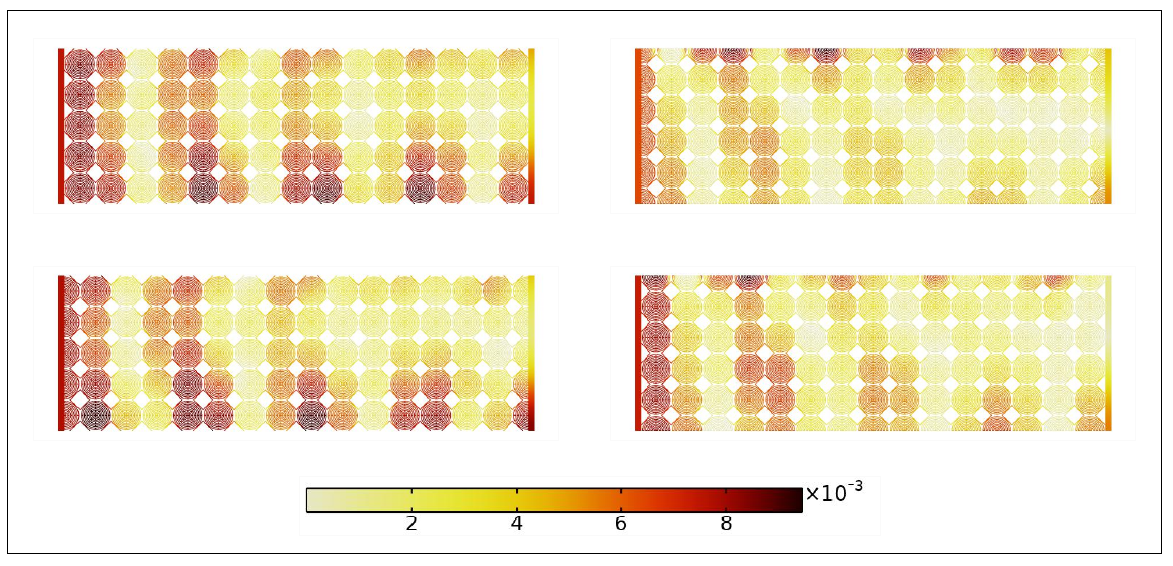}
\caption{Displacement field for the four 15x10 specimens at the frequency of 200 Hz. Specimen made out of cut 
$\Alpha$ (Top Left), cut $\Beta$ (Top Right), cut $\Gamma$ (Bottom Left) and cut $\Delta$ (Bottom Right).}
\label{fig:15x10_disp_test}
\end{figure}
Even with a sample consisting of 150 unit-cells overall, boundary effects still produce significant differences between the different ``cuts" that are prominent also in the bulk of the material.
This means that the specimen with an overall length of 75 cm is still too small compared to the wavelength 35 cm. 

\subsubsection{45x30 cells, $L_{\rm specimen}$=225 cm, wavelength $\lambda$=35 cm}
\begin{figure}[H]
\centering
\includegraphics[width=0.90\textwidth]{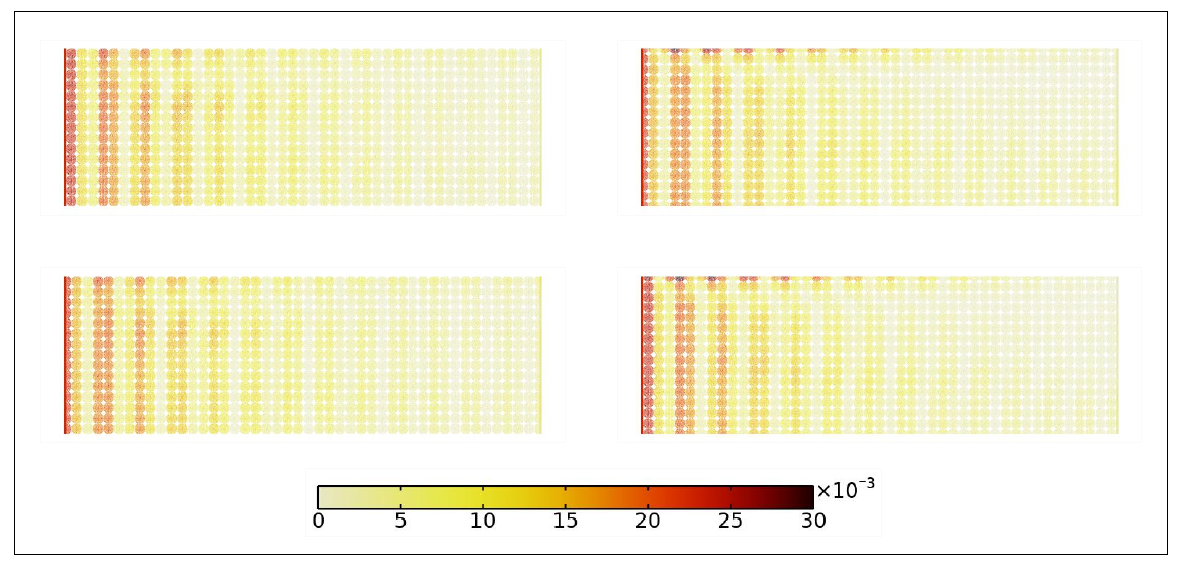}
\caption{Displacement field for the four 45x30 specimens at the frequency of 200 Hz. 
Specimen made out of cut  $\Alpha$ (Top Left), cut $\Beta$ (Top Right), cut $\Gamma$ (Bottom Left) and cut $\Delta$ (Bottom Right).}
\label{fig:45x30_disp_test}
\end{figure}
For our biggest samples consisting of 1350 unit cells, all four solutions converge regarding bulk response. 
However, some boundary effects are still visible.
\subsubsection{Summary}
The main assumption of Bloch-Floquet analysis is that the material is extended to infinity by periodic boundaries, i.e.\ the actual size of the specimen is infinitely big, and thus the metamaterial has no boundaries.
However, the dispersion curves coming from Bloch-Floquet analysis are inevitably used as a design tool in finite size metamaterial applications, often disregarding boundary effects that we showed can become predominant as soon as reducing the specimen's size to finite-sized problems.
Thus the question arises, what is the minimum size of a finite sized metamaterial so that we are allowed to neglect boundary effects?

Usually, the answer given is that a very big number of unit cells must be used to approximate well a metamaterial extended to infinity often unfeasible for finite-sized applications..
By using more unit cells, the length of the boundary of the metamaterial scales linearly while the area of the bulk scales quadratically, and thus the ratio of boundary to bulk tends to zero.

Another more interesting answer coming from the results in this section is the following:
Boundary effects can be neglected if the size of the specimen is ``big enough", i.e.\ bigger than a certain threshold, so that finite-sized specimens constructed from different cell's cuts show no qualitative difference in their behavior. 
This implies that, the RRM modeling of specimens that are bigger of this threshold would not need any more non-vanishing interface forces to provide the correct solution.
From the above results, we can see that even for a 45x30 specimen, there are still some different boundary behaviors for the different cuts. 
Therefore, we have not yet found the appropriate size for which boundary effects can be fully neglected.
At this size, our enriched continuum would not need surface forces to reproduce the different response associated to the four different cell's cuts.
On the other hand, it can be noticed that the ``bulk" response already becomes very similar in the 4 specimen's types as soon as the number of unit cells is increased.
%
%
%
\section{Conclusions}
In this paper, we have demonstrated that introducing the concept of interface forces is essential for modeling the response of finite-size metamaterials within a homogenized framework. 
By leveraging the RRMM which has proved excellent performance in describing the bulk behavior of metamaterials, we enhanced the model by incorporating the concept of interface forces that must be considered at the interfaces of all metamaterials. 

We elucidated the significance of this new concept through an in-depth analysis of a benchmark compression/extension test as well as a shear test. We showed that for the considered benchmark case the interface forces should exhibit the form
\begin{equation}
    f^{\rm interface}=(\alpha-1)\.t_{(RRMM)} + \beta.
\end{equation}
where $t_{\rm RRMM}=\left(\widetilde{\sigma} + \widehat{\sigma} \right) n$, is the RRM traction on the considered interface, $\alpha$ is a dimensionless parameter and $\beta$ is a surface force.

Our findings explicitly indicate that boundary effects and, consequently, interface forces can significantly impact finite-sized specimens. This paper conclusively establishes that homogenization schemes that do not account for the interface forces arising at the "homogenized" interfaces may result in substantial errors when these models are applied to model real structures which are inherently finite in size.


\small
\subsubsection*{Acknowledgements}
Angela Madeo, Jendrik Voss and Plastiras Demetriou acknowledge support from the European Commission through the funding of the ERC Consolidator Grant META-LEGO, N$^\circ$ 101001759.

\printbibliography

\appendix
\section{Boundary conditions on a symmetry plane for a relaxed micromorphic medium using Curie's Symmetry Principle}\label{symmetry}

Similarly to the work in \cite{demore2022unfolding} and \cite{ramirez2024effective}, we shall use Curie's Symmetry Principle to define the symmetry conditions of the reduced relaxed micromorphic model.
We suppose that our problem has a symmetry with respect to the plane $\mathcal {N}$ of normal $n\in\mathbb{R}^3$, we shall apply Curie's Symmetry Principle on our kinematical fields $u$ and $P$
\begin{align}
\begin{cases}
    u(x^{\star}) = u^{\star}(x)\,, \\
    P(x^{\star}) = P^{\star}(x)
    \label{curie}
\end{cases}
\end{align}
with $x^{\star}$ being the symmetric coordinate of $x$ with respect to $\mathcal {N}$. Let's define the corresponding orthonormal bases $\{t_{\alpha},t_{\beta},n\}$ and $\{t_{\alpha}^{*},t_{\beta}^{*},n^{*}\}$ where for every base we define two tangent vectors and one normal vector with respect to the symmetry plane $\mathcal {N}$.
Since $\mathcal {N}$ is a symmetry plane, it holds
\begin{equation}\label{equiv}
    t_{\alpha}^\star = t_{\alpha}\,,\quad t_{\beta}^\star = t_{\beta}\quad\text{and}\quad n^\star = -n\,.
\end{equation}
Let's express $u$ and $P$ in their corresponding base
\begin{align}
    u &= u_{\alpha}t_{\alpha}+u_{\beta}t_{\beta}+u_nn\,,\\
    u^\star &= u_{\alpha}t_{\alpha}^\star+u_{\beta}t_{\beta}^\star+u_nn^\star\label{eq:ustar1}
\end{align}
and
\begin{align}
    P &= P_{\alpha\alpha} t_{\alpha} \otimes t_{\alpha}+
    P_{\alpha\beta} t_{\alpha} \otimes t_{\beta}+
    P_{\alpha n} t_{\alpha} \otimes n+
    P_{\beta\alpha} t_{\beta} \otimes t_{\alpha}+
    P_{\beta\beta} t_{\beta} \otimes t_{\beta}+
    P_{\beta n} t_{\beta} \otimes n\\
    &\phantom=\;+
    P_{n\alpha} n \otimes t_{\alpha}+
    P_{n\beta} n \otimes t_{\beta}+
    P_{nn} n \otimes n\,,\notag\\
    P^\star &= P_{\alpha\alpha} t_{\alpha}^\star \otimes t_{\alpha}^\star+
    P_{\alpha\beta} t_{\alpha}^\star \otimes t_{\beta}^\star+
    P_{\alpha n} t_{\alpha}^\star \otimes {n}^\star+
    P_{\beta\alpha} t_{\beta}^\star \otimes t_{\alpha}^\star+
    P_{\beta\beta} t_{\beta}^\star \otimes t_{\beta}^\star+
    P_{\beta n} t_{\beta}^\star \otimes {n}^\star\label{eq:pstar1}\\
    &\phantom=\;+
    P_{n\alpha} {n}^\star \otimes t_{\alpha}^\star+
    P_{n\beta} {n}^\star \otimes t_{\beta}^\star+
    P_{nn} {n}^\star \otimes {n}^\star\notag
\end{align}
where the components of the displacement $u_{i}$ and the microdistortion tensor $P_{ij}$ are the same in both bases.
From Eq.\ (\ref{equiv}) it is apparent that we can rewrite equations (\ref{eq:ustar1}) and (\ref{eq:pstar1}) in the form
\begin{align}
    \label{ustar2}
    u^\star &= u_{\alpha}t_{\alpha}+u_{\beta}t_{\beta}-u_nn\,,\\
    \label{pstar2}
    P^\star &= P_{\alpha\alpha} t_{\alpha} \otimes t_{\alpha}+
    P_{\alpha\beta} t_{\alpha} \otimes t_{\beta}-
    P_{\alpha n} t_{\alpha} \otimes n+
    P_{\beta\alpha} t_{\beta} \otimes t_{\alpha}+
    P_{\beta\beta} t_{\beta} \otimes t_{\beta}-
    P_{\beta n} t_{\beta} \otimes n\\
    &\phantom=\; -P_{n\alpha} n \otimes t_{\alpha}-
    P_{n\beta} n \otimes t_{\beta}+
    P_{nn} {n} \otimes {n}\,.\notag
\end{align}
We then define the points $x$ and ${x}^*$ as $x=x_{0}+\epsilon n$ and ${x}^*=x_{0}+\epsilon{n}^*$ where $x_{0} \in \mathcal {N}$ and $\epsilon \in \mathbb{R}$. Points $x$ and ${x}^*$ represent symmetric points with respect to plane $\mathcal {N}$ and therefore it holds
\begin{equation}
    x^{\star} = x_0 - \epsilon  n \,.
\end{equation}
For the kinematical field $u$, substituting in (\ref{curie}), one can get
\begin{equation}
    u(x^{\star}) = u_{\alpha}(x_0-\epsilon n)\.t_{\alpha} + u_{\beta}(x_0-\epsilon n)\.t_{\beta} + u_{n}(x_0-\epsilon n)\.n = u_{\alpha}(x_0+\epsilon n)\.t_{\alpha} + u_{\beta}(x_0+\epsilon n)\.t_{\beta} -u_n(x_0+\epsilon n)\.n = u^{\star}(x)\,.
\end{equation}
By identification for each expression in $\{t_{\alpha},t_{\beta},n\}$ we have
\begin{align}
\begin{cases}
u_{\alpha}(x_0-\epsilon n)\.t_{\alpha}  =  u_{\alpha}(x_0+\epsilon n)\.t_{\alpha}\,, \\
u_{\beta}(x_0-\epsilon n)\.t_{\beta}  = u_{\beta}(x_0+\epsilon n)\.t_{\beta}\,, \\
u_{n}(x_0-\epsilon n)\.n  =  -u_{n}(x_0+\epsilon n)\.n\,.
\label{urec1}
\end{cases}
\end{align}
In the same way, for the kinematical field $P$ we arrive at
\begin{align}
\begin{cases}
P_{\alpha\alpha}(x_0-\epsilon n)\.t_{\alpha} \otimes t_{\alpha}  =  P_{\alpha\alpha}(x_0+\epsilon n)\.t_{\alpha} \otimes t_{\alpha}\,, \\
P_{\alpha\beta}(x_0-\epsilon n)\.t_{\alpha} \otimes t_{\beta}  =  P_{\alpha\beta}(x_0+\epsilon n)\.t_{\alpha} \otimes t_{\beta}\,, \\
P_{\alpha n}(x_0-\epsilon n)\.t_{\alpha} \otimes n =  -P_{\alpha n}(x_0+\epsilon n)\.t_{\alpha} \otimes n\,, \\
P_{\beta\alpha}(x_0-\epsilon n)\.t_{\beta} \otimes t_{\alpha} =  P_{\beta\alpha}(x_0+\epsilon n)\.t_{\beta} \otimes t_{\alpha}\,,\\
P_{\beta\beta}(x_0-\epsilon n)\.t_{\beta} \otimes t_{\beta}  =  P_{\beta\beta}(x_0+\epsilon n)\.t_{\beta} \otimes t_{\beta}\,, \\
P_{\beta n}(x_0-\epsilon n)\.t_{\beta} \otimes n =  -P_{\beta n}(x_0+\epsilon n)\.t_{\beta} \otimes n\,, \\
P_{ n\alpha}(x_0-\epsilon n)\.n \otimes t_{\alpha}  =  -P_{n\alpha}(x_0+\epsilon n)\.n \otimes t_{\alpha}\,, \\
P_{ n\beta}(x_0-\epsilon n)\.n \otimes t_{\beta} =  -P_{n\beta}(x_0+\epsilon n)\.n \otimes t_{\beta}\,, \\
P_{n n}(x_0-\epsilon n)\.n \otimes n =  P_{n n}(x_0+\epsilon n)\.n \otimes n\,.
\label{prec1}
\end{cases}
\end{align}
These conditions allow to reconstruct the displacement and microdistorsion fields at any point in space with respect to the symmetry plane, when their value is known on the opposite side of the symmetry plane. In the limit of $\epsilon \to 0$, equations \eqref{urec1} and \eqref{prec1} simplify to
\begin{align}
\begin{cases}
u_{\alpha}(x_0)\.t_{\alpha}  =  u_{\alpha}(x_0)\.t_{\alpha}\,, \\
u_{\beta}(x_0)\.t_{\beta}  = u_{\beta}(x_0)\.t_{\beta}\,, \\
u_{n}(x_0)\.n  =  -u_{n}(x_0)\.n
\end{cases}
\qquad\text{and}\qquad
\begin{cases}
P_{\alpha\alpha}(x_0)t_{\alpha} \otimes t_{\alpha}  =  P_{\alpha\alpha}(x_0)t_{\alpha} \otimes t_{\alpha}\,, \\
P_{\alpha\beta}(x_0)t_{\alpha} \otimes t_{\beta}  =  P_{\alpha\beta}(x_0)t_{\alpha} \otimes t_{\beta}\,, \\
P_{\alpha n}(x_0) t_{\alpha} \otimes n =  -P_{\alpha n}(x_0)t_{\alpha} \otimes n\,, \\
P_{\beta\alpha}(x_0) t_{\beta} \otimes t_{\alpha} =  P_{\beta\alpha}(x_0) t_{\beta} \otimes t_{\alpha}\,,\\
P_{\beta\beta}(x_0)t_{\beta} \otimes t_{\beta}  =  P_{\beta\beta}(x_0)t_{\beta} \otimes t_{\beta}\,, \\
P_{\beta n}(x_0) t_{\beta} \otimes n =  -P_{\beta n}(x_0)t_{\beta} \otimes n\,, \\
P_{ n\alpha}(x_0)n \otimes t_{\alpha}  =  -P_{n\alpha}(x_0)n \otimes t_{\alpha}\,, \\
P_{ n\beta}(x_0) n \otimes t_{\beta} =  -P_{n\beta}(x_0)n \otimes t_{\beta}\,, \\
P_{n n}(x_0) n \otimes n =  P_{n n}(x_0)n \otimes n\,.
\label{rec}
\end{cases}
\end{align}
Therefore it holds:
$u_{n}=P_{\alpha n}=P_{\beta n}=P_{n\alpha}=P_{n\beta}=0$.
Thus, the following conditions must be satisfied at $\mathcal N$
\begin{align}
\begin{cases}
\langle u , n \rangle = 0\,, \\
\langle P , t_{\alpha} \otimes n \rangle  =  0\,, \\
\langle P , t_{\beta} \otimes n \rangle  =  0\,,\\
\langle P , n \otimes t_{\alpha} \rangle  =  0\,, \\
\langle P , n \otimes t_{\beta} \rangle  =  0\,,
\label{final_cond}
\end{cases}
\qquad\forall x \in \mathcal N\,.
\end{align} 
For our problem where $ u_{3}=P_{13}=P_{23}=P_{31}=P_{32}=P_{33}=0$, $n=\{0,-1,0\}$, $t_{\alpha}=\{1,0,0\}$ and $t_{\beta}=\{0,0,1\}$ the conditions of equation (\ref{final_cond}) states
\begin{align}
\begin{cases}
\langle u , n \rangle = -u_{2}= 0\,, \\
\langle P , t_{\alpha} \otimes n \rangle  =  -P_{12}= 0\,, \\
\langle P , t_{\beta} \otimes n \rangle  =  0\,,\\
\langle P , n \otimes t_{\alpha} \rangle  = -P_{21}= 0\,, \\
\langle P , n \otimes t_{\beta} \rangle  =  0\,,
\label{final_cond2}
\end{cases}
\qquad\forall x \in \mathcal N\,.
\end{align}
where now sub-indices $1,2$ and $3$ refer to coordinates $x,y$ and $z$ respectively, since tangent vector $t_{\alpha}$ points in the $x-$direction, tangent vector $t_{\beta}$ points in the $z-$direction and our normal vector $n$ points in the negative $y-$direction.

\section{Scalability of the interface forces with the applied external load}\label{sec:scalability}
In the last section, we showed that by using a jump of traction we were able to capture well the behavior of our finite size specimen, that is, bulk behavior and boundary effects, in most cases. However, to find the numerical values of the coefficients in each of the expressions for the jump of traction, we always need the corresponding microstructured simulation itself to compare and calibrate the interface forces to be used in the RRM framework that produce the correct solution. 
Using a jump of traction would prove even more useful, if the found interface forces can be scaled according to some physical/mathematical law, when a parameter of our simulation is also scaled. 
Some possible parameters could be the magnitude of the excitation, the direction of the excitation, the number of unit cells in the $x$ or $y$ direction, etc. 

In this section, we will show that such a law exists for the scaling of the magnitude of the excitation.
We repeat that the expression of the interface forces that we used in the compression-like tests of this paper is
\begin{equation}
    f^{\rm interface}_{i_{n}}=(\alpha_{i_{n}}-1)\.t_{({\rm RRMM})_{i_n}} + \beta_{i_n}
\end{equation}
where $\alpha$ and $\beta$ are constants which take different values for different ``cuts" at the boundaries and $t_{\rm RRMM}$ is the value of the RRM traction at the considered interface. 
We will show in the remainder of this section that changing the intensity of the externally applied load scales the interface force introduced in eq.~(\ref{eq:interface_force_a-1}) via a basic relation.
We recall that the external excitation is of the form: $f^{\rm ext}=\overline{F}=10\. \widehat{e_1}+ 0\. \widehat{e_2}$ ]N/m$^2$].
We will show that when scaling the applied external force by a quantity $c\in\mathbb R$, i.e.\ applying $f^{\rm ext}=c\.\overline{F}$, then the interface force scales with the law
\begin{equation}
    f^{\rm interface}_{i_{n}}=(\alpha_{i_{n}}-1)\.t_{({\rm RRMM})_{i_n}} + c\.\beta_{i_n}
    \label{eq:scaling_law}
\end{equation}
The scaling law for the interface forces remains the same notwithstanding the cut ($\Alpha, \Beta, \Gamma, \Delta$), or the frequency which is considered.
This result is of paramount importance  in view of the use of the RRMM to optimize acoustic metastructural design. 
Indeed, it is sufficient to run only the 4 simulations for the RRMM with a given applied external load once to calibrate the constants $\alpha$ and $\beta$ appearing in the interface force $f^{\rm interface}$.
Then the RRMM simulations corresponding to the 4 different cases can be scaled with the scalability law \eqref{eq:scaling_law} alone without additional comparison to the microstructured simulations when changing the intensity of the externally applied load.
It is clear that this brings enormous design simplification in terms of computational time.

In the following three subsections, we show some explicit numerical results concerning the scalability law for externally applied load for some specific frequencies \rev{(We will only use the $\Beta$ Cut here, but the same is true for every cut).}
However, as already stated, this scalability law \eqref{eq:scaling_law} does not depend either on the ``cut" or the frequency.
It is clear that, when using a linear-elastic model, scaling the load will scale the solution of a certain amount.
However, this does not mean that the result found here about the ``scalability" of the introduced interface forces is a trivial result.
Indeed, this is an independent confirmation that the interface forces ansatz~(\ref{eq:scaling_law}) is the correct one to be used for problems of this type presented in this paper \rev{because it maintains the linear scaling of the magnitude of the excitation}.
The scalability of the solution is a true advancement when one is interested to study the transmissibility of the considered specimens under arbitrary loading even if being more or less intense than the given reference load used to fit the jump of traction initially.
\subsection{``Cut" $\Beta$ at 140 Hz with an excitation of scaled magnitude}\label{sec:140_beta_scale}
\begin{figure}[h!]
\centering
\includegraphics[width=0.9\textwidth]{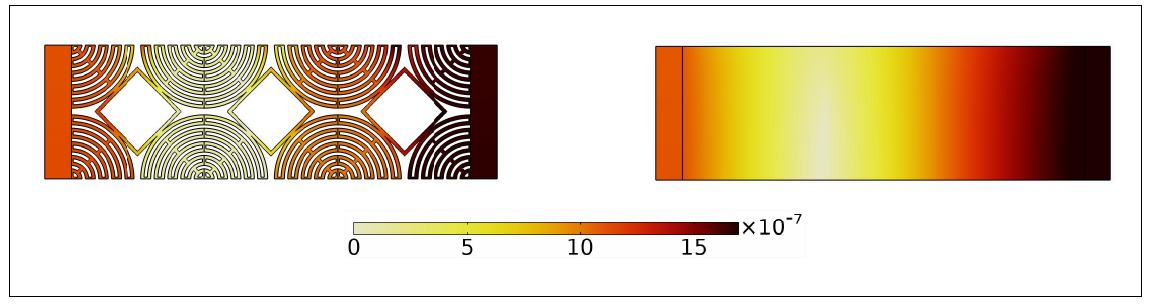}
\caption{Displacement field for the Microstructured (left) and for the RRMM with $\alpha_{L_x}=0.8, \alpha_{R_x}=0.75, \alpha_{L_y}=1, \alpha_{R_y}=1$ and $\beta_{L_x}=\beta_{R_x}= \beta_{L_y}=\beta_{R_y}=0$ (right). The magnitude of the excitation is \textbf{three times} the one in Figure \ref{fig:disp140_beta}.}
\label{fig:disp140betaFx3}
\end{figure}

We recall that according to Fig.~\ref{fig:disp140_beta}, a jump of traction is needed to capture the behavior of our metamaterial constructed from ``Cut" $\Beta$ already for the low frequency of 140 Hz. We recall once again the values of these coefficients here:
\[
    \alpha_{L_x}=0.8\,,\quad\beta_{L_x}=0\,,\quad\alpha_{L_y}=1\,,\quad\beta_{L_y}=0\qquad\text{and}\qquad\alpha_{R_x}=0.75\,,\quad\beta_{R_x}=0\,,\quad\alpha_{R_y}=1\,,\quad\beta_{R_y}=0\,.
\]
From the values of these coefficients it is apparent that, for both the left and right interfaces, a jump of traction is only needed in the $x-$direction. 
Here, we run the same simulation as the one corresponding to Fig.~\ref{fig:disp140_beta} with the only difference being that the magnitude of our excitation is now scaled. 
We then search again for those coefficients that will give us the correct solution.
The results can be seen in Fig.~\ref{fig:disp140betaFx3} for an excitation magnitude scaled three times.

The value of the interface force is given by $f^{int}_{i_{n}}=(\alpha-1)\.t_{{\rm RRMM}_{i_n}} + \beta_{i_n}$ and since the calibrated force for this case is $\beta_{i_{n}}=0$, it follows the scaling law $\beta\to c\.\beta$ in this specific case of $\beta=0$. Thus the interface force remains the same when scaling the externally applied load.
 
\subsection{``Cut" $\Beta$ at 700 Hz with an excitation of scaled magnitude}\label{sec:700_beta_scale}
\begin{figure}[h!]
\centering
\includegraphics[width=0.9\textwidth]{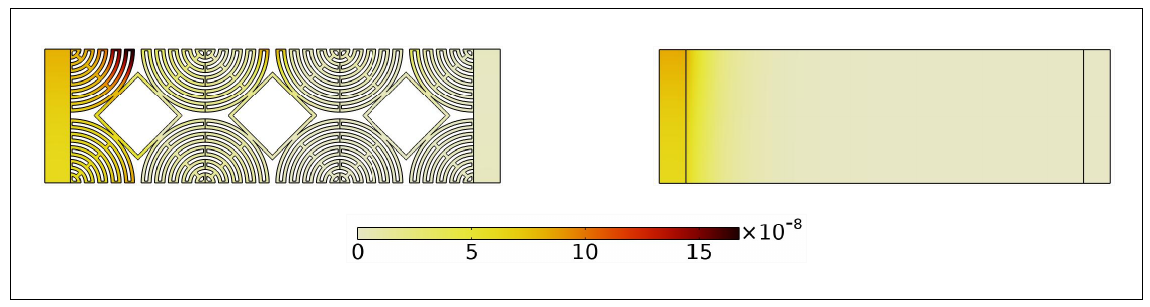}
\caption{Displacement field for the Microstructured (left) and for the RRMM with $\alpha_{L_x}=1.6, \alpha_{R_x}=1, \alpha_{L_y}=1, \alpha_{R_y}=1$ and $\beta_{L_x}=\beta_{R_x}=0, \beta_{L_y}=-6, \beta_{R_y}=0$ (right). The magnitude of the excitation is \textbf{three times} the one in Figure \ref{fig:disp700_beta}.}
\label{fig:disp700betaFx3}
\end{figure}

We now investigate what happens for the same ``Cut" $\Beta$ but for the frequency of 700 Hz which is in the lower part of the band-gap. According to Fig.~\ref{fig:disp700_beta}, a jump of traction is needed to capture the behavior of our metamaterial and we recall here, once again, the values of the coefficients used:
\[
    \alpha_{L_x}=1.6\,,\quad\beta_{L_x}=0\,,\quad\alpha_{L_y}=1\,,\quad \beta_{L_y}=-2\qquad\text{and}\qquad\alpha_{R_x}=1\,,\quad\beta_{R_x}=0\,,\quad\alpha_{R_y}=1\,,\quad\beta_{R_y}=0\,.
\]
From the values of these coefficients it is apparent that a jump of traction is only needed on the left interface for both directions $x$ and $y$. 
Here, we run the same simulation as the one in corresponding to Fig.~\ref{fig:disp700_beta} with the only difference being that the magnitude of our excitation is scaled. We then search for those coefficients that will give us the correct solution.
The result can be seen in Figure \ref{fig:disp700betaFx3} for an excitation magnitude scaled three times.

We use coefficients $\alpha_{L_x}$ and $\beta_{L_y}$ for implementing the jump of traction (the other $\beta_{i_{n}}$ are still zero), and in this case, we can see that according to the results, if the magnitude of the excitation is scaled by $c$ then the value of $\alpha_{L_x}$ remains unchanged while the value of $\beta_{L_y}$ must be multiplied by the scaling factor of the magnitude of the excitation $c$ as well. 
This means that the complete scalability law for ``cut" $\Beta$ also holds according to eq.\ \eqref{eq:scaling_law}.
\subsection{Scalability law for a shear test}
\begin{figure}[h!]
\centering
\includegraphics[width=0.9\textwidth]{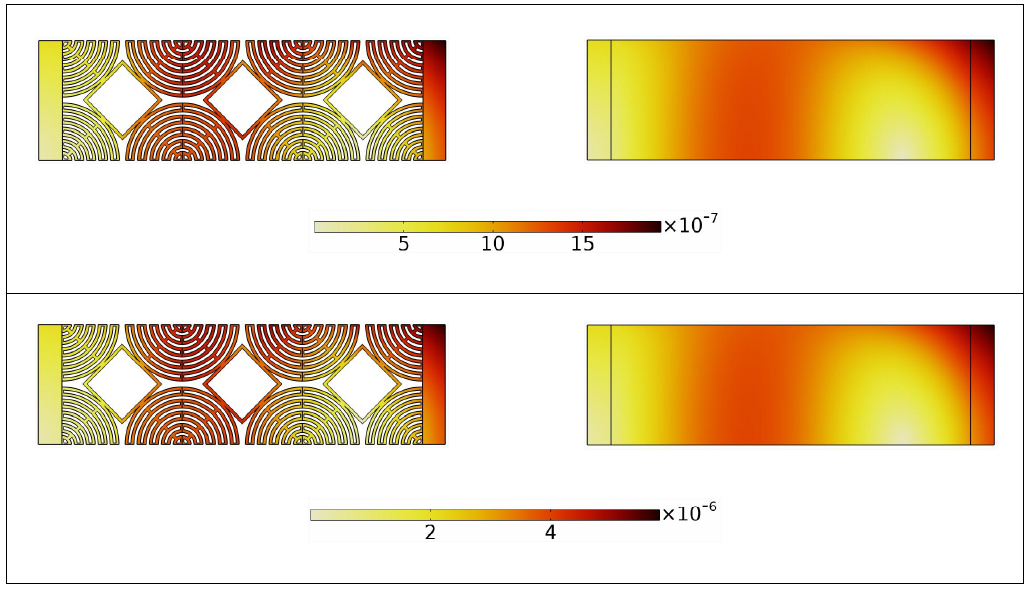}

\caption{Top: Displacement field for the Microstructured (left) and for the RRMM with $\alpha_{L_x}=1, \alpha_{R_x}=1, \alpha_{L_y}=1, \alpha_{R_y}=1$ and $\beta_{L_x}=-14,\beta_{R_x}=9, \beta_{L_y}=0, \beta_{R_y}=0$ (right).
Bottom: Displacement field for the Microstructured (left) and for the RRMM with $\alpha_{L_x}=1, \alpha_{R_x}=1, \alpha_{L_y}=1, \alpha_{R_y}=1$ and $\beta_{L_x}=-42,\beta_{R_x}=27, \beta_{L_y}=0, \beta_{R_y}=0$ (right). The magnitude of the excitation on the bottom is \textbf{three times} the one at the top.}
\label{fig:disp80betashearFx}
\end{figure}

We now change the direction of our excitation force by 90 degrees, so that the test represents a shear test, in order to show that the scalability law also holds in this case. 
Thus our excitation force is now given by: $\overline{F}=0\. \widehat{e_1}+ 10\. \widehat{e_2}$ [N/m$^2$].
Interface forces are used to capture the response of specimen $\Beta$ at 80 Hz and then the magnitude of the excitation is scaled by a factor of 3 (see Fig.~\ref{fig:disp80betashearFx}).

We thus observe that the same law as in eq.~(\ref{eq:scaling_law}) holds. 




\end{document}